\documentclass[fleqn]{article}
\usepackage[english]{babel}
\usepackage{graphicx}
\usepackage{amsfonts}
\usepackage{mathtext}
\usepackage[left=1.5cm, right=1cm]{geometry}
\usepackage{chngpage}
\usepackage{amsmath}
\usepackage{float}
\usepackage{hyperref}
\usepackage{physics}
\usepackage{gensymb}
\usepackage{bm}
\usepackage[absolute,overlay]{textpos}
\usepackage{xcolor}
\usepackage{multirow}
\usepackage{amssymb}
\usepackage[title]{appendix}
\usepackage{times}
\usepackage{epsfig}
\usepackage{dcolumn}
\usepackage{booktabs}
\usepackage{revsymb}
\usepackage{mathrsfs}
\usepackage{authblk}

	\title{Line Profile Asymmetry in Precision Spectroscopy}
	\author[1]{A. Anikin}
	\author[1,2]{T. Zalialiutdinov \thanks{Corresponding Author: zalialiutdinov@gmail.com}}
	\author[1]{D. Solovyev}
	\author[1,2]{L. Labzowsky}

\affil[1]{Department of Physics, St. Petersburg State University, St. Petersburg, 198504, Russia}
\affil[2]{Petersburg Nuclear Physics Institute named by B.P.
Konstantinov of National Research Centre "Kurchatov Institut", St. Petersburg, Gatchina 188300, Russia}
\date{}

\begin{document}	
	\maketitle
	\tableofcontents
	
	\newpage
	\section{Introduction}
	\label{intro}

    The problem of natural line profile in atomic physics was introduced in the context of quantum mechanics by Weisskopf and Wigner \cite{Weisskopf1930}. With the development of relativistic quantum field theory, it was first formulated in the framework of quantum electrodynamic theory (QED) and the S-matrix approach for one-electron atoms in the pivotal paper of F. Low \cite{Low}. Later, the QED theory of the line profile was also modified for atoms with many electrons \cite{lab1983} and applied to the overlapping resonances in highly charged ions with two electrons \cite{Karasiev_1992}. 
    The combination of these approaches with the corresponding development of the methods initiated in \cite{gellmanlow,sucher} has been successfully applied to theoretical calculations of radiative QED corrections to energy levels and transition rates in atoms and ions \cite{ANDREEV2008135,ZALIALIUTDINOV20181}.
	
	One of the most important consequences of line profile theory is the emergence of nonresonant (NR) corrections. For the first time, nonresonant (NR) corrections were introduced by F. Low within the S-matrix formalism \cite{Low}. Later the calculation of NR corrections were performed within the Line Profile Approach (LPA) for various atoms and ions, see \cite{ANDREEV2008135} and references therein. It was shown that the effects leading to asymmetry of the line profile set the limit to which the concept of energy has a physical meaning for the excited atomic states \cite{Labzowsky1994,2001}. This limit corresponds to the resonance approximation. If the distortion of the observed line profile is small, one can still formally consider the NR correction as contributing to the energy shift \cite{ANDREEV2008135}. Unlike all other energy corrections, the NR corrections depend on the specific process used to measure the energy difference. 
	
	The NR effects have attracted new interest in the last decade and have been discussed in a number of theoretical papers \cite{doi:10.1139/p02-094,2001,Labzowsky_1993,Jent-NR,PhysRevA.65.054502,Labzowsky_2004}. This was primarily due to advances in high-precision spectroscopy experiments \cite{PhysRevLett.84.5496,1s2sprecise,PhysRevLett.86.5679}. In particular, the experiment \cite{1s2sprecise} to measure the $1s-2s$ transition frequency is the most accurate ever performed in the optical domain, with a resulting uncertainty of about 10 Hz. Evaluation of the corresponding NR corrections within the QED theory and the LPA approach was performed in \cite{PhysRevA.65.054502,doi:10.1139/p07-014}. Later, an important experimental result was also obtained in \cite{PhysRevLett.86.5679}, where the $1s-2p$ Ly$_{\alpha}$ transition was measured. The corresponding evaluations for the NR corrections to the $1s-2p$ transition frequencies were considered in \cite{doi:10.1139/p02-094,Jent-NR,2001,PhysRevA.65.054502,Labzowsky_2004}. Later, similar calculations were performed taking into account the interference between neighboring hyperfine components of the $2p$ level \cite{PhysRevLett.98.203003}. 
	
    The ongoing development of theory and experiment leads to new challenges. One of the most striking examples is the well-known "proton radius puzzle", which arose as a result of  spectroscopic experiments on muonic hydrogen \cite{Pohl}. 
	The first success in solving this problem was achieved in \cite{Science}, where the asymmetry of the observed $2s-4p$ line profile in hydrogen was taken into account and has brought the proton charge radius almost into agreement with the muonic hydrogen value. Later experiments on proton-electron scattering and measurements of the Lamb shift also approached these results \cite{Cui2021,Bezginov}. 
	
	Although indispensable progress has been achieved in \cite{Science} by taking into account nonresonant effects in the scattering cross section, and quantum interference as their most significant part, the question of its influence on other spectroscopic precision experiments is currently being discussed in \cite{QIEpulses,https://doi.org/10.1002/andp.201900044,Matveev_2019, atoms5040048}. In particular, the results of recent work on the measurement of the $2s-8d$ transition energy in hydrogen again point to a discrepancy in proton charge radius \cite{matveevPRLnew}. Similar problems remain in two-photon spectroscopy when measuring the $1s-3s$ interval \cite{Fleurbaey,ediss27002}. The influence of quantum interference effects (QIE) on two-photon frequency spectroscopy of the $1s-3s$ transition in hydrogen was studied in \cite{PhysRevA.90.012512,PhysRevA.103.022833}. A similar analysis was also performed for one-photon spectroscopy in \cite{https://doi.org/10.1002/andp.201900044} and its application to measurements of the Lamb shift and helium triplet fine structure. 
	A detailed analysis of QIE as part of NR effects can also be addressed to \cite{PhysRevA.91.062506,PhysRevA.97.022510,PhysRevA.92.022514}. 
	In particular, in \cite{PhysRevA.92.022514} it was shown that these effects were either negligible or far below the level of experimental accuracy at the time. Later, the analysis of the asymmetry of line profiles performed in \cite{PhysRevA.92.062506} showed that there are "magic angles" at which quantum interference vanishes (see also \cite{PhysRevLett.107.023001,PhysRevA.87.032504}). Recently, the quantum interference effect was also considered in \cite{PhysRevA.97.022510} for spectroscopy of lithium-like HCI. 
	
	Concerning the spectroscopy of many-electron systems, the light two-electron atomic systems are worth highlighting. The recent accuracy in measuring the transition frequencies between the energy levels of the helium atom gave impetus to the study of QED corrections up to the order of $m\alpha^7$ \cite{PhysRevA.101.062516,PhysRevA.103.042809}. As a result of a comparative analysis of the theoretical and experimental values, a significant discrepancy between the theoretical and experimental values for the $2^3S_1-3^3D_1$ transition frequency was found in \cite{PhysRevA.103.042809}. The corresponding NR corrections for $2^3S_1-n^3D_1$ ($n=3,\,4,\,5$) transition energies were evaluated in \cite{PhysRevA.103.042809}, where it was shown that the quantum interference effect, previously unaccounted for, may partially eliminate the current imbalance between theory and experiment \cite{PhysRevLett.78.3658}.
	
	The study of multiphoton scattering processes and line profile asymmetry as applied to astrophysical problems is also of particular interest \cite{Brasken, solarassymetry, Jackson_2018}. Being a powerful tool for examining the dynamics of the universe evolution at an early stage, accurate calculations of radiation transfer in the interstellar and intergalactic medium with an appropriate evaluation of the scattering cross sections and line profiles are extremely important \cite{Seager_1999,Seager_2000}. Until recently, the solution of problems related to radiative transfer equations was considered only in the resonance approximation \cite{Lee_2013,leeheewon2016,Carswell}. In \cite{Lee_2013} it was shown that taking into account the asymmetry of the Ly$_{\alpha}$ profile can lead to an underestimation of the redshift of some astrophysical sources by $\Delta z \sim 10^{-3}-10^{-4}$, which is on the level of current experimental observations \cite{Wang_2021}. The effect of electromagnetically induced transparency, which leads to a distortion of the absorption profile, was discussed in this context in \cite{EIT-1,EIT-2,EIT-3}.
	
	The works \cite{Low,Labzowsky1994,2001,PhysRevA.65.054502,KARASIOV1992453, Jentschura2001,PhysRevA.79.052506} has opened a whole new branch of research dealing with NR effects and their role in modern spectroscopy. Since then, a number of investigations have been carried out in this field by various authors and research groups \cite{PhysRevLett.98.203003,Jent-NR}. The most important results concern not only spectroscopy of hydrogen and helium atoms, but also of highly charged ions and mesoatoms. Subsequently, numerous papers \cite{doi:10.1139/p02-094,PRL-LSSP,doi:10.1139/p07-014} confirmed all features of the previously predicted NR corrections. Later, it was shown in \cite{PhysRevA.79.052506} that there is an influence of the detection method on the measured frequency value. The dependence of the NR corrections on the arrangement of the experimental setup was studied in \cite{Solovyev2020,PhysRevA.92.022514,PhysRevA.97.022510} and also in \cite{PhysRevA.103.022833} for the case of two-photon hydrogen spectroscopy.
	
	Since precision spectroscopy of atomic systems is of great meaning in modern physics in connection with the goal of precise determination of fundamental constants, it becomes increasingly important to take into account not only QED corrections but also NR effects. To this end, in the present review we discuss recent achievements in the study of nonresonant effects and asymmetry of line profiles using a rigorous quantum electrodynamics theory.
	
	\subsection{Outline}
	
	The outline of this review is as follows. We begin with the notations and glossary compiled in section \ref{notations}. Section \ref{sectionfirst} briefly recalls some basic concepts of the relativistic description of photon scattering by the atomic electron in the framework of the bound state QED. In this section we also consider successively the derivation of the differential and total cross sections for the one-photon scattering process and various types of nonresonant corrections to the spectral line profile. The latter lead to an asymmetry of the line profile and are indispensable for determining the level of accuracy at which the concept of energy levels itself becomes inadequate for the analysis of experimental data.
	
	Section \ref{angleone} is devoted to the description of angular correlations of quantum interference effects. Particular emphasis will be made on investigations of spectroscopy of hydrogen, muonic hydrogen and helium-3 isotopes. Two-photon spectroscopy of light atomic systems is also of interest. In section \ref{twophotonH} the analysis of angular correlations of QIE in $2s-ns/nd$ and $2^3S-n^3D$ transitions in hydrogen and helium-4 is performed. 
	
	In section \ref{BBR} we briefly discuss the influence of thermal equilibrium radiation field on the line profile broadening and leading order NR effects. Finally, we conclude with a brief summary and an outlook in section \ref{theendfinal}.
	
	This review includes three appendices. In Appendix \ref{appendix1} the derivation of equation for the one-photon scattering amplitude in nonrelativistic limit and dipole approximation is presented. In Appendix \ref{ang2} we give a detailed derivation of angular correlations in two-photon scattering cross section discussed in section \ref{twophotonH}. Results of analytical expressions for NR corrections to $2s-ns/nd$ transition energies in hydrogen are collected in Appendix \ref{ang2explicit}.

	\subsection{Notations and glossary}
	\label{notations}
	
	Throughout this review relativistic units are employed, where the velocity of light $ c=1 $, Planck's constant $ \hbar=1 $, mass of electron $ m=1 $, respectively. The electron charge $ e=-\left|e\right| $ is related to the fine structure constant $ \alpha $ via
	\begin{eqnarray}
		\alpha=\frac{e^2}{\hbar c}.
	\end{eqnarray} 
	For $4-$vectors and tensor a standard notations are used for covariant (lower index) and contravariant (upper index) components, which are related to each other by metric tensor $g_{\mu\nu}=g^{\mu\nu}=(1,-1,-1,-1)$, so that, for example, for vector components $A_{\mu}$: $A_{\mu}=g_{\mu\nu}A^{\nu}$ (implying here and further in review Einstein's sum convention). Greek indices run over set $(0,\,1,\,2,\,3)$ and Latin indices run over set $(1,\,2,\,3)$.
	
	
	Bold letters denote 3-vector: for example $ \textbf{r}=(r^1, r^2, r^3) $. For the spatial position $ \textbf{r} $ we also use notations $ r=\left|\textbf{r}\right| $ and $ \textbf{r}_{ij}=\textbf{r}_i-\textbf{r}_j $, $ r_{ij}=\left|\textbf{r}_{ij}\right| $. The unit radius-vector is defined as $ {\bf n}={\bf r}/|{\bf r}| $. The coordinate 4-vector is $ x^{\mu}=(t,\textbf{r}) $, the momentum 4-vector is $ p^{\mu}=(\epsilon,\textbf{p}) $, $ \epsilon $ is the energy. The photon wave 4-vector is $ k^{\mu}=(\omega,\textbf{k}) $, where $ \omega $ is the photon frequency, $ \textbf{k} $ is the photon momentum (wave vector). The corresponding volume elements are $ d^4x=dt d\textbf{r} $, $ d^4p=d\epsilon d\textbf{p} $ and $ d^4k=d\omega d\textbf{k} $. Accordingly, $ x^2=t^2-\left|\textbf{r}\right|^2 $, $ p^2=\epsilon^2-\left|\textbf{p}\right|^2 $,  $ k^2=\omega^2-\left|\textbf{k}\right|^2 $. The photon wave function is written as $ A^{\mu}=(\Phi,\textbf{A}) $, $ \Phi $ is the scalar potential, $ \textbf{A} $ is the vector potential of the electromagnetic field. The 4-vector components of the Dirac matrices read $ \gamma^{\mu}=(\gamma^{0},\bm{\gamma}) $, where $ \gamma^{0}=\beta $, $ \bm{\gamma}=\beta\bm{\alpha} $. The standard representation of these matrices which are employed throughout this review
	\begin{eqnarray}
	\label{hereisbeta}
		\bm{\alpha}=
		\left(
		\begin{matrix}
			0 & \bm{\sigma} \\
			\bm{\sigma} & 0
		\end{matrix}
		\right),\;\;\;\;\;\;\;\;\;
		\beta=
		\left(
		\begin{matrix}
			1 & 0 \\
			0 & -1
		\end{matrix}
		\right),
	\end{eqnarray}
	where $ \bm{\sigma} $ are the Pauli matrices. The scalar and vector  product of two 3-vectors $ \textbf{a} $ and $ \textbf{b} $ are denoted as $ \textbf{a}\,\textbf{b} $ and $ \textbf{a}\times\textbf{b} $ respectively.  The scalar product of two 4-vectors is $ a_{\mu}b^{\mu}=a_0b_0-\textbf{a}\,\textbf{b} $. 
	
	For the matrix elements of certain operator two types of notations are used. A symbol $ \left(\dots\right)_{A'A} $ denotes the integral over spatial variables taken with the Dirac wave functions $ \overline{\psi}_{A'}\equiv \psi_{A'}^\dagger \gamma_0 $ and $ \psi_{A} $ (unless otherwise specified), where $ A' $ and $ A $ are the sets of quantum numbers. For the S-matrix element between states $ A' $ and $ A $ the subscripts are used $ \hat{S}^{(i)}_{A'A} $, where superscript $ (i) $ indicates the order of perturbation theory for the described process. 
	
	The use of variety of abbreviations for technical terminology can hardly be circumvented. We compile the following glossary of abbreviations used in this review
	\begin{table}[H]
		\begin{center}
			\begin{tabular}{c|c}
				\hline
				\hline
				HCI &  Highly Charged Ions \\
				LPA & Line Profile Approach  \\
				PT   &  Perturbation Theory  \\
				QED & Quantum Electrodynamics  \\
				SE  &   Self-Energy   \\
				NR  &   Nonresonant  \\
				QIE & Quantum interference effects\\
				BBR &  Black Body Radiation \\
				\hline
				\hline
			\end{tabular}
		\end{center}
	\end{table}
	
	\section{Line profile assymetry and nonresonant effects}
	\label{sectionfirst}
	
	One of the important features of the line profile theory in QED is the emergence of NR corrections. For the first time they were introduced in the seminal work \cite{Low} where the modern QED theory of the natural (Lorentz) line profile in atomic physics was formulated. These corrections indicate the limit up to which the concept of the energy for an excited atomic state has a physical meaning \cite{ANDREEV2008135}. This limit corresponds to the resonance approximation, when only the dominant (resonant) term remains in the amplitude of the process, see \cite{ANDREEV2008135,ZALIALIUTDINOV20181} for details. If the Lorentz profile distortion caused by other contributions is small, one can still consider the NR correction as an additional energy shift. Unlike all other energy corrections, this correction depends on the particular process used to measure the energy difference between the levels, and therefore it should be studied theoretically independently each time according to the experimental setup. 
	One can state that nonresonant corrections set the limit for the accuracy of all atomic frequency standards. One reason for this is that if the line profile (the shape of the resonance) becomes asymmetric with respect to frequency, there is no way to uniquely determine the position of the resonance.
	
	Below we consider a fully relativistic derivation of the differential and total photon scattering cross sections in the framework of the QED theory and the S-matrix formalism, which gives an exhaustive exposition of nonresonant effects and their influence on the determination of the transition frequency.

	\subsection{Amplitude of photon scattering on an atom 
	}
	\label{subsection1}
	
\begin{figure}[H]
		\center{\includegraphics[width=0.8\linewidth]{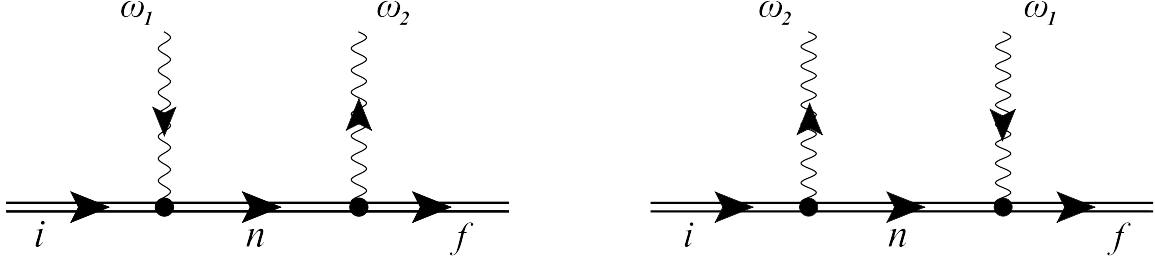}}
		\caption{Photon scattering on bound electron. Wavy line denotes absorption (arrow towards vertex) or emission (arrow outwards vertex) of photon and double solid line means bound electron in the field of nucleus (Furry picture); $\omega_1, \omega_2$ are frequencies of absorbed and emitted photons, respectively, $ i $, $ n $ and $ f $ denotes initial, intermediate and final states of the electron, respectively.}
		\label{onephoton}
	\end{figure}

	Consider first the process of photon scattering on a one-electron atom. The corresponding Feynman diagrams are presented in Fig. \ref{onephoton}. 
	Let $ i $, $ n $ and $ f $ denote initial, intermediate and final electron states, and $ \{\omega j_{\gamma} m_{\gamma}s\} $ and $ \{\omega' j_{\gamma}' m_{\gamma}'s'\} $ are the quantum numbers of the initial and final photons, respectively, where $ \omega $ is the frequency, $ j_{\gamma} m_{\gamma} $ denote the photon angular momentum and its projection and $ s $ defines the parity of the photon state. 
	
	Then the S-matrix matrix element of scattering process is 
	\begin{eqnarray}
		\label{eq1}
		S^{(2)}_{fi}=(-ie)^2\int d^4x d^4y\Big{[} \overline{\psi}_f(x)\gamma^{\mu}A^{(\boldsymbol{k}_2,\boldsymbol{e}_2)*}_{\mu}(x)S(x,y)\gamma^{\nu}A^{(\boldsymbol{k}_1,\boldsymbol{e}_1)}_{\nu}(y)\psi_i(y) + \overline{\psi}_f(x)\gamma^{\nu}A^{(\boldsymbol{k}_1,\boldsymbol{e}_1)}_{\nu}(x)S(x,y)\gamma^{\mu}A^{(\boldsymbol{k}_2,\boldsymbol{e}_2)*}_{\mu}(y)\psi_i(y)\Big{]}.
	\end{eqnarray}
	Here  $\psi_{A}(x)=e^{-iE_{A}t}\psi(\boldsymbol{x})$ is the solution of Dirac equation for bound electron in state $A$ and
	\begin{eqnarray}
		\label{eq2}
		A^{(\boldsymbol{k},\boldsymbol{e})}_{\mu}(x)=\sqrt{\frac{2\pi}{\omega}}e^{(\lambda)}_{\mu}e^{-ik x},
	\end{eqnarray}
	is the photon wave-function in the coordinate representation,
	where $\omega=|\boldsymbol{k}|$, $e^{(\lambda)}_{\mu}$ are the components of the photon polarization $4-$vector and $ x^{\mu}=(t_{x},\textbf{x}) $ and $ y^{\mu}=(t_{y},\textbf{y}) $ are the coordinate 4-vectors. Complex conjugation of the photon wave-functions in Eq. (\ref{eq1}) means emission of the photon. Normalization factor in Eq. (\ref{eq2}) $\sqrt{2\pi/\omega}$ is chosen to obtain the Coulomb interaction between electron and nucleus in the form $ Ze^2/r$. 
	In Eq. (\ref{eq1}) $S(x,y)$ denotes the Feynman propagator of an atomic electron, which can be represented using the eigenmode expansion \cite{A-B-QED} as follows
	\begin{eqnarray}
		\label{eq3}
		S(x,y)=\frac{i}{2\pi}\int^{-\infty}_{\infty}d\Omega e^{i\Omega (t_x-t_y)}\sum_n \frac{\psi_n(\boldsymbol{x})\overline{\psi}_n(\boldsymbol{y})}{E_n(1-i0)+\Omega},
	\end{eqnarray}
	where the summation runs over the entire Dirac spectrum of an electron in the field of the nucleus. Consideration of real photons leads to the transversality condition and $\gamma^{\mu}e^{(\lambda)}_{\mu}=\boldsymbol{e}\boldsymbol{\alpha}$ with the transverse photon polarization, 3-vector $\boldsymbol{e}$, and the corresponding wave-function of transversal photons
	\begin{eqnarray}
		\label{eq6trans}
		\boldsymbol{A}_{\boldsymbol{k}, \boldsymbol{e}}=\sqrt{\frac{2\pi}{\omega}}\boldsymbol{e}e^{-i\boldsymbol{k}\boldsymbol{r}}.
	\end{eqnarray}
	
	The integration over the time variables and frequency $\Omega$ in Eq. (\ref{eq1}) yields   
	\begin{eqnarray}
		\label{eq4}
		S^{(2)}_{fi}=-2\pi i \delta(E_i + \omega_1 - E_f - \omega_2)U^{(2)}_{fi},
	\end{eqnarray}
	where $U^{(2)}_{fi}$ is the amplitude of one-photon scattering process $i + \gamma \rightarrow f - \gamma$ \cite{ANDREEV2008135,ZALIALIUTDINOV20181} 
	\begin{eqnarray}
		\label{eq5}
		U^{(2)}_{fi}=e^2\Bigg{[} \sum_n \frac{\Big{(} \boldsymbol{\alpha}\boldsymbol{A}^*_{\boldsymbol{k}_2, \boldsymbol{e}_2} \Big{)}_{fn} \Big{(} \boldsymbol{\alpha}\boldsymbol{A}_{\boldsymbol{k}_1, \boldsymbol{e}_1} \Big{)}_{ni}}{E_n(1-i0) - E_i - \omega_1} + \sum_n \frac{\Big{(} \boldsymbol{\alpha}\boldsymbol{A}_{\boldsymbol{k}_1, \boldsymbol{e}_1} \Big{)}_{fn} \Big{(} \boldsymbol{\alpha}\boldsymbol{A}^*_{\boldsymbol{k}_2, \boldsymbol{e}_2} \Big{)}_{ni}}{E_n(1-i0) - E_f + \omega_1} \Bigg{]},
	\end{eqnarray}
	where $  \boldsymbol{\alpha}A_{\boldsymbol{k}, \boldsymbol{e}} $ and $  \boldsymbol{\alpha}A^{*}_{\boldsymbol{k}, \boldsymbol{e}} $ are the operators of absorption and emission of photons, respectively. Then the differential cross section of the process is
	\begin{eqnarray}
		\label{eq7}
		d\sigma_{fi}=2\pi \big{|}U^{(2)}_{fi}\big{|}^2\delta(E_i + \omega_1 - E_f - \omega_2)\frac{d \boldsymbol{k}_2}{(2\pi)^3}.
	\end{eqnarray}
	where $ d \boldsymbol{k}_2 = \omega_2^2d\omega_2 d\boldsymbol{n}_{k_2}$, $\omega_2=|\boldsymbol{k}_2|$ is the photon frequency and $ \boldsymbol{n}_{k_2}=\boldsymbol{k}_2/|\boldsymbol{k}_2|$ is the photon propagation unit vector. 
	
	The matrix elements (\ref{eq5}) can be evaluated with the use of the partial-wave expansions 
	\begin{eqnarray}
		\label{eq1s}
		\boldsymbol{e}e^{-i\boldsymbol{k}\boldsymbol{r}}=\sum_{j_{\gamma}m_{\gamma}s}\big{[} \boldsymbol{e} \boldsymbol{Y}^{(s)}_{j_{\gamma}m_{\gamma}}(\boldsymbol{n}_{k}) \big{]} \boldsymbol{A}^{(s)*}_{j_{\gamma}m_{\gamma}}(\boldsymbol{n}_{r}), 
	\end{eqnarray}
	where $\boldsymbol{n}_{r}=\boldsymbol{r}/|\boldsymbol{r}|$ and $\boldsymbol{A}^{(s)}_{j_{\gamma}m_{\gamma}}$ are the components of vector-potential 
	\begin{eqnarray}
		\label{eq6}
		\boldsymbol{A}^{(-1)}_{j_{\gamma}m_{\gamma}}(\boldsymbol{n}_r)=\sqrt{\frac{j_{\gamma}}{2j_{\gamma}+1}}g_{j_{\gamma}-1}(kr)\boldsymbol{Y}_{j_{\gamma}j_{\gamma}-1m_{\gamma}}(\boldsymbol{n}_r) +  \sqrt{\frac{j_{\gamma}+1}{2j_{\gamma}+1}}g_{j_{\gamma}+1}(kr)\boldsymbol{Y}_{j_{\gamma}j_{\gamma}+1m_{\gamma}}(\boldsymbol{n}_r), 
	\end{eqnarray}
	\begin{eqnarray}
		\boldsymbol{A}^{(0)}_{j_{\gamma}m_{\gamma}}(\boldsymbol{n}_r)=g_{j_{\gamma}}(kr)\boldsymbol{Y}_{j_{\gamma}j_{\gamma}m_{\gamma}}(\boldsymbol{n}_r), 
	\end{eqnarray}
	\begin{eqnarray}
		\label{eq8}
		\boldsymbol{A}^{(+1)}_{j_{\gamma}m_{\gamma}}(\boldsymbol{n}_r)=\sqrt{\frac{j_{\gamma}+1}{2j_{\gamma}+1}}g_{j_{\gamma}-1}(kr)\boldsymbol{Y}_{j_{\gamma}j_{\gamma}-1m_{\gamma}}(\boldsymbol{n}_r) + \sqrt{\frac{j_{\gamma}}{2j_{\gamma}+1}}g_{j_{\gamma}+1}(kr)\boldsymbol{Y}_{j_{\gamma}j_{\gamma}+1m_{\gamma}}(\boldsymbol{n}_r),
	\end{eqnarray}
	and $ \boldsymbol{Y}^{(s)}_{jm}$ are the components of spherical tensor:
	\begin{eqnarray}
		\label{eq2s}
		\boldsymbol{Y}^{(-1)}_{j_{\gamma}m_{\gamma}}(\boldsymbol{n}_r)=\sqrt{\frac{j_{\gamma}}{2j_{\gamma}+1}}\boldsymbol{Y}_{j_{\gamma}j_{\gamma}-1m_{\gamma}}(\boldsymbol{n}_r) - \sqrt{\frac{j_{\gamma}+1}{2j_{\gamma}+1}}\boldsymbol{Y}_{j_{\gamma}j_{\gamma}+1m_{\gamma}}(\boldsymbol{n}_r), 
	\end{eqnarray}
	\begin{eqnarray}
		\label{eq3s}
		\boldsymbol{Y}^{(0)}_{j_{\gamma}m_{\gamma}}(\boldsymbol{n}_{k})=\boldsymbol{Y}_{j_{\gamma}j_{\gamma}m_{\gamma}}(\boldsymbol{n}_r), 
	\end{eqnarray}
	\begin{eqnarray}
		\label{eq4s}
		\boldsymbol{Y}^{(+1)}_{j_{\gamma}m_{\gamma}}(\boldsymbol{n}_{k})=-\sqrt{\frac{j_{\gamma}+1}{2j_{\gamma}+1}}\boldsymbol{Y}_{j_{\gamma}j_{\gamma}-1m_{\gamma}}(\boldsymbol{n}_{k}) - \sqrt{\frac{j}{2j+1}}\boldsymbol{Y}_{j_{\gamma}j_{\gamma}+1m_{\gamma}}(\boldsymbol{n}_{k}).
	\end{eqnarray}
	The spherical vector $\boldsymbol{Y}_{jlm}$ in Eqs. (\ref{eq2s})-(\ref{eq4s}) is defined as follows
	\begin{eqnarray}
		\label{eq5s}
		\boldsymbol{Y}_{j_{\gamma}l_{\gamma}m_{\gamma}}(\boldsymbol{n}_{k})=\sum_{m_{\gamma}\mu}C^{j_{\gamma}m_{\gamma}}_{l_{\gamma}m_{\gamma}1\mu}Y_{l_{\gamma}m_{\gamma}}(\boldsymbol{n}_{k})\boldsymbol{\chi}_{\mu}
		.
	\end{eqnarray}
	Here $\boldsymbol{\chi}_{\mu}$ is the spin function of a particle with spin one and $Y_{lm}$ is the spherical harmonic. 
	The function $ g_{j_{\gamma}}(kr) $ in Eqs. (\ref{eq6})-(\ref{eq8}) is related to the spherical Bessel function $ j_{j_{\gamma}}(kr) $
	\begin{eqnarray}
		\label{ph5}
		g_{j_{\gamma}}(kr)=4\pi i^{j_{\gamma}}j_{j_{\gamma}}(kr).
	\end{eqnarray}
	Using Eqs. (\ref{eq6}-\ref{eq8}) operator in matrix elements of Eq. (\ref{eq5}) can be written in the form of multipole decomposition
	\begin{eqnarray}
		\label{eq16}
		\boldsymbol{\alpha}\boldsymbol{A}^*_{\boldsymbol{k}, \boldsymbol{e}}=\sqrt{\frac{2\pi}{\omega}}\sum_{j_{\gamma}m_{\gamma}s}\big{[} \boldsymbol{e} \boldsymbol{Y}^{(s)}_{j_{\gamma}m_{\gamma}}(\boldsymbol{n}_{\boldsymbol{k}}) \big{]}\boldsymbol{\alpha}\boldsymbol{A}^{(s)*}_{j_{\gamma}m_{\gamma}}.
	\end{eqnarray}   
	Finally, introducing notation
	\begin{eqnarray}
		\label{eq17}
		C^{j_{\gamma_1}m_{\gamma_1}s_1}_{j_{\gamma_2}m_{\gamma_2}s_2}(\boldsymbol{e}_1, \boldsymbol{n}_{\boldsymbol{k}_1}; \boldsymbol{e}_2, \boldsymbol{n}_{\boldsymbol{k}_2})=\big{[} \boldsymbol{e} \boldsymbol{Y}^{(s_1)}_{j_{\gamma_1}m_{\gamma_1}}(\boldsymbol{n}_{\boldsymbol{k}_1}) \big{]} \big{[} \boldsymbol{e}_2 \boldsymbol{Y}^{(s_2)}_{j_{\gamma_2}m_{\gamma_2}}(\boldsymbol{n}_{\boldsymbol{k}_2}) \big{]}^*,
	\end{eqnarray}
	the transition amplitude (\ref{eq5}) can be written in the form \cite{PhysRevA.24.183}:
	\begin{eqnarray}
		\label{eq18}
		U^{(2)}_{fi}=e^2\frac{2\pi}{\sqrt{\omega_1 \omega_2}}\Bigg{[} \sum_{\substack{j_{\gamma_1}m_{\gamma_1}s_1\\j_{\gamma_2}m_{\gamma_2}s_2}}
		C^{j_{\gamma_1}m_{\gamma_1}s_1}_{j_{\gamma_2}m_{\gamma_2}s_2}(\boldsymbol{e}_1, \boldsymbol{n}_{\boldsymbol{k}_1}; \boldsymbol{e}_2, \boldsymbol{n}_{\boldsymbol{k}_2}) \times \\
		\nonumber
		\times \Bigg{\{} \sum_n \frac{\Big{(} \boldsymbol{\alpha}\boldsymbol{A}^{(s_2)*}_{j_{\gamma_2}m_{\gamma_2}} \Big{)}_{fn} \Big{(} \boldsymbol{\alpha}\boldsymbol{A}^{(s_1)}_{j_{\gamma_1}m_{\gamma_1}} \Big{)}_{ni}}{E_n(1-i0) - E_i - \omega_1} + \sum_n \frac{\Big{(} \boldsymbol{\alpha}\boldsymbol{A}^{(s_1)}_{j_{\gamma_1}m_{\gamma_1}} \Big{)}_{fn} \Big{(} \boldsymbol{\alpha}\boldsymbol{A}^{(s_2)*}_{j_{\gamma_2}m_{\gamma_2}} \Big{)}_{ni}}{E_n(1-i0) - E_i - \omega_2} \Bigg{\}} \Bigg{]},
	\end{eqnarray}
	where photons of a particular type are given by the external sums in Eq. (\ref{eq18}). 
	
	In the resonant approximation it is assumed that in the transition process $i + \gamma_1 \rightarrow f + \gamma_2$ there is an intermediate state $r$, for which the frequency of the absorbed photon $\omega_1$ is equal to the energy difference $E_{r} - E_i$ and the leading contribution to the scattering cross section comes from the term with $n=r$ in the first sum in curly brackets of Eq. (\ref{eq18}). The resulting divergent contribution should be regularized by including an infinite set of Feynman graphs representing the one-loop self-energy correction for bound electron \cite{Low,ANDREEV2008135}. Thus, the natural level width $\Gamma_{r}$ appears in the corresponding energy denominator, and the square of the resonant regularized contribution gives the line profile of the corresponding process.

	\subsection{QED derivation of Lorentz spectral line profile}
	\label{subsection3}
	
	To derive the standard Lorentz form for the line profile from QED, we should consider the resonant process of elastic photon scattering on atomic electron in the ground state $a$. Using the S-matrix formalism in the Furry picture \cite{Furry, A-B-QED, LabKlim}, within the resonant approximation, the amplitude of the process shown in Fig. \ref{onephoton}, Eq. (\ref{eq18}), can be reduced to the form:
	\begin{eqnarray}
		\label{1}
		U^{\mathrm{sc}}_{aa}=\frac{\langle a | \boldsymbol{\alpha}\boldsymbol{A}_{\boldsymbol{k}_2, \boldsymbol{e}_2} | r\rangle \langle r| \boldsymbol{\alpha}\boldsymbol{A}^{*}_{\boldsymbol{k}_1, \boldsymbol{e}_1} | a\rangle}{E_{r}-E_{a}-\omega}.
	\end{eqnarray}
	 From Eq. (\ref{1}) the emission amplitude, $ U^{\mathrm{em}}_{ra} $, 
	 can be defined as follows
	\begin{eqnarray}
		\label{2}
		U^{\mathrm{em}}_{ra} = \frac{\langle r | \boldsymbol{\alpha} \boldsymbol{A}^{*}_{\boldsymbol{k}_2, \boldsymbol{e}_2} | a\rangle }{E_{r}-E_{a}-\omega}
		.
	\end{eqnarray}
	Here it is assumed that the absorption and emission processes are separated within the framework of the resonance approximation, see the discussion in \cite{ANDREEV2008135}.
	
	For a resonant excitation process this expression has a singularity at $ \omega = E_{r}-E_{a} $.  In order to avoid the resulting divergence, an infinite number of the electron self-energy "loop after loop" should be inserted into the internal electron line in Fig. \ref{onephoton}, see \cite{Low}. Using the standard QED derivations in the Furry picture one can arrive at a geometric progression  \cite{LabKlim}, which finally gives the amplitude $ U^{\mathrm{em}}_{ra} $ as
	\begin{eqnarray}
		\label{3}
		U^{\mathrm{em}}_{ra} = \frac{\langle r |  \boldsymbol{\alpha}\boldsymbol{A}^{*}_{\boldsymbol{k}_2, \boldsymbol{e}_2} | a\rangle }{E_{r}-E_{a}-\omega +\langle r | \hat{\Sigma}\left(E_{r} \right) | r\rangle }
	\end{eqnarray}
	where $ \langle r | \hat{\Sigma}\left(E_{r} \right) | r\rangle  $ is the diagonal matrix element of the electron self-energy operator, $ \hat{\Sigma}\left(E_{r} \right) $:
	\begin{gather}
		\label{4}
		\langle r| \hat{\Sigma}\left(E_{r} \right) |r\rangle =\frac{e^2}{2\pi \mathrm{i}}
		\sum\limits_{n}\langle rn | \frac{1-\boldsymbol{\alpha}_1\boldsymbol{\alpha}_2}{r_{12}}I_{n}\left(r_{12};E_{r} \right) | nr\rangle,
	\end{gather}
	\begin{eqnarray}
		\label{5}
		I_{n}\left(r_{12};E_{r} \right)=\int\limits^{\infty}_{-\infty}\frac{e^{\mathrm{i}|\Omega|r_{12}}d\omega}{E_{n}(1-\mathrm{i}0)-E_{r}-\Omega}.
	\end{eqnarray}
	The matrix element $ \langle ab \left| X \right| ba \rangle $ should be understood as 
	\begin{eqnarray}
		\label{6}
		\langle ab \left| X \right| ba \rangle = \langle a(1)b(2) \left| X(1,2) \right| b(1)a(2) \rangle
	\end{eqnarray}
	where $1$, $2$ are two different  electron variables, $ \boldsymbol{\alpha}_{1,2} $ are Dirac matrices acting on the wave functions with variables $1$, $2$ respectively, $ r_{12}=\left| \boldsymbol{r}_1-\boldsymbol{r}_2\right| $. Summation in Eq. (\ref{4}) is extended over entire Dirac spectrum of an electron. 
	
	The real part of the matrix element $ \langle r |  \hat{\Sigma}\left(E_{r} \right) | \rangle $ diverges and have to be renormalized \cite{lindgren}. The renormalized real part of the matrix element Eq. (\ref{4}) represents the lowest order electron self-energy contribution to the Lamb shift of the level $ r $:
	\begin{eqnarray}
		\label{7}
		\mathrm{Re}\langle r| \hat{\Sigma}_{\mathrm{REN}}\left(E_{r} \right)| r\rangle= L^{\mathrm{SE}}_{r}.
	\end{eqnarray}
	However, as will be shown below, the real part plays no role in the nonresonant continuation of the Lorentz profile, and we will focus on the imaginary part of the matrix element Eq. (\ref{4}). It can be shown analytically \cite{LabKlim} that 
	\begin{eqnarray}
		\label{8}
		\mathrm{Im}\langle r| \hat{\Sigma}_{\mathrm{REN}}\left(E_{r} \right)| r\rangle=-\frac{\Gamma_{r}}{2},
	\end{eqnarray}
	where $\Gamma_{r}$ is the width of the resonant atomic state $r$. Unlike the real part, the imaginary part of $ \langle r | \hat{\Sigma}_{\mathrm{REN}}\left(E_{r} \right)| r\rangle $ does not diverge and does not require renormalization.
	
	Inserting Eqs. (\ref{7}), (\ref{8}) into Eq. (\ref{3}), we find
	\begin{eqnarray}
		U^{\mathrm{em}}_{ra}=\frac{\langle r |  \boldsymbol{\alpha} A^{*}_{\boldsymbol{k}_2, \boldsymbol{e}_2} | a \rangle}{E_{r}+L^{\mathrm{SE}}_{r}-E_{a}-\frac{\mathrm{i}}{2}\Gamma_{r}-\omega}.
		\label{9}
	\end{eqnarray}
	Now the expression for $ U^{\mathrm{em}}_{ra} $ is regularized, the pole is shifted to the complex plane. In principle, the electron self-energy correction to the energy $ E_{a} $ should also be included in Eq. (\ref{9}). If $ a $ corresponds to the ground state, then the width $ \Gamma_{a} $ is absent. Otherwise, the line profile contains the sum of the widths \cite{ZALIALIUTDINOV20181}. The Lamb shift $ L^{\mathrm{SE}}_{a} $ can be obtained by inserting the electron self-energy loops into the outer electron line in Fig. \ref{onephoton}, see \cite{ANDREEV2008135}, and is also unimportant for our purposes.
	
	To obtain the emission line profile, the amplitude modulus (\ref{9}) should be squared, multiplied by the phase volume $ d\boldsymbol{k}'/(2\pi)^3 $, integrated over the photon emission directions and summed over photon polarization. In the nonrelativistic limit which is certainly valid to the hydrogen atom $ \langle r |   \boldsymbol{\alpha}\boldsymbol{A}^{*}_{\boldsymbol{k}, \boldsymbol{e}} | a \rangle = \frac{e}{\sqrt{\omega}}\langle r |  \boldsymbol{e}\boldsymbol{p} | a \rangle $, where $ \boldsymbol{p} $ is the electron momentum operator, $ m $ is the electron mass. The factor $ \frac{1}{\sqrt{\omega}} $  comes from the normalization of the electromagnetic field potentials \cite{LabKlim}. Then the standard expression for the Lorentz profile arises as
	\begin{eqnarray}
		\label{lorS}
		\phi_{\mathrm{L}}(\omega)d\omega=\frac{1}{N}\sum\limits_{\boldsymbol{e}}\int \frac{\omega d\omega d\boldsymbol{n}_{k}}{(2\pi)^3} | U^{\mathrm{em}}_{ra} | 
		=\frac{1}{N}\frac{W_{ra}d\omega}{(\omega_{0}-\omega)^2+\frac{1}{4}\Gamma_{r}^2},
	\end{eqnarray}
	where 
	\begin{eqnarray}
		\label{11}
		W_{ra}=\frac{4}{3}e^2\omega_{0}|\langle r| \boldsymbol{p} | a \rangle|^2
	\end{eqnarray}
	is the emission rate for the transition $ r\rightarrow a+\gamma $, $ \omega_{0} $ is the corresponding resonant frequency $ \omega_{0}=E_{r}-E_{a} $ and $ N $ is the normalization factor.
	
	In Eqs. (\ref{lorS}), (\ref{11}) we have set $ \omega=\omega_{0} $ in the expressions for $ W_{ra} $ and $ \Gamma_{r} $ and neglected $ L^{\mathrm{SE}}_{r} $. Obviously, the difference of the Lamb shifts $L^{\mathrm{SE}}_{a}$ and $L^{\mathrm{SE}}_{r}$ can simply be included in the definition of $\omega_0$. The transition rate coincides with the partial width of the $r$ level: $W_{ra}=\Gamma_{ra}$. If there are no other decay channels for the state $r$ than $r\rightarrow a$, $\Gamma_{ra}=\Gamma_{r}$. The normalization factor should be chosen from the condition:
	\begin{eqnarray}
		\label{12}
		\int \phi_{\mathrm{L}}\left(\omega\right)=1.
	\end{eqnarray}
	In the resonance approximation, the integration interval in Eq. (\ref{12}) can be extended from $ \omega=-\infty $ to $ \omega=\infty $. Then integration in the complex plane results in $ N=N^{(0)}\equiv 1/2\pi $.

	\subsection{Nonresonant extension of Lorentz spectral line profile}
	\label{subsection3ext}
	
	Three types of nonresonant contributions that distort the standard Lorentz profile Eq. (\ref{lorS}) can be distinguished. The first type (not discussed in this review) corresponds to distortion effects arising in external fields, see \cite{ZALIALIUTDINOV20181} and references therein, when there is a mixing of states with opposite parity. For example, an external electric field leads to overlapping of the $ 1s_{1/2}\rightarrow 2p_{1/2}$ and $ 1s_{1/2}\rightarrow 2s_{1/2}$ transitions. The second type is represented by the remaining terms in the scattering amplitude Eq. (\ref{eq18}). 
	 The study of this group of nonresonant effects is of particular interest in modern experiments \cite{Science,science2020} and is discussed in the following sections.
	
	 Here we will be interested in nonresonant corrections of the third type arising for the picked out resonant term. These corrections arise when the frequency dependence in Eq. (\ref{lorS}) is taken into account for both the partial transition rate $ W_{ra} $ and the total width $ \Gamma_{r} $. 
	The nonresonant extension of $ W_{ra} $ is trivial: Eq. (\ref{11}) should be replaced by
	\begin{eqnarray}
		\label{13}
		W_{ra}(\omega)=\frac{4}{3}e^2\omega \left|\langle r| \boldsymbol{p}  | a\rangle\right|^2.
	\end{eqnarray} 
	
	To derive the dependence $ \Gamma_{r}(\omega) $ we will follow \cite{ANDREEV2008135,LabKlim} and start with a nonresonant extension of the expression (\ref{4}). Accurate QED evaluation of self-energy contribution leads to the matrix element $ \langle r | \hat{\Sigma}\left(E_{a}\right)| r\rangle $ replaced by $ \langle r | \hat{\Sigma}\left(E_{a}+\omega\right)| r\rangle $:
	\begin{gather}
		\label{14}
		\langle r | \hat{\Sigma}\left(E_{a}+\omega\right)| r\rangle=\frac{e^2}{2\pi}\sum\limits_{n}
		\langle rn | \frac{1-\boldsymbol{\alpha}_1\boldsymbol{\alpha}_2}{r_{12}}I_{n}\left(r_{12}, E_{a}+\omega \right)
		| n r\rangle ,
	\end{gather}
	 resulting in the Lorentz profile 
	\begin{eqnarray}
		\label{15}
		\phi_{\mathrm{L}}(\omega)=\frac{1}{N}\frac{W_{ra}(\omega)}{(\omega_{0}-\omega)^2+\frac{1}{4}\Gamma^2_{r}(\omega)}.
	\end{eqnarray}
	
	To evaluate $ \Gamma_{r}(\omega) $, the integral $ I_{n}\left(r_{12};E_{a}+\omega\right) $ can be presented in the form:
	\begin{eqnarray}
		\label{16}
		I_{n}\left(r_{12};E_{a}+\omega\right)=\int\limits_{-\infty}^{\infty}\frac{e^{\mathrm{i}|\Omega|r_{12}}d\Omega}{E_{n}(1-\mathrm{i}0)-E_{a}-\omega+\Omega}
		=\int\limits_{-\infty}^{\infty}\frac{e^{\mathrm{i}\Omega r_{12}}d\Omega}{E_{n}(1-\mathrm{i}0)-E_{a}-\omega+\Omega}
		-
		2\mathrm{i}\int\limits_{-\infty}^{0}\frac{\mathrm{sin}(\Omega r_{12})d\Omega}{E_{n}(1-\mathrm{i}0)-E_{a}-\omega+\Omega}
	\end{eqnarray}
	for the case when $ E_{n}-E_{a}-\omega <0 $. The first integral on the right-hand side of Eq. (\ref{16}) can be evaluated in the complex plane variable $ \Omega $. The contour of integration can be closed in the upper half plane. Note that energies $ E_{n} $ in Eq. (\ref{16}) are the Dirac energy values and may be positive and negative. The poles corresponding to the negative energy values are located in the lower half-plane and do not contribute to the integral. 
	
	It follows from physical considerations that the wings of the Lorentz profile for a certain emission resonance to the ground state should not extend too far into the region where neighbour resonances begin to dominate. Therefore, the integration interval in Eq. (\ref{12}) should be restricted by $\omega\in[0, \omega_{\mathrm{max}}]$, where $\omega_{\mathrm{max}} =\omega_{0}+\frac{1}{2}\Delta_{{FS}}$, $\Delta_{\mathrm{FS}}$ is the fine structure interval (for example), changing the normalization coefficient $N$ accordingly. The choice $ E_{n}-E_{a}-\omega <0 $ and $\omega\leqslant \omega_{\mathrm{max}}$ results in a nonzero result only for positive-energy states with the energies lower than $ E_{r} $. For the hydrogen atom, with $ r=2p_{1/2} $ only the ground state $ n=1s_{1/2} $ contributes. In the second integral on the right-hand side of Eq. (\ref{16}), under such restrictions, the energy denominator is always nonzero, and the imaginary part can be omitted. Therefore, this integral is purely imaginary and, accordingly, does not contribute to $ \Gamma(\omega) $. In the same way re-combining the integrals in Eq. (\ref{16}) it is easy to see that for $ E_{n}-E_{a}-\omega>0 $ the result of the integration is always zero. Accurate evaluation of $ I_{n}\left(r_{12};E_{a}+\omega\right) $, see \cite{LabKlim}, using Cauchy theorem yields 
	\begin{gather}
		\label{17}
		\Gamma_{r}(\omega)=-2e^2\sum\limits_{E_{n}<E_{r}}
		\langle rn | \frac{1-\boldsymbol{\alpha}_1\boldsymbol{\alpha}_2}{r_{12}}\mathrm{sin}\left((E_{a}-E_{n}+\omega)r_{12} \right)
		| nr\rangle .
	\end{gather}
	
	In the nonrelativistic limit, the following estimates are valid: $ E_{n}\approx m(\alpha Z)^2 $, $ r_{12}\approx 1/m\alpha Z $ (in relativistic units), where $ Z $ is the nuclear charge, $\alpha$ is the fine structure constant. 
	In the case of resonance decay only to the ground state, the term with $ n=a $ remains in the sum over $ n $ in Eq. (\ref{17}). Then, $ \mathrm{sin}\left((E_{a}-E_{n}+\omega)r_{12} \right)=\mathrm{sin}\left(\omega r_{12} \right) $, where $ \omega \leqslant \omega_{\mathrm{max}} \approx m(\alpha Z)^2$. Applying the Taylor series expansion to $ \mathrm{sin}\left(\omega r_{12} \right) $, we find the parts that depend and do not dependent on $\boldsymbol{\alpha}$-matrices. The first term of the Taylor expansion in the $ \boldsymbol{\alpha} $-independent part gives zero due to the orthogonality of the wave functions. The next term after using $ r_{12}^2=r_1^2+r_2^2-2(\boldsymbol{r}_1\boldsymbol{r}_2) $ results in 
	\begin{eqnarray}
		\label{18}
		\Gamma_{r}(\omega)=-\frac{2}{3}e^2\omega^3 \left|\langle r |\boldsymbol{r}| a\rangle  \right|^2.
	\end{eqnarray}
	In the part of Eq. (\ref{17}) containing $ \boldsymbol{\alpha} $-matrices it is enough to keep the first term of Taylor expansion. In the nonrelativistic limit with the help of equality $ \langle r  \left| \boldsymbol{p}  \right| a  \rangle = -\mathrm{i}\omega_{0}\langle r  \left| \boldsymbol{r}  \right| a  \rangle$ we find
	\begin{eqnarray}
		\label{19}
		\Gamma^{''}_{r}(\omega)=2e^2\omega\omega_{0}^2\left| \langle r  \left| \boldsymbol{r} \right|  a \rangle \right|^2.
	\end{eqnarray}
		Combining Eqs. (\ref{18}) and (\ref{19}) we get
	\begin{eqnarray}
		\label{20}
		\Gamma_{r}(\omega)=\Gamma'_{r}(\omega)+\Gamma^{''}_{r}(\omega)=
		2e^2\omega\left[\omega_{0}^2-\frac{1}{3}\omega^2   \right] \left| \langle r  \left| \boldsymbol{r} \right|  a \rangle \right|^2 = \frac{3}{2} \frac{\omega}{\omega_0}\left(1-\frac{1}{3}\frac{\omega^2}{\omega_0^2}\right)\Gamma^{(0)}_{r},
	\end{eqnarray}
	where $ \Gamma^{(0)}_{r} =\frac{4}{3}\omega_0^3 \left| \langle r  \left| \boldsymbol{r} \right|  a \rangle \right|^2$ is the level width at the point of the resonance. 
	
	Finally, introducing the variable $ x=\omega/\omega_{0} $ ($ \omega_{0}=E_{r}-E_{a}$ ) we arrive at the extended Lorentz profile of the form:
	\begin{eqnarray}
		\label{lorE}
		\phi_{\mathrm{E}}(x)=\frac{1}{N}\frac{x\Gamma_{ra}^{(0)}}{\omega_{0}^2(x-1)^2+\frac{9}{16}x^2\left(1-\frac{1}{3}x^2 \right)^2\left( \Gamma_{r}^{(0)} \right)^2}.
	\end{eqnarray} 
	The normalization factor $ 1/N $ is defined by
	\begin{eqnarray}
		\label{22}
		\omega_{0}\int\limits_{0}^{\omega_{0}x_{\mathrm{max}}}\phi_{\mathrm{E}}(x)dx=1,
	\end{eqnarray}
	where $ x_{\mathrm{max}}=\omega_{\mathrm{max}}/\omega_{0} $. In principle, the radiative correction to the energy $ L^{\mathrm{SE}}_{r}(\omega) $ also depend on $ \omega $ and contributes to the distortion of the Lorentz profile. The expression for $ L^{\mathrm{SE}}_{r}(\omega) $ is more complicated (see \cite{ANDREEV2008135}), although its $\omega$-dependent part does not diverge and does not require additional renormalization. However, in this paper, we neglect the corresponding modification. 
	
	The expression (\ref{lorE}) is a rigorous QED result for obtaining a nonresonant extension of the Lorentz line profile. The analytical dependence of $\Gamma_{r}$ on $\omega$ outside of resonance was recently discussed in \cite{friction} in connection with blackbody radiation friction due to the finite lifetime of atomic levels. Other forms of expression (\ref{lorE}) are also presented in the literature \cite{gunnpeterson, leeheewon, leeheewon2016}.  Basically, they are obtained within the framework of quantum mechanics, i.e. phenomenological approaches. In the next section, we will compare them in detail.

	\subsection{Red damping wing of Ly$_{\alpha} $ transition in the expanding universe}
	\label{subsection4}
	
	In this section, we examine one of the possible fields of significance for NR effects - cosmological recombination of H atoms in the early universe. This recombination occurs when the equilibrium between the emission and absorption of photons by H atoms is violated and emission begins to dominate. More and more photons are absorbed (radiation escapes from the matter) and more and more H atoms appear to be in the ground state. One of the most important channels for the escape of radiation is the shift of the photon frequency towards the red wing of the Ly$_{\alpha} $ Lorentz profile due to the expansion of the universe. When the frequency of the photon reaches the critical value $ \omega_c $ below which the photon with this frequency cannot be absorbed (the probability of absorption becomes too low), the photon escapes from matter. Since all transitions between bound states in atoms, including cascades, occur in a quantum jump (i.e. do not require time \cite{ZALIALIUTDINOV20181}), the redshift of the photon frequency only occurs during the photon's journey from one atom to another. Therefore, the escape probability should depend on the density of hydrogen atoms within the period of cosmological recombination. In this section, we use a simulated description of the radiation escape without introducing the Sobolev escape probability. Our aim is to compare the escape probability during the travel of a photon between two H atoms with the standard Lorentz profile and the extended nonresonant profile. This comparison should show the importance of nonresonant effects for cosmological recombination theory.
	
	The Lorentzian as an excellent approximation for the resonance scattering can not provide an accurate description of the radiation damping.  When a myriad of absorbers are involved in scattering process, the line centre is significantly saturated and the broad damping wing becomes important. Such a situation arises for the absorption of a continuous radiation from high-redshift objects by the clouds of neutral hydrogen in the intergalactic medium (IGM) \cite{gunnpeterson,inproc}. 
	
	The detailed analysis of photon scattering in Ly$_{\alpha} $ and the 21 cm lines gives invaluable information about the evolution of the universe \cite{Santos_reionization, Hirata_lya}. An important characteristic of photon scattering on hydrogen clouds in IGM is optical depth $ \tau $ which directly depend on a line profile and proportional to the column density of neutral hydrogen. In the standard analysis of Ly$ _{\alpha} $ radiation transfer in the expending universe from a source at redshift $ z_{s} $ to the to the observator at redshift $ z_{0}=0 $, the following expression for optical depth is widely used \cite{gunnpeterson, Peebles_book}
	\begin{eqnarray}
		\label{tau}
		\tau(z) = N_{\mathrm{HI}}\int\limits_{0}^{z_{\mathrm{s}}}dz \sigma (n)(1+z)^2
		\left[\frac{H(z)}{H_{0}} \right]^{-1}.
	\end{eqnarray}
	Here $ N_{\mathrm{HI}} $ is the characteristic mean column density of atomic hydrogen, $ \sigma(n) $ is the scattering cross section, $n= c(1+z)/\lambda_{\mathrm{obs}}$, $ H_{0} $ is the Hubble constant and $ H(z) $ is the Hubble factor. In case of Ly$_{\alpha} $ scattering the cross section $ \sigma(n) $ in the vicinity of resonance  is given by
	\begin{eqnarray}
		\label{sgm1}
		\sigma(\omega) = \frac{3\lambda_{\alpha}^2\Gamma_{2p}}{4}\phi(\omega),
	\end{eqnarray}
	where $ \phi $ is the line profile and $ \lambda_{\alpha} = c/\omega_{0} $ is the wavelength of $ 2p\rightarrow 1s $ transition.  To apply the broad damped profile of the high-redshift hydrogen clouds various approximations for $ \phi $ were introduced during last decades \cite{Lee_2013,leeheewon2016, Carswell,leeheewon,Peebles,leeold}. Below we compare optical depths $ \tau $ calculated with different forms of damped profiles. 
	
	In the classical theory, natural line broadening is described by the damped harmonic oscillator with the periodic dipole driven force. Then, the angular integration of the time-averaged intensity yields well-known $ \omega^4 $- dependence of the Rayleigh scattering
	\begin{eqnarray}
		\label{lorR}
		\phi_{\mathrm{R}}(\omega)=\frac{1}{2\pi}\frac{\Gamma_{2p}(4\omega^4/\omega_{0}^2)}{(\omega_{0}^2-\omega^2)^2+\Gamma^{2}_{2p}\omega^2}.
	\end{eqnarray}
	Another form for $ \phi $  based on the two-level approximation was introduced in \cite{Peebles}:
	\begin{eqnarray}
		\label{lorP}
		\phi_{\mathrm{P}}(\omega)=\frac{1}{2\pi}\frac{\Gamma_{2p}(\omega/\omega_{0})^4}{(\omega_{0}-\omega)^2+\Gamma_{2p}^2(\omega/\omega_{0})^6/4}.
	\end{eqnarray}
	
	Recently, in \cite{leeheewon2016,leeheewon} the Ly$_{\alpha} $ scattering cross section deduced from the quantum mechanical time-dependent second-order perturbation theory was considered. For practical applications, the following analytic function that corrects Eq. (\ref{lorP}) was introduced in \cite{leeheewon}
	\begin{gather}
		\label{lorQM}
		\phi_{\mathrm{QM}}(\omega)=\phi_{\mathrm{R}}(1+f(\omega))
		=\phi_{\mathrm{L}}\frac{4(\omega/\omega_{0})^4}{(1+\omega/\omega_{0})^2}(1+f(\omega)), 
	\end{gather}
	where
	\begin{eqnarray}
		\nonumber
		f(\omega)= a(1-e^{-bx})+cx+dx^2,
	\end{eqnarray}
	with $ a=0.376 $, $ b=7.666 $, $ c=1.922 $ and $ d=-1.036 $.
	
\begin{figure}
		\centering
		\includegraphics[scale=0.9]{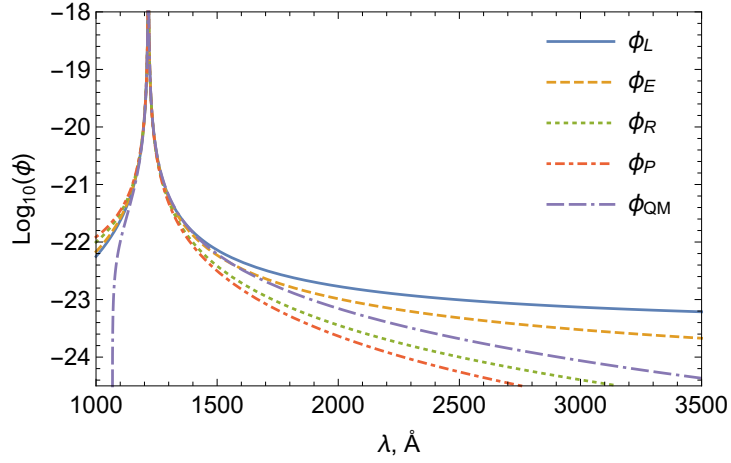}
		\caption{Comparison of red damped wings for absorption profiles given by Eqs. (\ref{lorS}), (\ref{lorE}), (\ref{lorR}), (\ref{lorP}), (\ref{lorQM}) as a functions of wavelength $ \lambda $. Normalization factor $ N=2\pi $ is chosen for Eqs. (\ref{lorS}) and (\ref{lorE}). The peak corresponds to the resonance wavelength $ \lambda_{\alpha}=1216 $ \r{A}.}
		\label{fig3a}
	\end{figure}

	The absorption profiles Eqs. (\ref{lorS}), (\ref{lorE}) (both with normalization factor $ N=2\pi $) and profiles Eqs. (\ref{lorR}), (\ref{lorP}), (\ref{lorQM}) are presented in Fig. \ref{fig3a}. 

	\section{Nonresonant effects in the total and differential scattering cross sections}
	
		\subsection{Nonresonant corrections to the total cross section}
	\label{subsection2}
	
	In this section we consider nonresonant corrections to the total cross section for one-photon scattering process in the hydrogen atom. 
	Neglecting the hyperfine structure, the set of quantum numbers for a particular atomic state is $nljm$, where $n$ is the principal quantum number, $l$ is the angular momentum of electron, $j$ is total angular momentum of electron ($j=l+s$) and $m_{j}$ its projection.
	
The total cross section for the one-photon scattering process is obtained by integrating Eq. (\ref{eq18}) over the emission directions of the outgoing photon $\boldsymbol{n}_{{\boldsymbol{k}_2}}$ and summing over the final photon polarizations $\boldsymbol{e}^*_2 $. If the incident radiation is isotropic and unpolarized then additional integration over $\boldsymbol{n}_{{\boldsymbol{k}_1}}$ and averaging over polarization $\boldsymbol{e}_1 $ should be performed. Then, integrating over the photon directions, summing over the photon polarizations and projections of the final state $m_{j_f}$ and averaging over the projections of the initial state $m_{j_f}$, we find
	\begin{eqnarray}
		\label{eq24}
		\sigma_{fi} = \frac{e^4}{(2\pi)^3}\frac{\omega(\omega_{0}-\omega)}{(2j_i + 1)(2j_{\gamma_2} + 1)}\sum_{m_{j_f}m_{j_i}}\sum_{\substack{j_{\gamma_1}m_{\gamma_1}s_1\\j_{\gamma_2}m_{\gamma_2}s_2}} \Bigg{|}\sum_{m_{r}}\frac{\Big{(} \boldsymbol{\alpha}\boldsymbol{A}^{(s_2)*}_{j_{\gamma_2}m_{\gamma_2}} \Big{)}_{fr} \Big{(} \boldsymbol{\alpha}\boldsymbol{A}^{(s_1)}_{j_{\gamma_1}m_{\gamma_1}} \Big{)}_{ri}}{E_{r} - E_i - \omega - \frac{i}{2}\Gamma_{r}} + \\
		\nonumber 
		+ \sum_{n\neq r} \frac{\Big{(} \boldsymbol{\alpha}\boldsymbol{A}^{(s_2)*}_{j_{\gamma_2}m_{\gamma_2}} \Big{)}_{fn} \Big{(} \boldsymbol{\alpha}\boldsymbol{A}^{(s_1)}_{j_{\gamma_1}m_{\gamma_1}} \Big{)}_{ni}}{E_n - E_i - \omega} + \sum_{n} \frac{\Big{(} \boldsymbol{\alpha}\boldsymbol{A}^{(s_1)}_{j_{\gamma_1}m_{\gamma_1}} \Big{)}_{fn} \Big{(} \boldsymbol{\alpha}\boldsymbol{A}^{(s_2)*}_{j_{\gamma_2}m_{\gamma_2}} \Big{)}_{ni}}{E_n - E_f + \omega} \Bigg{|}^2
		,
	\end{eqnarray}
	where the orthogonality property for $C^{j_{\gamma_1}m_{\gamma_1}s_1}_{j_{\gamma_2}m_{\gamma_2}s_2}$ \cite{ZALIALIUTDINOV20181, PhysRevA.24.183} is used and $\omega_0 \equiv E_{r} - E_i$. Here and below, we also neglect the frequency dependence in $\Gamma_r$, assuming the insignificance of Lorentz wings in atomic spectroscopy experiments. Then the total cross section can be written as a sum of two contributions 
	\begin{eqnarray}
		\label{eq30}
		\sigma_{fi} =\sigma^{(0)}_{fi} + \sigma^{(1)}_{fi},
	\end{eqnarray}
	where $\sigma^{(0)}_{fi}$ is the resonant contribution 
	\begin{eqnarray}
		\label{eq28}
		\sigma^{(0)}_{fi}(\omega js)=e^4\frac{\omega_{ri}\omega_{rf}}{(2\pi)^4}\frac{2j_{r}+1}{(2j_{\gamma}+1)(2j_i+1)}\frac{\Gamma_{fr}W_{ir}(js)}{(\omega_0 - \omega)^2 + \frac{\Gamma^2_{r}}{4}},
	\end{eqnarray}
	and $\sigma^{(1)}_{fi}$ is the nonresonant contribution to $\sigma^{(0)}_{fi}$
	\begin{eqnarray}
		\label{eq29}
		\sigma^{(1)}_{fi}(\omega js)=2e^4\frac{\omega_{ri}\omega_{rf}}{(2\pi)^4}\frac{2j_{r}+1}{(2j_{\gamma}+1)(2j_i+1)}\Re\Bigg{[}\sum_{n\neq r} \frac{\Gamma_{rf;fn}W_{ir;ni}(js)}{(\omega_0 - \omega - \frac{i}{2}\Gamma_{r})(E_n - E_{r})} + \sum_{n}\frac{\Gamma_{rf;ni}W_{ir;fn}(js)}{(\omega_0 - \omega -\frac{i}{2}\Gamma_{r})(E_n - E_f + \omega_0)} \Bigg{]}.
	\end{eqnarray}
	In Eqs. (\ref{eq28}) and (\ref{eq29}) we introduced the following notations \cite{2001}
	\begin{eqnarray}
		\label{eq25}
		W_{AB;CD}(j_{\gamma}s)=\frac{2\pi}{2j_{D}+1}\sum_{\substack{m_Am_B \\ m_Cm_D}}\sum_{m_{\gamma}}\big{(} \boldsymbol{\alpha}\boldsymbol{A}^{(s)*}_{j_{\gamma}m_{\gamma}} \big{)}_{AB} \big{(} \boldsymbol{\alpha}\boldsymbol{A}^{(s)}_{j_{\gamma}m_{\gamma}} \big{)}_{CD},
	\end{eqnarray}
	\begin{eqnarray}
		\label{eq26}
		W_{AB}\equiv W_{AB;BA}(j_{\gamma}s),
	\end{eqnarray}
	\begin{eqnarray}
		\label{eq27}
		\Gamma_{AB;CD}=\sum_{j's'}W_{AB;CD}(j_{\gamma}'s'),
	\end{eqnarray} 
	so that $\Gamma_{AB;BA}\equiv\Gamma_{AB}$ is a partial width of level $A$.
	
	Defining the resonant frequency of the $i+\gamma_{1}\rightarrow r$ transition as the maximum of the cross section Eq. (\ref{eq30}), the nonresonant correction is
	\begin{eqnarray}
		\label{eq31}
		\delta_{\mathrm{NR}}=-\frac{1}{4}\frac{\Gamma^2_{r}}{\Gamma_{fr}W_{ir}}\Re\Bigg{[} \sum_{n\neq r}\frac{\Gamma_{rf;fn}W_{ir;ni}(j_{\gamma}s)}{E_n - E_{r}} + \sum_{n}\frac{\Gamma_{rf;ni}W_{ir;fn}(j_{\gamma}\lambda)}{E_n - E_f + \omega_0} \Bigg{]}.
	\end{eqnarray}
	
For the Ly$_{\alpha}$ transition in a neutral hydrogen atom ($i = f = 1s$, $r =2p$), numerical calculations of Eq. (\ref{eq25}) were performed in \cite{2001}. The result is
	\begin{eqnarray}
		\label{eq32}
		\delta_{\mathrm{NR}}\equiv \delta^{(2p)}_{1s,1s}=-2.93\text{ Hz}.
	\end{eqnarray}
	It should be noted that  in the scattering experiment the Ly$_{\alpha}$ resonance consists of two peaks corresponding to fine structure components. In the nonrelativistic approximation the distortion of these two peaks is the same and is determined by Eq. (\ref{eq32}). 
	
	 Similar calculations were performed in \cite{2001} for the $1s_{1/2}\rightarrow 2p_{3/2} \rightarrow 1s_{1/2}$ transition. The NR correction turned out to be
	\begin{eqnarray}
		\label{eq33}
		\delta^{(2p_{1/2})}_{2s,1s}=-1.51\text{ Hz}.
	\end{eqnarray}
	Values given by Eqs. (\ref{eq32}) and (\ref{eq33}) are an order of magnitude less than the accuracy of modern experiments. For a two-photon $2s-1s$ transition in a neutral hydrogen atom, the experimental error is about 11 Hz with a relative uncertainty of few parts of $10^{-15}$ \cite{1s2sprecise}, while the Ly$_\alpha$ transition was measured with much larger error (about 1 MHz) in \cite{Eikema2001}. Nevertheless, it is the nonresonant correction to the total cross section that sets the principal limit in measurements of photon scattering experiments, determining the asymmetry of the line profile which can not be vanished, see the discussion in \cite{doi:10.1139/p07-014,PhysRevA.79.052506,PRL-LSSP}.

	\subsection{Differential scattering cross section: two neighboring resonances}
		It is convenient to represent the amplitude of photon scattering (see Eq. (\ref{eq18})) as
	\begin{eqnarray}
		\label{eq20}
		U^{(2)}_{fi} = e^2\frac{2\pi}{\sqrt{\omega_1 \omega_2}}\Bigg{[} \sum_{\substack{j_{\gamma_1}m_{\gamma_1}s_1\\\j_{\gamma_2}m_{\gamma_2}s_2}}   
		C^{j_{\gamma_1}m_{\gamma_1}s_1}_{j_{\gamma_2}m_{\gamma_2}s_2}(\boldsymbol{e}_1, \boldsymbol{n}_{\boldsymbol{k}_1}; \boldsymbol{e}_2, \boldsymbol{n}_{\boldsymbol{k}_2}) \Bigg{\{}\sum_{m_{r}}\frac{\Big{(} \boldsymbol{\alpha}\boldsymbol{A}^{(s_2)*}_{j_{\gamma_2}m_{\gamma_2}} \Big{)}_{fr} \Big{(} \boldsymbol{\alpha}\boldsymbol{A}^{(s_1)}_{j_{\gamma_1}m_{\gamma_1}} \Big{)}_{ri}}{E_{r} - E_i - \omega_1 - \frac{i}{2}\Gamma_{r}} + \\
		\nonumber 
		+ \sum_{n\neq r} \frac{\Big{(} \boldsymbol{\alpha}\boldsymbol{A}^{(s_2)*}_{j_{\gamma_2}m_{\gamma_2}} \Big{)}_{fn} \Big{(} \boldsymbol{\alpha}\boldsymbol{A}^{(s_1)}_{j_{\gamma_1}m_{\gamma_1}} \Big{)}_{ni}}{E_n - E_i - \omega_1} + \sum_{n} \frac{\Big{(} \boldsymbol{\alpha}\boldsymbol{A}^{(s_1)}_{j_{\gamma_1}m_{\gamma_1}} \Big{)}_{fn} \Big{(} \boldsymbol{\alpha}\boldsymbol{A}^{(s_2)*}_{j_{\gamma_2}m_{\gamma_2}} \Big{)}_{ni}}{E_n - E_i - \omega_2} \Bigg{\}} \Bigg{]}.
	\end{eqnarray}
	Here, in nonresonant terms, the infinitesimal parts $(1-i0)$ in the denominators are omitted, since the divergent contributions are now absent. In the resonant approximation it is assumed that $\Gamma_{r}$ is independent on $\omega_1$ and the photon emission operators given by Eq. (\ref{eq16}) are taken at fixed transition energies, i.e. with $\omega_{1}=E_{r}-E_{i}$ and $\omega_{2}=E_{r}-E_{f}$.  Retaining only the first (resonant) term in Eq. (\ref{eq20}), and after the integration over $\omega_2$ in Eq. (\ref{eq7}), the corresponding scattering cross section (line profile) becomes symmetric with respect to the resonant frequency $\omega_{0}\equiv E_{r}-E_{i}$:
	\begin{eqnarray}
		\label{eq22}
		\frac{d\sigma_{fi}(\omega_1)}{d\boldsymbol{n}_{k_2}}=\mathrm{const}\frac{f^{(1\gamma)}_{fi}(r,r)}{(\omega_1 - \omega_0)^2 + \frac{\Gamma^2_{r}}{4}},
	\end{eqnarray}
	where
	\begin{eqnarray}
		\label{fr}
		f^{(1\gamma)}_{fi}(r,r') = 
		\left( \sum_{\substack{j_{\gamma_1}m_{\gamma_1}s_1\\ j_{\gamma_2}m_{\gamma_2}s_2}}   
		C^{j_{\gamma_1}m_{\gamma_1}s_1}_{j_{\gamma_2}m_{\gamma_2}s_2}(\boldsymbol{e}_1, \boldsymbol{n}_{\boldsymbol{k}_1}; \boldsymbol{e}_2, \boldsymbol{n}_{\boldsymbol{k}_2}) \sum_{m_{r}} 
		\Big{(} \boldsymbol{\alpha}\boldsymbol{A}^{(s_2)*}_{j_{\gamma_2}m_{\gamma_2}} \Big{)}_{fr} \Big{(} \boldsymbol{\alpha}\boldsymbol{A}^{(s_1)}_{j_{\gamma_1}m_{\gamma_1}} \Big{)}_{ri}
		\right )
		\\\nonumber\times
		\left( \sum_{\substack{j'_{\gamma_1}m'_{\gamma_1}s'_1\\ j'_{\gamma_2}m'_{\gamma_2}s'_2}}   
		C^{j'_{\gamma_1}m'_{\gamma_1}s'_1}_{j'_{\gamma_2}m'_{\gamma_2}s'_2}(\boldsymbol{e}_1, \boldsymbol{n}_{\boldsymbol{k}_1}; \boldsymbol{e}_2, \boldsymbol{n}_{\boldsymbol{k}_2}) \sum_{m_{r'}} 
		\Big{(} \boldsymbol{\alpha}\boldsymbol{A}^{(s'_2)*}_{j'_{\gamma_2}m'_{\gamma_2}} \Big{)}_{fr'} \Big{(} \boldsymbol{\alpha}\boldsymbol{A}^{(s'_1)}_{j'_{\gamma_1}m'_{\gamma_1}} \Big{)}_{r'i}
		\right )^*
	\end{eqnarray}
	and "$\mathrm{const}$" is a constant unnecessary for further analysis of the cross section. Thus, when only the resonant term is taken into account, the spectral line profile has a Lorentzian shape, with a maximum at $\omega=\omega_0$. Then the resonant transition frequency $ \omega_{\mathrm{res}} $ can be determined as the extreme of Eq. (\ref{eq22}).
	
	Accounting for the remaining (nonresonant) terms in Eq. (\ref{eq20}) leads to asymmetry of the line profile and the search for the maximum becomes inconsistent. However, if the asymmetry is minor, then the resonant transition frequency $ \omega_{\mathrm{res}} $ can still be determined from $ d\sigma_{if} $ as $\omega_{\mathrm{res}}=\omega_{\mathrm{max}}$, where $ \omega_{\mathrm{max}} $ corresponds to the maximum value of $ \sigma_{if}(\omega) $:
	\begin{eqnarray}
		\label{def18}
		\left.\frac{d\sigma_{if}(\omega)}{d\omega}\right|_{\omega = \omega_{\mathrm{max}}}=0.
	\end{eqnarray}
	In the resonant approximation, see Eq. (\ref{eq22}), we immediately find
	\begin{eqnarray}
		\label{def19}
		\omega_{\mathrm{res}}=\omega_{\mathrm{max}}=\omega_{0}.
	\end{eqnarray}
	As long as the line profile is symmetrical with respect to $\omega=\omega_{\mathrm{max}}$, the definition (\ref{def19}) remains equivalent to any other way of extracting $\omega_{\mathrm{res}}$ from the line profile. In the case of imperceptible asymmetry, the condition (\ref{def18}) leads to the shifted value, i.e. $\omega_{\mathrm{max}} = \omega_{\mathrm{res}} + \delta_{\mathrm{NR}}$, where $\delta$ is determined by the nonresonant terms in Eq. (\ref{eq20}). The restriction imposed on the definition of the resonant frequency from (\ref{def18}) in the presence of asymmetry follows from the relation $\delta_{\mathrm{NR}}\ll \Gamma_{r}$. Otherwise, even for $\delta_{\mathrm{NR}}\sim \Gamma_{r}$, the transition frequency definition becomes ambiguous.
	
	Let us assume that in the process of resonant absorption $i+\gamma_1\rightarrow r$ of incident photon another closest transition $i+\gamma_1\rightarrow r'$ is also allowed by the selection rules \cite{PhysRevA.91.033417,PhysRevA.93.012510,zalialiutdinov2017generalized}. Then similar to Eq. (\ref{eq20}), the two resonant terms in the amplitude can be written explicitly as follows
	\begin{eqnarray}
		\label{eq20-2}
		U^{(2)}_{fi} = e^2\frac{2\pi}{\sqrt{\omega_1 \omega_2}}\Bigg{[} \sum_{\substack{j_{\gamma_1}m_{\gamma_1}s_1\\ j_{\gamma_2}m_{\gamma_2}s_2}}   
		C^{j_{\gamma_1}m_{\gamma_1}s_1}_{j_{\gamma_2}m_{\gamma_2}s_2}(\boldsymbol{e}_1, \boldsymbol{n}_{\boldsymbol{k}_1}; \boldsymbol{e}_2, \boldsymbol{n}_{\boldsymbol{k}_2}) \Bigg{\{}\sum_{m_{r}}\frac{\Big{(} \boldsymbol{\alpha}\boldsymbol{A}^{(s_2)*}_{j_{\gamma_2}m_{\gamma_2}} \Big{)}_{fr} 
			\Big{(} \boldsymbol{\alpha}\boldsymbol{A}^{(s_1)}_{j_{\gamma_1}m_{\gamma_1}} \Big{)}_{ri}}{E_{r} - E_i - \omega_1 - \frac{i}{2}\Gamma_{r}} 
	    \\
		\nonumber 
		+
		\sum_{m_{r'}}
		\frac{\Big{(} \boldsymbol{\alpha}\boldsymbol{A}^{(s_2)*}_{j_{\gamma_2}m_{\gamma_2}} \Big{)}_{fr'} 
			\Big{(} \boldsymbol{\alpha}\boldsymbol{A}^{(s_1)}_{j_{\gamma_1}m_{\gamma_1}} \Big{)}_{r'i}}{E_{r'} - E_i - \omega_1 - \frac{i}{2}\Gamma_{r'}}
		+ \sum_{n\neq r,r'} \frac{\Big{(} \boldsymbol{\alpha}\boldsymbol{A}^{(s_2)*}_{j_{\gamma_2}m_{\gamma_2}} \Big{)}_{fn} \Big{(} \boldsymbol{\alpha}\boldsymbol{A}^{(s_1)}_{j_{\gamma_1}m_{\gamma_1}} \Big{)}_{ni}}{E_n - E_i - \omega_1} + \sum_{n} \frac{\Big{(} \boldsymbol{\alpha}\boldsymbol{A}^{(s_1)}_{j_{\gamma_1}m_{\gamma_1}} \Big{)}_{fn} \Big{(} \boldsymbol{\alpha}\boldsymbol{A}^{(s_2)*}_{j_{\gamma_2}m_{\gamma_2}} \Big{)}_{ni}}{E_n - E_i - \omega_2} \Bigg{\}} \Bigg{]}
		,
	\end{eqnarray}
	where $\Gamma_{r'}$ is the natural width of the state $r'$. The last two terms in Eq. (\ref{eq20-2}) represent an off-resonant contribution to the amplitude and can be neglected in the resonant approximation, i.e. assuming that $\omega_{1}$ is close to the energy difference $E_{r}-E_{i}$ or $E_{r'}-E_{i}$. After integration over $\omega_2$ in Eq. (\ref{eq7}), the corresponding cross section can be reduced to
	\begin{eqnarray}
		\label{eq22nr}
		\frac{d\sigma_{fi}(\omega_1)}{d\boldsymbol{n}_{k_2}}=\mathrm{const}
		\left(
		\frac{f^{(1\gamma)}_{fi}(r,r)}{(\omega_0 - \omega_1)^2 + \frac{\Gamma^2_r}{4}} 
		+ \frac{2f^{(1\gamma)}_{fi}(r,r')(\omega_0 - \omega_1)(\omega_0 - \omega_1 + \Delta)}{\big{(}(\omega_0 - \omega_1)^2 + \frac{\Gamma^2_{r'}}{4}\big{)}\big{(}(\omega_0 - \omega_1 + \Delta)^2 + \frac{\Gamma^2_{r'}}{4}\big{)}}
		+
		\frac{f^{(1\gamma)}_{fi}(r',r')}{(\omega_0 - \omega_1 + \Delta)^2 + \frac{\Gamma^2_{r'}}{4}} 
		\right)
		,
	\end{eqnarray}
	where $\Delta\equiv E_{r'}-E_{r}$. The first and the third terms in Eq. (\ref{eq22nr}) describe the Lorentzian profile for the transitions $i+\gamma_1\rightarrow r$ and $i+\gamma_1\rightarrow r'$, respectively, while the second term corresponds to the interference (in the resonant approximation) between them. 
	
	The nonresonant leading-order correction to the transition frequency $i+\gamma_1\rightarrow r$ can be obtained from Eq. (\ref{eq22nr}) by neglecting the last term. Then using the definition Eq. (\ref{def18}), we find 
	\begin{eqnarray}
		\label{add1}
		\frac{d}{d\omega_1}\sigma_{if}(\omega_1)
		=
		-\frac{8
			\left(
			f^{(1\gamma)}_{\mathrm{fi}}(r,r')
			\left(
			\Gamma_r^2-4 (\omega_1-\omega_{0})^2
			\right)+4 \Delta  f^{(1\gamma)}_{fi}(r,r) (\omega_1-\omega_{0})
			\right)
		}
		{\Delta  \left(\Gamma_r^2+4 (\omega_1-\omega_{0})^2\right)^2}
		=0
		.
	\end{eqnarray}
	Expansion of Eq. (\ref{add1}) into a Taylor series in the vicinity of $ \omega_{0} $ yields
	\begin{eqnarray}
		\label{add2}
		-\frac{8 f^{(1\gamma)}_{fi}(r,r')}{\Gamma_r^2\Delta}
		-\frac{32 f^{(1\gamma)}_{fi}(r,r)(\omega_1 -\omega_{0})}{\Gamma_r^4}
		+O\left((\omega_1 -\omega_{0})^2\right)
		=0
		.
	\end{eqnarray}
	Finally, solving Eq. (\ref{add2}) for $ \omega_1 $, we arrive at
	\begin{eqnarray}
		\label{23}
		\omega_{\mathrm{max}}=\omega_{0}-\delta_{\mathrm{NR}},
	\end{eqnarray}
	where
	\begin{eqnarray}
		\label{LEAD}
		\delta_{\mathrm{NR}} = \frac{f^{(1\gamma)}_{fi}(r,r')}{f^{(1\gamma)}_{fi}(r,r)}\frac{\Gamma_{r}^2}{4\Delta}.
	\end{eqnarray}
	Eq. (\ref{LEAD}) is obtained as the lowest order term of an expansion of the result in terms of $ \Gamma /\Delta $ and represents the part of NR correction called quantum interference effect (QIE). The approximations that were used to derive Eq. (\ref{24}) are valid up to higher orders of the $ \Gamma /\Delta $ parameter, which is usually small (e.g. for two neighbouring fine structure components, see specific examples below). The parameter $\Gamma/\Delta$ may not be small for two neighboring hyperfine sublevels and in this case requires a special study \cite{PhysRevA.79.052506}. 
	
	The next order contribution can be obtained from Eq. (\ref{eq22nr}) by keeping the last term corresponding to the Lorentzian line shape of second resonance $i+\gamma_1\rightarrow r'$. Then repeating the calculations above, the nonresonant correction to the transition frequency takes the form:
	\begin{eqnarray}
		\label{eqNLO}
		\delta_{\mathrm{NR}} = \frac{f^{(1\gamma)}_{fi}(r,r') }{
			f^{(1\gamma)}_{fi}(r,r)}\frac{\Gamma_r^2}{4 \Delta }
		\\\nonumber
		+\frac{\left(\left[f^{(1\gamma)}_{fi}(r,r')\right]^3+2 \left[f^{(1\gamma)}_{fi}(r,r')\right]^2 f^{(1\gamma)}_{fi}(r,r)+f^{(1\gamma)}_{fi}(r,r')\left[(f^{(1\gamma)}_{fi}(r,r)\right]^2+\left[f^{(1\gamma)}_{fi}(r,r)\right]^2 f^{(1\gamma)}_{fi}(r',r')\right)}{\left[f^{(1\gamma)}_{fi}(r,r)\right]^3}\frac{\Gamma_r^4 }{16 \Delta ^3 }.
	\end{eqnarray}
	The first term in Eq. (\ref{eqNLO}) is exactly the same as in Eq. (\ref{LEAD}), while the second is proportional to the ratio $\frac{\Gamma_r^4 }{16 \Delta ^3 }$. In most cases considered in this review, the second contribution is a small addition to the leading order correction  Eq. (\ref{LEAD}). 
	
	As follows from Eq. (\ref{23}), the NR correction depends on the arrangement of the experiment, that is, on the angular and polarization correlations between the incident and outgoing photons. All information about such correlations is given by the ratio $f^{(1\gamma)}_{fi}(r,r')/f^{(1\gamma)}_{fi}(r,r)$. Since the coefficients Eq. (\ref{fr}) depend on the angular quantum numbers of particular atomic states and the mutual orientation of vectors $\boldsymbol{e}_1,\, \boldsymbol{n}_{\boldsymbol{k}_1},\, \boldsymbol{e}_2,\, \boldsymbol{n}_{\boldsymbol{k}_2}$, the resulting value of the NR correction is determined by the experimental geometry and the photon detection method \cite{Jent-NR,Solovyev2020}.
	
	The NR effects are not limited to QIE that arises due to resonant scattering at neighbouring atomic levels. In particular, the nonresonant terms in Eq. (\ref{eq20-2}), representing the scattering on virtual states, may be relevant for highly charged ions, see, for example, \cite{Labzowsky1994}. Furthermore, the expressions Eq. (\ref{LEAD}) and (\ref{eqNLO}) were obtained in the resonant approximation. However, for light atomic systems and particular examples considered below, they are insignificant at the present level of experimental accuracy.
	
	Before proceeding to partial examples, it is necessary to write down the scattering amplitude. In the nonrelativistic limit and dipole approximation \cite{A-B-QED, LabKlim} (see Appendix \ref{appendix1}) it is
	\begin{eqnarray}
		\label{eq23}
		U^{(2)}_{fi}=2\pi e^2\sqrt{\omega_1\omega_2}\Bigg{[} \sum_n \frac{\Big{(}\boldsymbol{r}\boldsymbol{e}^*_2 \Big{)}_{fn} \Big{(} \boldsymbol{r}\boldsymbol{e}_1 \Big{)}_{ni}}{E_n(1-i0) - E_i - \omega_1} + \sum_n \frac{\Big{(} \boldsymbol{r}\boldsymbol{e}_1 \Big{)}_{fn} \Big{(} \boldsymbol{r}\boldsymbol{e}^*_2 \Big{)}_{ni}}{E_n(1-i0) - E_f + \omega_1}\Bigg{]},
	\end{eqnarray} 
	where now the summation runs over the Schr\"odinger spectrum.

	\section{Angular correlations and the limits of precision measurements: applications to one-photon spectroscopy}
	\label{angleone}
	
	\subsection{Quantum interference effect}
	\label{QIE}
	Unlike the resonant value of the transition frequency, the NR corrections depend on the processes of excitation and de-excitation of the atomic level, the type of experiment and the method of extracting the transition frequency value from the experimental data. Therefore, refinement of the transition frequency value may have sense until the NR corrections are smaller than the accuracy of the experiment \cite{Labzowsky1994, 2001}. For all cases studied in \cite{Labzowsky1994,2001, Jent-NR} and later works on this topic, the NR corrections appeared to be negligible. In particular, according to \cite{PhysRevA.79.052506}, this was also the case for the measurement  the two-photon $1s-2s$ transition frequency in hydrogen in \cite{1s2sprecise}. The situation changed when the results of the high-precision measurement of the $2s_{1/2}^{F=0}\rightarrow 4p_{1/2}^{F=1}$ and $2s_{1/2}^{F=0}\rightarrow 4p_{3/2}^{F=1}$ transition frequencies in \cite{Science} were reported. The uncertainty of these measurements was quoted to be much smaller than the observed interference effects. According to line profile theory, these interference effects manifest the existence of NR corrections. 
	
	In the present section we investigate the problem from this point of view. We derive expressions for the cross section of resonant photon scattering on hydrogen atom, taking into account the fine and hyperfine structure of atomic levels. The corresponding expressions for amplitudes contain dependencies on the directions and polarizations of the incident (absorbed) and outgoing (emitted) photons. This can be used to describe different types of experiments with different correlations between the directions and polarizations of the two photons. All this is used to derive NR corrections to the photon scattering cross sections and to determine the transition frequencies. 

	Following \cite{Solovyev2020} we focus on NR corrections originating from the neighbouring components of the fine structure levels, as in \cite{Science}, where the mutual influence of the transitions $ 2s_{1/2}^{F=0}\rightarrow 4p_{1/2}^{F=1}$ and $ 2s_{1/2}^{F=0}\rightarrow 4p_{3/2}^{F=1} $ was observed. First, we consider NR corrections to $2s_{1/2}^{F=0}\rightarrow 4p_{1/2}^{F=1}$ transitions due to quantum interference with $2s_{1/2}^{F=0}\rightarrow 4p_{3/2}^{F=1}$ transitions. Corresponding corrections to another transition $2s_{1/2}^{F=0}\rightarrow 4p_{3/2}^{F=1}$ are similar but have opposite sign. 
	We demonstrate that the NR corrections to these transitions in this particular case do not depend on the type of experiment and experimental geometry. However, they do depend on the choice of the detected decay branch: they are different if the detection process ends in the states with $ F=0,\,1$ or $2$. When the frequency of the outgoing photon is not fixed at all, the result of the measurement begins to depend on both the type of experiment and the experimental arrangement (geometry).
	
	We will distinguish two types of experiments of that sort. In an experiment of the first type the directions of photon propagation are fixed: the incident photon direction $ \boldsymbol{n}_{k_1} $ coincides with the direction of the laser beam and the outgoing photon direction $ \boldsymbol{n}_{k_2} $ is defined by the detector position. In the second type of experiment the incident photon polarization $ \boldsymbol{e}_{1} $ and the outgoing photon direction $ \boldsymbol{n}_{k_2} $ are fixed; this is exactly the situation in experiment \cite{Science}. 
	
	In the nonrelativistic limit the matrix elements in the scattering amplitude given by Eq. (\ref{eq23}) do not depend explicitly on the photon directions $ \boldsymbol{n}_{k_1} $ and $ \boldsymbol{n}_{k_2} $. Implicitly this dependence enters via the transversality condition. Since in the laser beam the transversality condition will be satisfied automatically, the direction of this beam in an experiment of type 2 can be chosen arbitrary. Dependence on $ \boldsymbol{n}_{k_1} $, $ \boldsymbol{n}_{k_2} $ becomes explicit after summation over photon polarizations. Then for the type 1 experiment we have to evaluate $ \sum\limits_{\boldsymbol{e}_1, \boldsymbol{e}_2}d\sigma_{if} $, for the the type 2 experiment it is necessary to evaluate $ \sum\limits_{\boldsymbol{e}_2}d\sigma_{if} $, where the differential cross section is given by Eq. (\ref{eq22nr}). In the nonrelativistic limit the corresponding factors $f_{fi}(r,r')$ in cross section Eq. (\ref{eq22nr}) for the experiment of the first and second types are given by the following expressions \cite{Solovyev2020}:
	\begin{eqnarray}
		\label{eq54}
		f^{(1)}_{fi}(r,r') = 36\sum_{xy}(-1)^{F_{r'} - F_r +x+y} \Pi^2_x\Pi_y 
		\begin{Bmatrix}
			1 & x & 1 \\
			F_{r'} & F_f & F_r
		\end{Bmatrix}
		\begin{Bmatrix}
			1 & x & 1 \\
			F_{r'} & F_i & F_r
		\end{Bmatrix}
		\begin{Bmatrix}
			1 & 1 & y \\
			1 & 1 & x
		\end{Bmatrix}
		\begin{Bmatrix}
			1 & 1 & x \\
			1 & 1 & 1
		\end{Bmatrix}^2\times
		\\\nonumber
		\times
		\left\lbrace\left\lbrace a_{1}^{(1)}\otimes b_{1}^{(1)} \right\rbrace_{y} 
		\otimes
		\left\lbrace a_{1}^{(1)}\otimes b_{1}^{(1)} \right\rbrace_{y}\right\rbrace_{00}
		\\\nonumber
		\times
		\langle n_fl_fj_fF_f||r||n_rl_rj_rF_r\rangle \langle n_rl_rj_rF_r||r||n_il_ij_iF_i\rangle \langle n_il_ij_iF_i||r||n_{r'}l_{r'}j_{r'}F_{r'}\rangle  \langle n_{r'}l_{r'}j_{r'}F_{r'}||r||n_fl_fj_fF_f\rangle,
	\end{eqnarray}
	\begin{eqnarray}
		\label{eq55}
		f^{(2)}_{fi}(r,r') = 6\sum_{xy}(-1)^{F_{r'} - F_r + y}\Pi^2_x\Pi_y 
		\begin{Bmatrix}
			1 & x & 1 \\
			F_{r'} & F_f & F_r
		\end{Bmatrix}
		\begin{Bmatrix}
			1 & x & 1 \\
			F_{r'} & F_i & F_r
		\end{Bmatrix}
		\begin{Bmatrix}
			1 & 1 & y \\
			1 & 1 & x
		\end{Bmatrix}
		\begin{Bmatrix}
			1 & 1 & x \\
			1 & 1 & 1
		\end{Bmatrix}
		\\\nonumber
		\times
		\left\lbrace\left\lbrace a_{1}^{(2)}\otimes b_{1}^{(2)} \right\rbrace_{y} 
		\otimes
		\left\lbrace a_{1}^{(2)}\otimes b_{1}^{(2)} \right\rbrace_{y}\right\rbrace_{00}
		\\\nonumber
		\times
		\langle n_fl_fj_fF_f||r||n_rl_rj_rF_r\rangle \langle n_rl_rj_rF_r||r||n_il_ij_iF_i\rangle \langle n_il_ij_iF_i||r||n_{r'}l_{r'}j_{r'}F_{r'}\rangle  \langle n_{r'}l_{r'}j_{r'}F_{r'}||r||n_fl_fj_fF_f\rangle,
	\end{eqnarray}
	where $ a^{(1)}_1=\boldsymbol{n}_{k_1} $, 
	$ a^{(2)}_1=\boldsymbol{e}_{1} $, $b^{(1)}_1=b^{(2)}_1=\boldsymbol{n}_{k_2}$ and $\Pi_{a}=\sqrt{2a+1}$. In Eqs. (\ref{eq54}), (\ref{eq55}) factor $\left\lbrace a_{1}^{(1,2)}\otimes b_{1}^{(1,2)} \right\rbrace_{y} $ denotes the tensor product of rank $y$ of two tensors $a_{1}^{(1,2)}$ and $b_{1}^{(1,2)}$ with rank $1$ \cite{VMK}. This tensor product completely defines the angular correlations in the scattering cross section.
	
	Defining the transition frequency via the maximum of cross section according to the condition Eq. (\ref{def18}) the corresponding NR leading order correction takes the form:
	\begin{eqnarray}
		\label{24}
		\delta_{\mathrm{NR}}^{(1,2)} = \frac{f^{(1,2)}_{fi}(r,r')}{f^{(1,2)}_{fi}(r,r)}\frac{\Gamma_r^2}{4\Delta}.
	\end{eqnarray}
	With our definition of $ \Delta = E_{r'}-E_{r}$ this value corresponds to the lower component of the fine structure of the resonant level $ nl $. For the upper sublevel of two neighboring components of energy level we would arrive to the same expression as Eq. (\ref{24}) but with the opposite sign of $ \Delta $ and with $ \Gamma=\Gamma_{nlj_{r'}F_{r'}} $. NR correction in Eq. (\ref{24}) can depend on the arrangement of the experiment, i.e., on the angles between the vectors $ \boldsymbol{n}_{k_1} $ and $ \boldsymbol{n}_{k_2} $ in the experiment of type 1 or on the angles between the vectors $ \boldsymbol{e}_{k_1} $ and $ \boldsymbol{n}_{k_2} $ in the experiment of type 2.

	\subsection{Application to spectroscopy of hydrogen}
	\label{subsection5}
	
	Now we turn to evaluation of $ 2s_{1/2}^{F=0}\rightarrow 4p_{1/2}^{F=1} $ transition frequency with account for NR corrections originating from the neighboring $ 4p_{3/2}^{F=1} $ level \cite{Science,Solovyev2020}. For this purpose we set in all equations $ n_i l_i=2s $, $ j_i=1/2 $, $ F_i=0 $, $ nl=4p $, $ j=1/2 $, $ F=1 $, $ j'=3/2 $, $ F'=1 $. 
	The results of evaluations of nonresonant corrections in the $ 2s_{1/2}^{F=0}\rightarrow 4p_{1/2}^{F=1}(4p_{3/2}^{F=1}) \rightarrow f $ photon scattering process with a fixed final state, $f$, are presented in Table \ref{tab1phNR}. 
	\begin{table}
		\centering
		\caption{The NR corrections in kHz to the transitions frequency $ 2s_{1/2}^{F=0}\rightarrow 4p_{1/2}^{F=1} $ with account for the neighbouring $ 4p_{3/2}^{F=1} $ state for the experiment of type 2 ($ \boldsymbol{e}_1\boldsymbol{n}_{k_2} $ correlation). The same values are obtained for the experiment of type 1 ($ \boldsymbol{n}_{k_1}\boldsymbol{n}_{k_2} $ correlation).}
		\begin{tabular}{|c c|}
			\hline
			Final state & $ \delta_{\mathrm{NR}}^{(2)} $\\
			\hline
			$ 1s_{1/2}^{F=0} $ &   61.2355\\
			$ 1s_{1/2}^{F=1} $ &  -30.6178 \\
			$ 2s_{1/2}^{F=0} $ &   61.2357 \\
			$ 2s_{1/2}^{F=1} $ &  -30.6178 \\
			$ 3s_{1/2}^{F=0} $ &   61.2362 \\
			$ 3s_{1/2}^{F=1} $ &  -30.6181 \\
			$ 3d_{3/2}^{F=1} $ &   30.6174\\
			$ 3d_{3/2}^{F=2} $ &    6.1236  \\
			\hline
		\end{tabular}
		\label{tab1phNR}
	\end{table}
	For evaluation of NR corrections according to Eqs. (\ref{eq54}), (\ref{eq55}) and (\ref{24}) we use theoretical values given in \cite{HorbHess}, which incorporate relativistic, QED, nuclear size, the hyperfine structure corrections. The same concerns the value of the widths $ \Gamma=\Gamma_{4p_{1/2}^{F=1}}=1.2941\times 10^{7} $ Hz and the fine structure interval $ \Delta = E_{4p_{3/2}^{F=1}}-E_{4p_{1/2}^{F=1}}= 1367433.30(28)$ kHz \cite{HorbHess}. These values give a sufficiently accurate result for $ \delta \omega $ up to four digits after the decimal point. The parameter $ \Gamma /\Delta $ in this case is equal to $ 0.00946 $, so the expansion in powers of this parameter works very well.
	
	As can be seen, the NR corrections to the transition frequency $ 2s_{1/2}^{F=0}\rightarrow 4p_{1/2}^{F=1} $ do not depend on the type of experiment and consequently on the geometry of this experiment. However, these NR corrections appear to depend strongly on the method of the frequency detection, i.e. on the choice of the state to which the excited $ 4p_{1/2}^{F=1} $ level finally decays. Moreover, this dependence concerns only the quantum numbers of this final state, and the result is nearly independent on the frequency of the outgoing photon. The latter circumstance is understandable since according to Eq. (\ref{24}) the NR corrections are proportional to the ratio $ f_{\mathrm{nr}}/f_{\mathrm{res}}$ where the corresponding energy differences nearly cancel. 
	
	When the hyperfine structure of the final levels is resolved, the NR corrections differ only by the values of the total angular momentum $ F_{f} $ of the final hyperfine sublevel. This can be seen from the closed expressions (\ref{eq54}), (\ref{eq55}), (\ref{24}) for the NR corrections via $ 6j $-symbols. Therefore, for the transition frequency $ 2s_{1/2}^{F=0}\rightarrow 4p_{1/2}^{F=1} $, three different values of $ \omega_{\mathrm{res}}^{\mathrm{max}\;(1,2)} $ corresponding to $ F_f=0,\,1,\,2 $ can be derived for both types of experiment by using $ \omega_0 $ from \cite{HorbHess} and NR corrections from Table \ref{tab1phNR}:
	\begin{eqnarray}
		\label{three}
		F_f=0\;\;\;\omega_{\mathrm{res}}^{\mathrm{max}\;(1,2)}
		=616520152497.3(4)\;\mathrm{kHz}
		\\\nonumber
		F_f=1\;\;\;\omega_{\mathrm{res}}^{\mathrm{max}\;(1,2)}
		=616520152527.9(4)\;\mathrm{kHz}
		\\\nonumber
		F_f=2\;\;\;\omega_{\mathrm{res}}^{\mathrm{max}\;(1,2)}
		=616520152552.4(4)\;\mathrm{kHz}
	\end{eqnarray}
	
	These three values differ from each other by more than 50 kHz (the uncertainties for $\omega_{0}$ are taken from \cite{HorbHess}). This is 15 times larger than the accuracy of measurement quoted in \cite{Science} (3 kHz). Nevertheless, all 3 numbers in Eq. (\ref{three}) have equal rights to be interpreted as "$ 2s_{1/2}^{F=0}-4p_{1/2}^{F=1} $ transition frequency". If in the process of the frequency measurement only the emission of the outgoing photon is detected without fixing of its frequency, the summation over all the final states should be done. In the case of our interest this summation looks as follows
	\begin{eqnarray}
		\label{avr}
		\delta_{\mathrm{NR}}^{(1,2)} = \frac{\sum\limits_{n_fl_fj_fF_f}f^{(1,2)}_{\mathrm{nr}}}{\sum\limits_{n_fl_fj_fF_f}f^{(1,2)}_{\mathrm{res}}}\frac{\Gamma^2}{4\Delta}
		.
	\end{eqnarray}
	
\begin{figure}[hbtp]
		\caption{The NR correction for the transition frequency $ 2s_{1/2}^{F=0}\rightarrow 4p_{1/2}^{F=1} $ as a function of the angle between the vectors $ \boldsymbol{n}_{k_1} $, $ \boldsymbol{n}_{k_2} $ for the experiment of type 1 (solid line) and as a function of the angle between the vectors $ \boldsymbol{e}_1 $, $ \boldsymbol{n}_{k_2} $ in the experiment of type 2 (dashed line) according to Eq. (\ref{avr}).}
		\centering
		\includegraphics[scale=0.85]{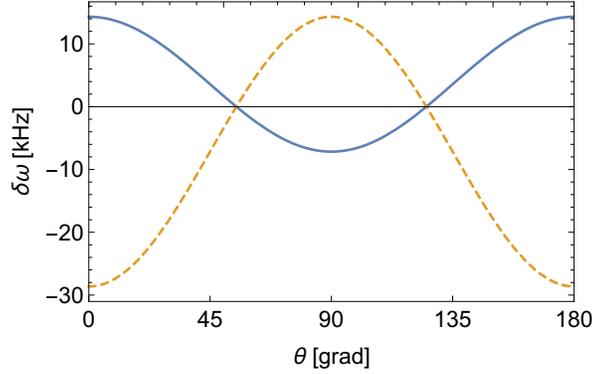}
		\label{fig2}
	\end{figure}

	Now the NR correction begins to depend on the type of the experiment and on the angles between the vectors $ \boldsymbol{n}_{k_1} $, $ \boldsymbol{n}_{k_2} $ in the experiment of the first type or between the vectors $ \boldsymbol{e}_1 $, $ \boldsymbol{n}_{k_2} $ in the experiment of the second type. The results for $ 2s_{1/2}^{F=0}\rightarrow 4p_{1/2}^{F=1}  $ transition are presented in Fig. \ref{fig2}.
	
	According to Eq. (\ref{avr}) and Fig. \ref{fig2}, the NR correction vanishes for certain angles $ \theta_1=54.7^{\circ}  $ and $ \theta_2=125.3^{\circ}  $, which are the same for both types of experiment. The possible use of the "magic angles" for determination of transition frequencies in atoms was mentioned in \cite{HorbHess,https://doi.org/10.1002/andp.201900044}. In \cite{https://doi.org/10.1002/andp.201900044} it was noted that the method of extracting the transition frequency value from the experimental data used in \cite{Science} is actually equivalent to the use of "magic angles". The "magic angles" are connected with the roots of equation $ P_{2}(x) =0$ where $ P_{2} $ is the Legendre polynomial.
	
	Recently, evaluation of atomic transition frequencies with the use of "magic angles" was considered in \cite{PhysRevA.92.062506}. Values for "magic angles" in \cite{PhysRevA.92.062506} coincide with quoted above for similar transitions. Evaluation of transition frequency $ 2s_{1/2}^{F=0}-4p_{1/2}^{F=1} $ with the use of Eq. (\ref{avr}) for "magic angles" with the theoretical values of $ \omega_0 $, $ \Gamma $ and $ \Delta $ from \cite{HorbHess} gives  
	\begin{eqnarray}
		\label{new1}
		\omega_{\mathrm{res}}^{\mathrm{max}\;(1,2)}=616520152558.5(4)\;\mathrm{kHz}.
	\end{eqnarray}
	A similar evaluation of the $2s_{1/2}^{F=0}-4p_{3/2}^{F=1}  $ transition frequency yields
	\begin{eqnarray}
		\label{new2}
		\omega_{\mathrm{res}}^{\mathrm{max}\;(1,2)}=616521519991.8(4)\;\mathrm{kHz}.
	\end{eqnarray}

	\subsection{Application to spectroscopy of muonic hydrogen}
	
	Following the analysis given in sections \ref{QIE} and \ref{subsection5}, we consider the QIE in the one-photon fluorescence spectroscopy of muonic hydrogen on the example of the transitions $2s_{1/2}^{F_i=0,\,1}\rightarrow 2p_{j_r}^{F_r}$. In this particular case, the nonresonant corrections to the transition frequencies arises due to the interference between the $2p_{j_r}^{F_r}$ sublevels (fine and hyperfine splitted states) in which the transitions $2s_{1/2}^{F_i=0,\,1}\rightarrow 2p_{j_{r}}^{F_{r}}$ and $2s_{1/2}^{F_i=0,\,1}\rightarrow 2p_{j_{r'}}^{F_{r'}}$ ($j_{r'}F_{r'}\neq j_{r}F_{r}$) are allowed by the electric dipole selection rules. The physical process considered here is the one-photon scattering described in section \ref{subsection1}. Hyperfine states of muonic hydrogen are separated by several hundred GHz \cite{HFS_muon,HFS_muon2, HFS_muon3} and have linewidths of a few tens of GHz \cite{MILOTTI1998137}. In particular, the natural width of $2p$ fine and hyperfine sublevels are approximately equivalent: $\Gamma_{2p_{3/2}^{F=1}}\approx\Gamma_{2p_{1/2}^{F=1}}\approx\Gamma_{2p_{3/2}^{F=2}}\approx\Gamma_{2p_{1/2}^{F=0}}=116.49\times 10^{9} $ s$ ^{-1} $ or $ 18.54 $ GHz. The used energies of atomic states for muonic hydrogen are listed in Table~\ref{tab:mu-energy}.
	\begin{table}[H]
	\label{tab:mu-energy}
		\centering
		\caption{Energies of atomic states in meV and Hz for muonic hydrogen. All the values are taken from \cite{MILOTTI1998137}.}
		\begin{tabular}{ c c c}
			\hline
			\hline
			state & meV & Hz \\
			\hline
			$ 1s_{1/2}^{F=0} $ & $ -2047.75 $ & $-4.95144\times 10^{14}  $\\
			$ 1s_{1/2}^{F=1} $ & $ -1865.3  $ & $-4.51028\times 10^{14 } $\\
			$ 2s_{1/2}^{F=0} $ & $ -244.37  $ & $-5.90804\times 10^{13 } $\\
			$ 2s_{1/2}^{F=1} $ & $ -221.532 $ & $-5.35662\times 10^{13 } $\\
			$ 2p_{1/2}^{F=0} $ & $ -30.9524 $ & $-7.48426\times 10^{12} $\\
			$ 2p_{1/2}^{F=1} $ & $ -23.3505 $ & $-5.64613\times 10^{12 } $\\
			$ 2p_{3/2}^{F=1} $ & $ -18.7182 $ & $-4.52604\times 10^{12}  $\\
			$ 2p_{3/2}^{F=2} $ & $ -15.6775 $ & $-3.7908\times 10^{12 } $\\
			\hline
			\hline
		\end{tabular}
	\end{table}
	
	Following the results of section \ref{angleone} we will be interested in the particular geometry of experiment when the the polarization vector $\boldsymbol{e}_1$ of incident photon is fixed. Then the NR correction could depend on the angle $\theta$ between vector $\boldsymbol{e}_1$ and propagation direction of outgoing photon $\boldsymbol{n}_{k_2}$. In Table \ref{tabMUH} the results for NR correction to the transition frequency $2s_{1/2}^{F_i=0,\,1}\rightarrow 2p_{j_r}^{F_r}$ are summarized for different final atomic states and angles $\theta = 0,\,\pi/2$. 
	\begin{table}[H]
		\centering
		\caption{Partial contributions $ \delta_{\mathrm{NR}}(i-r[r']) $ to the total NR correction to singlet ($F_{i}=0$) and triplet ($F_{i}=1$) transition line $i\rightarrow r$ arising due to the interference with the $i\rightarrow r'$ transition branch. The angle between the polarization vector of the incident photon and the propagation vector of the outgoing photon is denoted by $\theta$, which corresponds to the experiment of the second type according to section \ref{QIE}, the energy splitting is defined as $ \Delta\equiv E_{n_{r}l_{r}j_{r'}F_{r'}}-E_{n_rl_rj_rF_r} $. The results are presented for two different cases: 1) when the final states are assumed to be fixed; 2) the summation over all allowed final states is performed. Cases independent of angular correlation are given without specifying the angle $\theta$.}
		\begin{tabular}{ c c c l c r r}
			\hline
			\hline
			$ i $ & $ r $ & $ r' $ & $ f $ & $ \Delta $, Hz & $ \delta_{\mathrm{NR}}(i-r[r']) $, Hz & $ \delta_{\mathrm{NR}}(i-r[r']) $, meV\\
			\hline
			$ 2s_{1/2}^{F=0} $ & $ 2p_{3/2}^{F=1} $ & $ 2p_{1/2}^{F=1} $ & $ 1s_{1/2}^{F=0} $ & $ -1.12009\times 10^{12} $ & $ -3.71\times 10^{7} $ & $-1.53\times 10^{-4}$\\
			$ -              $ & $-               $ & $ -              $ & $ 1s_{1/2}^{F=1} $ & $  $ & $  7.41\times 10^{7} $ & $3.06\times 10^{-4}$\\
			$ -              $ & $-               $ & $ -              $ & $ 2s_{1/2}^{F=0} $ & $  $ & $ -3.60\times 10^{7} $ & $-1.49\times 10^{-4}$\\
			$ -              $ & $-               $ & $ -              $ & $ 2s_{1/2}^{F=1} $ & $  $ & $  7.18\times 10^{7} $ & $2.97\times 10^{-4}$\\
			$ -              $ & $-               $ & $ -              $ & $ \sum\limits_{\substack{n=1,2\\F=0,1}} ns_{1/2}^{F} $ and $ \theta=0 $ & $ $ & $  7.40\times 10^{7} $ & $3.06\times 10^{-4}$\\
			$ -              $ & $-               $ & $ -              $ & $ \sum\limits_{\substack{n=1,2\\F=0,1}} ns_{1/2}^{F} $ and $ \theta=\frac{\pi}{2} $ & $  $ & $ -1.94\times 10^{7} $ & $-8.02\times 10^{-5}$\\
			\hline
			$ 2s_{1/2}^{F=1} $ & $ 2p_{3/2}^{F=2} $ & $ 2p_{3/2}^{F=1} $ & $ 1s_{1/2}^{F=1} $ and $ \theta=0 $     & $ -7.35238\times 10^{11} $ & $ 2.63\times 10^{7} $ & $1.09\times 10^{-4}$\\
			$ -              $ & $-               $ & $ -              $ & $ 1s_{1/2}^{F=1} $ and $ \theta=\frac{\pi}{2} $ & $  $ & $ -7.28\times 10^{6} $  & $-3.01\times 10^{-5}$\\
			$ -              $ & $-               $ & $ -              $ & $ 2s_{1/2}^{F=1} $ and $ \theta=0     $ & $  $ & $ 2.58\times 10^{7} $ & $1.07\times 10^{-4}$\\
			$ -              $ & $-               $ & $ -              $ & $ 2s_{1/2}^{F=1} $ and $ \theta=\frac{\pi}{2} $ & $  $  & $     -7.13\times 10^{7}$ & $-2.95\times 10^{-4}$\\
			$ -              $ & $-               $ & $ -              $ & $ \sum\limits_{n=1,2} ns_{1/2}^{F=1} $ and $ \theta=0     $ & $  $  & $ 2.63\times 10^{7} $ & $1.09\times 10^{-4}$\\
			$ -              $ & $-               $ & $ -              $ & $ \sum\limits_{n=1,2} ns_{1/2}^{F=1} $ and $ \theta=\frac{\pi}{2} $ & $  $ & $ -7.27\times 10^{6} $ & $-3.01\times 10^{-5}$\\
			$ 2s_{1/2}^{F=1} $ & $ 2p_{3/2}^{F=2} $ & $ 2p_{1/2}^{F=1} $ & $ 1s_{1/2}^{F=1} $ and $ \theta=0 $ & $ -1.85532\times 10^{12} $ & $ 2.01\times 10^{7} $ & $8.31\times 10^{-5}$\\
			$ -              $ & $-               $ & $ -              $ & $ 1s_{1/2}^{F=1} $ and $ \theta=\frac{\pi}{2} $ & $  $  & $ -5.55\times 10^{6}  $ & $ -2.29\times 10^{-5} $\\
			$ -              $ & $-               $ & $ -              $ & $ 2s_{1/2}^{F=1} $ and $ \theta=0 $ & $  $ & $ 1.91\times 10^{7} $ & $7.90\times 10^{-5}$\\
			$ -              $ & $-               $ & $ -              $ & $ 2s_{1/2}^{F=1} $ and $ \theta=\frac{\pi}{2} $ & $  $  & $ -5.28\times 10^{6} $ & $2.18\times 10^{-5}$\\
			$ -              $ & $-               $ & $ -              $ & $ \sum\limits_{n=1,2} ns_{1/2}^{F=1} $ and $ \theta=0 $ & $  $ & $ 2.00\times 10^{7} $ & $8.27\times 10^{-5}$\\
			$ -              $ & $-               $ & $ -              $ & $ \sum\limits_{n=1,2} ns_{1/2}^{F=1} $ and $ \theta=\frac{\pi}{2} $ & $  $ & $ -5.54\times 10^{6} $ & $2.29\times 10^{-5}$\\
			$ 2s_{1/2}^{F=1} $ & $ 2p_{3/2}^{F=2} $ & $ 2p_{1/2}^{F=0} $ & $ 1s_{1/2}^{F=1} $ and $ \theta=0 $ & $ -3.69345\times 10^{12} $ & $ 6.30\times 10^{6} $ & $2.61\times 10^{-5}$\\
			$ -              $ & $-               $ & $ -              $ & $ 1s_{1/2}^{F=1} $ and $ \theta=\frac{\pi}{2} $ & $  $ & $ -1.74\times 10^{6} $ & $-7.20\times 10^{-6}$\\
			$ -              $ & $-               $ & $ -              $ & $ 2s_{1/2}^{F=1} $ and $ \theta=0 $ & $  $ & $ 5.68\times 10^{6} $ & $2.35\times 10^{-5}$\\
			$ -              $ & $-               $ & $ -              $ & $ 2s_{1/2}^{F=1} $ and $ \theta=\frac{\pi}{2} $ & $  $ & $ -1.57\times     10^{6} $ & $-6.49\times 10^{-6}$\\
			$ -              $ & $-               $ & $ -              $ & $ \sum\limits_{n=1,2} ns_{1/2}^{F=1} $ and $ \theta=0 $ & $  $ & $     6.28\times 10^{6} $ & $2.60\times 10^{-5}$\\
			$ -              $ & $-               $ & $ -              $ & $\sum\limits_{n=1,2} ns_{1/2}^{F=1} $ and $ \theta=\frac{\pi}{2} $ & $  $     & $ -1.73\times 10^{6}  $ & $-7.15\times 10^{-6}$\\
			\hline
			\hline
		\end{tabular}
		\label{tabMUH}
	\end{table}

In the case of muonic hydrogen, the effect of quantum interference plays a small role and cannot be responsible for the so-called "proton radius puzzle" \cite{PhysRevA.92.022514}. Nevertheless, these systematics require careful evaluation and quantification as they are close to the sixth-order QED corrections \cite{BORIE2012733} and thus may affect the determination of fundamental physical constants.

	\subsection{Application to spectroscopy of helium-3 isotope}
	\label{subsection6}
	
	The interest in the two-electron $^3$He isotope system, as well as in helium-4, is primarily due to the evaluation of root-mean-square charge radii and the current discrepancy in experimental results and theory, see for example \cite{PhysRevLett.108.143001,PhysRevLett.119.263002,10.1093/nsr/nwaa216}. In this section we analyse the NR effects in application to the experiment \cite{PhysRevLett.108.143001}, where transitions between different hyperfine sublevels were observed.
	
	The measurements \cite{PhysRevLett.108.143001} of the $2^3S-2^3P$ transition frequencies were carried out by the method of fluorescence spectroscopy. Following this work, we denote the energy level of the helium isotope as $n^{\kappa}L_{J}^{F}$, where $n$ is the principal quantum number and $\kappa =2S+1$ is the multiplicity of the level ($S$ is the total spin momentum), $L$ is the total orbital momentum of two electrons, $J$ is the total angular momentum of the electrons and $F$ is the total angular momentum of the atom ($\boldsymbol{F}=\boldsymbol{J} + \boldsymbol{I}$, where $\boldsymbol{I}$ is a nuclear spin, which in the case of helium-3 is $1/2$.
	In the case of the $2^3S-2^3P$ transition, quantum interference arises due to neighbouring electric dipole transitions from the $2^3S$ state to the fine and hyperfine patterns of the $2^3P$ level. Since the experiment \cite{PhysRevLett.108.143001} concerns one-electron excitation, the use of the expressions (\ref{eq54}), (\ref{eq55}) and (\ref{24}) is justified for estimating the NR correction to the transition frequency $i\rightarrow r$ in $^3$He. In the following, we assume that in addition to the main excitation channel, the interference branch $i\rightarrow r'$ ($r' \neq r$) is also allowed by the selection rules for electric dipole transitions. Then, the cumulative NR correction is
	\begin{eqnarray}
		\label{eq65}
		\delta_{\mathrm{NR}}(2^3S_{1}^{F_i}- 2^3P_{J_r}^{F_r})=\sum_{J_{r'}F_{r'}\neq J_{r}F_{r} }\delta_{\mathrm{NR}}(2^3S_{1}^{F_i}- 2^3P_{J_r}^{F_r}[2^3P_{J_{r'}}^{F_{r'}}])
		,
	\end{eqnarray}
	where $\delta_{\mathrm{NR}}(2^3S_{1}^{F_i}- 2^3P_{J_r}^{F_r}[2^3P_{J_{r'}}^{F_{r'}}])$ denotes the partial contributions due to the interference of one-photon transitions to sublevels with $J_{r}F_{r} $ and $J_{r'}F_{r'} $.
	
	To calculate the NR corrections, we use the $^3$He energies from \cite{doi:10.1139/p06-009} and the natural level widths borrowed from \textit{The NIST Atomic Spectra Database} \footnote{https://www.nist.gov/pml/atomic-spectra-database}, both of which can be found in Table \ref{tab2}.
	\begin{table}[H]
		\centering
		\begin{tabular}{c|c|c}
			\hline
			\hline
			Level  & Energy, MHz & Natural width, Hz \\ 
			\hline
			$2^3P_{0}^{F=1/2}$& 5068832675.730  &1625926.899  \\ \hline
			$2^3P_{1}^{F=1/2}$& 5068804582.860  & \multicolumn{1}{c}{\multirow{2}{*}{1626002.179}} \\ 
			$2^3P_{1}^{F=3/2}$   & 5068800070.670     & \multicolumn{1}{c}{}    \\ \hline
			$2^3P_{2}^{F=3/2}$ & 5068805250.892    & \multirow{2}{*}{1625932.103}     \\ 
			$2^3P_{2}^{F=5/2}$ & 5068798289.789  &         \\ \hline\hline
		\end{tabular}
		\caption{Energies in MHz and natural widths in Hz for some $^3$He states.}
		\label{tab2}
	\end{table}
	
\begin{figure}[H]
		\caption{The NR corrections in kHz for the transitions studied in experiment \cite{PhysRevLett.108.143001} (see legend) as a function of the angle between the vectors of absorbed photon polarization and emitted photon propagation, $ \boldsymbol{e}_1 $, $ \boldsymbol{n}_{\boldsymbol{k}_2} $. Graphs corresponding to $2^3S^{3/2}_{1}\rightarrow 2^3P^{1/2}_{0}$ and $2^3S^{1/2}_{1}\rightarrow 2^3P^{1/2}_{0}$ transitions are omitted.}
		\centering
		\includegraphics[scale=1]{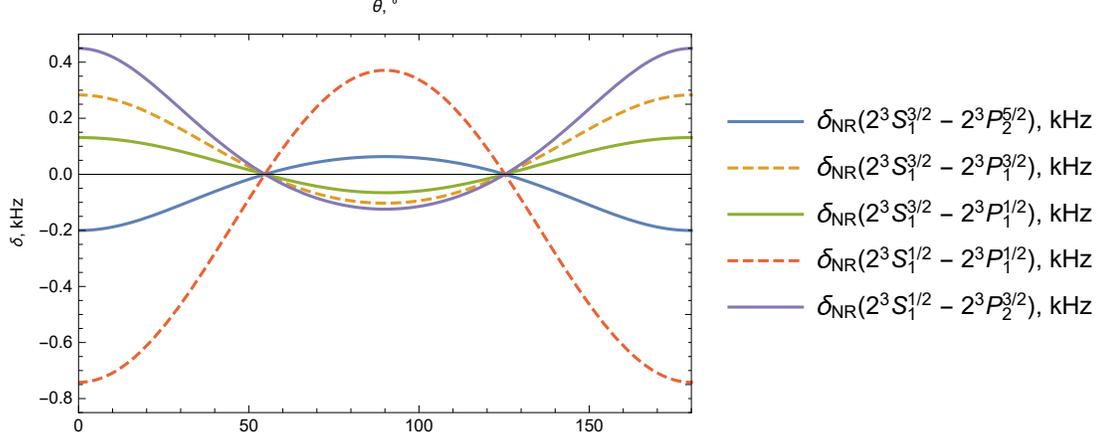}
		\label{fig4}
	\end{figure}

	Restricting ourselves to an experiment of the second type, see section \ref{subsection5}, below we consider all $2^3S-2^3P$ transitions and estimate the corresponding NR corrections using the formulas (\ref{eq54}), (\ref{eq55}) and (\ref{24}). The final results can be presented in the form of graphs depicted in Fig \ref{fig4}. 

    There are two transitions observed in the experiment, namely $2^3S^{F=3/2}_{1}\rightarrow 2^3P^{F=1/2}_{0}$ and $2^3S^{F=1/2}_{1}\rightarrow 2^3P^{1/2}_{0}$, for which NR corrections are negligible, so the corresponding plots are omitted. As before, all NR corrections are proportional to the factor $1 + 3\cos2\theta$, so the magic angles are preserved. Comparing the values of the NR correction, shown in the Fig.~\ref{fig4}, with the Zeeman shift and the corresponding uncertainties, around the $0.5$ and $2$ kHz, respectively, see \cite{PhysRevLett.108.143001}, one can find that the NR corrections are comparable and become even larger  than the latter for the transitions $2^3S^{1/2}_1 \rightarrow 2^3P^{F=1/2}_1$ and $2^3S^{F=1/2}_1 \rightarrow 2^3P^{F=3/2}_2$.
	
		Using the energy values of fine and hyperfine manifold for $2^3S $ and $ 2^3P$ levels, one can obtain so-called centroid energy by the use of the formula \cite{PhysRevLett.108.143001}:
	\begin{eqnarray}
		\label{eq87}
		E\left(n^{\kappa}L\right) = \frac{\sum_J(2J + 1)E\left( n^{\kappa}L_J \right)}{\left(2S + 1\right)\left( 2L + 1 \right)}=\frac{\sum_{J,F}\left( 2F + 1 \right)E\left( n^{\kappa}L^F_J \right)}{\left( 2I + 1 \right)\left( 2S + 1 \right)\left( 2L + 1 \right)}
		.
	\end{eqnarray}
	Then according to Eq. (\ref{eq87}) for the centroid $2^3P - 2^3S $ transition frequency we find
	\begin{eqnarray}
	\label{eq88}
		E^{^3\mathrm{He}}_{\mathrm{centr}}\equiv E\big{(} 2^3P \big{)} - E\big{(} 2^3S \big{)}=\frac{1}{6}\Bigg{\{} \frac{1}{3}\Big{[} \omega\left(  2^3P^{F=1/2}_1  -  2^3S^{F=1/2}_1  \right) + \omega\left( 2^3P^{F=1/2}_0 - 2^3S^{1/2}_1 \right) +\omega\left( 2^3P^{F=1/2}_1 - 2^3S^{F=3/2}_1 \right)  
	\\\nonumber
		+ \omega\left( 2^3P^{F=1/2}_0 - 2^3S^{F=3/2}_1 \right) \Big{]}+ \frac{4}{3} \Big{[} \omega\left( 2^3P^{F=3/2}_2 - 2^3S^{F=1/2}_1 \right) + \omega\left( 2^3P^{F=3/2}_1 - 2^3S^{F=3/2}_1 \right) \Big{]} + 2\omega\left( 2^3P^{F=5/2}_2 - 2^3S^{F=3/2}_1 \right) \Bigg{\}}
		\\\nonumber
		= 276\,702\,827\,204\,.8\,\mathrm{kHz},
	\end{eqnarray}
	where $\omega\left( 2^{3}P^{F_r}_{J_r} - 2^{3}S^{F_i}_{J_i} \right)$ is the $2^{3}P^{F_r}_{J_r} \rightarrow 2^{3}S^{F_i}_{J_i}$ transition frequency measured in the experiment.

    Assuming that each transition value in Eq. (\ref{eq88}) includes the NR correction, we can eliminate them from centroid energy by calculating the corresponding shift:
	\begin{eqnarray}
		\label{eq89}
		\delta^{(i)}_{\mathrm{centr}}\equiv \frac{1}{6}\Bigg{\{} \frac{1}{3}\Big{[} \delta_{\mathrm{NR}}\left(  2^3P^{F=1/2}_1  -  2^3S^{F=1/2}_1  \right) + \delta_{\mathrm{NR}}\left( 2^3P^{F=1/2}_0 - 2^3S^{F=1/2}_1 \right) 
		\\\nonumber
		+ \delta_{\mathrm{NR}}\left( 2^3P^{F=1/2}_1 - 2^3S^{F=3/2}_1 \right) +\delta_{\mathrm{NR}}\left( 2^3P^{F=1/2}_0 - 2^3S^{F=3/2}_1 \right) \Big{]} 
		\\\nonumber
		+ \frac{4}{3} \Big{[} \delta_{\mathrm{NR}}\left( 2^3P^{F=3/2}_2 - 2^3S^{F=1/2}_1 \right) + \delta_{\mathrm{NR}}\left( 2^3P^{F=3/2}_1 - 2^3S^{F=3/2}_1 \right) \Big{]} + 2\delta_{\mathrm{NR}}\left( 2^3P^{F=5/2}_2 - 2^3S^{F=3/2}_1 \right) \Bigg{\}}.
	\end{eqnarray}

    Eq. (\ref{eq89}) as well as partial and total NR corrections Eq. (\ref{eq65}) vanishes at "magic angles". Despite the fact that this value is smaller than all relativistic and QED corrections, it is close to kHz and approaches the level of the nuclear polarizability contribution ($-1.1$ kHz for the $2^3P-2^3S$ centroid energy, see \cite{PhysRevLett.108.143001}). It is important to note that the value of NR corrections can depend on experimental parameters such as pressure and blackbody radiation, see sections \ref{twophotonHe} and \ref{BBR}, respectively.

	\section{Two-photon spectroscopy of hydrogen and helium}
	\label{twophotonH}
	
	Following the analysis of nonresonant effects in one-photon spectroscopy described in the section \ref{angleone}, we will focus on the interference that occurs when measuring $ 2s\rightarrow ns/nd $ ($ n=4,\;6,\;8,\;12 $ is the principal quantum number) transition frequencies. In these experiments, hydrogen atoms are prepared in an atomic beam in the state $ 2s_{1/2}^{F=1} $ and then excited to the state $ ns_{1/2}^{F=1} $ or $ nd_{3/2}^{F=2} $ due to the absorption of two polarized laser photons propagating in opposite directions. Detection of the excited $ ns/nd$ fraction of the atoms can be observed via its fluorescence (i.e. decay to the $ 2p $ state) \cite{PhysRevA.52.2664} or the decrease in metastable $ 2s $ signal \cite{PhysRevLett.68.1120,PhysRevLett.69.2326}. In both cases, interference occurs between the different fine sublevels $ nd_{3/2}^{F=2} $ and $ nd_{5/2}^{F=2} $, leading to an asymmetry of the line profile. In \cite{PhysRevA.52.2664}, the detection of the excited $ ns/nd$ fraction of atoms via their fluorescence was shown to have much higher potential accuracy than the experiments monitoring the rate of quenching of metastable states \cite{PhysRevLett.82.4960,deB-2}. The latter is limited by a large background of unexcited $ 2s $ atoms. Recently, it was shown in \cite{Anikin2021} that the nonresonant corrections to the $ 2s_{1/2}^{F=1}-nd^{F=2}_{3/2(5/2)} $ transition frequencies, measured in the experiments of the type \cite{PhysRevLett.82.4960,deB-2}, reached the level of several kHz, which makes them important for the determination of the proton charge radius $ r_{p} $ and the Rydberg constant $ R_{\infty} $. In particular, taking into account the NR correction, the averaged values of $ r_{p} $ were found to agree with the results reported in \cite{Science}. This section is devoted to the theoretical description of an experimental method based on the detection of fluorescence from outgoing photons. 
	
	In section \ref{angleone} it was shown that for a certain geometry the influence of NR effects in one-photon spectroscopy can be significantly reduced \cite{Jent-NR,Solovyev2020,PhysRevA.92.062506}. Extending this approach, we introduce here expressions for the cross section for the corresponding resonant two-photon scattering on the hydrogen atom levels and take into account fine and hyperfine structure. As before, these expressions depend on the directions and polarizations of the incident (absorbed) and outgoing (emitted) photons. Therefore, various correlations between directions and polarizations should be considered within the framework of the three-photon scattering process (two photons are absorbed and one is emitted). The results of the evaluation are then used to obtain NR corrections to the two-photon absorption cross section and to determine the $2s-ns/nd$ transition frequencies.

\begin{figure}[H]
		\caption{Two-photon excitation process of a bound electron. The wavy line denotes the absorption or emission of the photon. The double solid line denotes the bound electron (Furry picture); $\omega_1$, $\omega_2$ are the frequencies of the absorbed photons, while $\omega_3$ is the frequency of the emitted photon. The indices $i,\,n,\,k,\,f$ correspond to the initial, two intermediate and final states of the electron, respectively. According to the Feynman rules, there are 5 more diagrams that relate to the permutations of the photons, which we omit here for brevity.}
		\center{\includegraphics[width=0.4\linewidth]{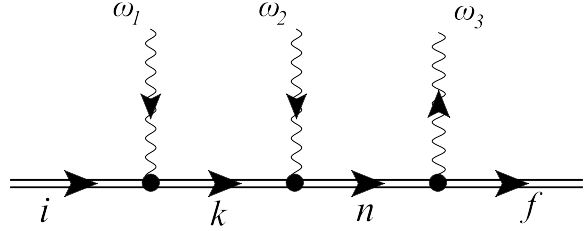}}
		\label{threescat}
	\end{figure}

	In complete analogy with the results of the section \ref{angleone}, the process we are interested in is described by the Feynman diagram in Fig. \ref{threescat}, which corresponds to the process of two-photon scattering by an atom with subsequent re-emission of a photon.
	
	The corresponding S-matrix element of the process is
	\begin{eqnarray}
		\label{90}
		S^{(3)}_{fi}=(-ie)^3\int d^4x_{3}d^4x_{2}d^4x_1
		\overline{\psi}_{f}(x_{3})\gamma_{\mu_{3}}A^*_{\mu_{3}}(x_3)
		S(x_3,x_2)\gamma_{\mu_2}A_{\mu_{2}}(x_2)S(x_2,x_1)\gamma_{\mu_1}A_{\mu_{1}}(x_1)\psi_{i}
		(x_1),
	\end{eqnarray} 
	with all the notations given in the section \ref{subsection1}. Integration over the time variables in Eq. (\ref{90}) yields
	\begin{eqnarray}
		\label{eq93}
		S_{fi}^{(3)}=-2\pi i\;\delta(E_{f}-E_{i}+\omega_f-\omega_1-\omega_2)U^{(3)}_{fi},
	\end{eqnarray}
	where the amplitude $U^{(3)}_{fi}$ of the process is given by
	\begin{eqnarray}
		\label{eq94}
		U^{(3)}_{fi}=e^3
		\sum\limits_{nk}
		\frac{
			\langle  f      |
			\bm{\alpha}\boldsymbol{A}^*_{\boldsymbol{k}_3,\boldsymbol{e}_3}
			|       n       \rangle
			\langle n       |
			\bm{\alpha}\boldsymbol{A}_{\boldsymbol{k}_2,\boldsymbol{e}_2}
			|       k       \rangle
			\langle k       |
			\bm{\alpha}\boldsymbol{A}_{\boldsymbol{k}_1,\boldsymbol{e}_1}
			|       i       \rangle
		}
		{(E_{n} - E_{f} - \omega_3) (E_k - E_{i} - \omega_2)}
		+(5\;\mbox{permutations}).
	\end{eqnarray}
	
	Permutations in Eq. (\ref{eq94}) should be understood as all possible rearrangement of indices $1,\,2,\,3$ denoting the corresponding photons, and the differential cross section of the scattering process is defined by
	\begin{eqnarray}
		\label{eq95}
		\frac{d\sigma_{fi}}{d\boldsymbol{n}_{\boldsymbol{k}_3}}=2\pi\delta(E_{f}-E_{i}+\omega_3-\omega_1-\omega_2)
		\left|U^{(3)}_{fi}\right|^2\omega^2_2\omega^2_3
		\frac{d\omega_2}{(2\pi)^3}
		\frac{d\omega_3}{(2\pi)^3}.
	\end{eqnarray}
	Here $\boldsymbol{n}_{\boldsymbol{k}}$ is a solid angle in $\boldsymbol{k}-$space of corresponding photon. Using the nonrelativistic limit and the dipole approximation, one obtains (in length form \cite{ZALIALIUTDINOV20181})
	\begin{eqnarray}
		\label{eq96}
		U^{(3)}_{fi}=e^3(2\pi)^{3/2}\sqrt{\omega_{1}\omega_{2}\omega_{3}}
		\sum\limits_{nk}\frac{
			\langle f |
			\boldsymbol{e}_3^*\boldsymbol{r}
			| n  \rangle
			\langle  n |
			\boldsymbol{e}_2\boldsymbol{r}
			| k \rangle
			\langle k |
			\boldsymbol{e}_1\boldsymbol{r}
			| i \rangle
		}
		{(E_{n} - E_{f} - \omega_3) (E_k - E_{i} - \omega_2)}
		+(5\;\mbox{permutations}).
	\end{eqnarray}
	
	We are interested in the case when two incident photons are absorbed in some intermediate state $ n $, i.e. $ \omega_1+\omega_2=E_{n}-E_{i} $. In the resonant approximation, this intermediate state makes the dominant contribution, and the remaining nonresonant terms in the scattering amplitude can be omitted. Such an approximation is justified by the fact that the corresponding nonresonant corrections go beyond the accuracy of experiments \cite{ZALIALIUTDINOV20181}. Then assuming that frequencies of two incident laser photons are equal, i.e.
	\begin{eqnarray}
		\label{equal}
		\omega_1=\omega_2\equiv\omega
		, 
	\end{eqnarray}
	the cross section Eq. (\ref{eq95}) with nonrelativistic scattering amplitude Eq. (\ref{eq96}) can be reduced to
	\begin{eqnarray}
		\label{eq97}
		\frac{d\sigma_{fi}}{d\boldsymbol{n}_{\boldsymbol{k}_3}}=\frac{e^6}{(2\pi)^5}\omega^6(E_i + 2\omega - E_f)^3\Bigg{|} \sum_{nk} \frac{\langle f | \boldsymbol{e}^*_3\boldsymbol{r}| n\rangle}{E_n - E_i - 2\omega - \frac{i}{2}\Gamma_n}\Bigg{(} \frac{\langle n | \boldsymbol{e}_2\boldsymbol{r}| k\rangle \langle k | \boldsymbol{e}_1\boldsymbol{r}| i\rangle}{E_n - E_i - \omega} + \frac{\langle n | \boldsymbol{e}_1\boldsymbol{r}| k\rangle \langle k | \boldsymbol{e}_2\boldsymbol{r}| i\rangle}{E_k - E_n + \omega} \Bigg{)} \Bigg{|}^2,
	\end{eqnarray}
	where the regularization procedure for the divergent denominator and the relation $ \langle a | \textbf{p} | b \rangle = \mathrm{i}(E_{a}-E_{b})\langle a | \textbf{r} | b \rangle$ have been applied \cite{ANDREEV2008135}. The appearance of the imaginary part leads to the formation of an absorption line profile \cite{ANDREEV2008135}. The regularization procedure should be carried out by summing an infinite number of one-loop self-energy insertions ("loop after loop") into the electron propagator. In the resonant approximation, this leads to the appearance of level widths in the divergent denominator. In addition, note that the regularization in the case of two-photon absorption repeats the "one-photon calculations", see section~\ref{subsection3}, and allows for nonresonant extension, see section~\ref{subsection3ext}, which we omit for brevity.
	
	To introduce the leading-order nonresonant correction from the cross section given by Eq. (\ref{eq97}), one can consider the terms closest in energy in the sum over $n$, i.e. in the case of two neighboring states, these are $n=r$ (the leading resonance term, to which the NR-correction is introduced) and $ n=r'$ (the closest in energy to the resonance term) \cite{2001,Jent-NR,PhysRevA.65.054502,Labzowsky1994}. The set of quantum numbers for the additional state $ r' $ should allow the connection with the initial state by the absorption of two electric dipole photons (as the for the resonant one) and, therefore, be permitted by the two-photon selection rules \cite{grynbergPHD}. Then, using the same approximations as for the one-photon correction, i.e., neglecting the quadratic nonresonant contribution (see Eq. (\ref{eqNLO})) and the level width in the energy denominator corresponding to the NR state, the dominant contribution can be found.
	
	We assume a standard set of quantum numbers for atomic states in the matrix elements of Eq. (\ref{eq97}): principal quantum number $n$, electron orbital momentum $l$, electron total angular momentum $j$, atomic angular momentum $F$, and its projection $M_{F}$. After summing over the projections of total momentum in the final state and averaging over the projections of the initial state, see \cite{ZALIALIUTDINOV20181}, the cross section becomes:
	\begin{eqnarray}
		\label{eq101}
		\frac{d\sigma_{fi}}{d\boldsymbol{n}_{\boldsymbol{k}_3}}=\frac{e^6}{2F_i+1}\Bigg{[} \frac{f^{(2\gamma)}_{fi}(r,r)}{(\omega_0 - 2\omega)^2 + \frac{\Gamma^2_r}{4}} + \frac{f^{(2\gamma)}_{fi}(r',r')}{(\omega_0 - 2\omega + \Delta)^2 + \frac{\Gamma^2_{r'}}{4}} + \frac{2f^{(2\gamma)}_{fi}(r,r')(\omega_0 - 2\omega)(\omega_0 - 2\omega + \Delta)}{\big{(}(\omega_0 - 2\omega)^2 + \frac{\Gamma^2_{r'}}{4}\big{)}\big{(}(\omega_0 - 2\omega + \Delta)^2 + \frac{\Gamma^2_{r'}}{4}\big{)}} \Bigg{]},
	\end{eqnarray}
	where $\Delta=E_{r'} - E_r$ and 
	\begin{eqnarray}
		\label{eq102}
		f^{(2\gamma)}_{fi}(r,r')=\sum_{M_{F_i}M_{F_f}}T_{fri}\Big{(}\frac{\omega_0}{2}\Big{)}T^*_{fr'i}\Big{(}\frac{\omega_0}{2}\Big{)},
	\end{eqnarray}
	together with the notation
	\begin{eqnarray}
		\label{eq103}
		T_{fni}(\omega)=\omega^3(E_i - E_f + 2\omega)^{3/2}\sum_{M_{F_n}}\langle f| \boldsymbol{e}^*_3\boldsymbol{r} | n\rangle\sum_{k}\Bigg{[} \frac{\langle n| \boldsymbol{e}_2\boldsymbol{r} | k\rangle\langle k| \boldsymbol{e}_1\boldsymbol{r} | i\rangle}{E_r - E_i - \omega} + \frac{\langle n| \boldsymbol{e}_1\boldsymbol{r} | k\rangle\langle k| \boldsymbol{e}_2\boldsymbol{r} | i\rangle}{E_r - E_n + \omega} \Bigg{]}.
	\end{eqnarray}
	The coefficients Eq. (\ref{eq102}) determine angular dependencies. Their analytical evaluation is given in Appendix \ref{ang2}. In the nonrelativistic limit, the matrix elements in Eq. (\ref{eq101}) do not explicitly depend on the photon directions $ \boldsymbol{n}_{\boldsymbol{k}_3} $, $ \boldsymbol{n}_{\boldsymbol{k}_2}$ and $ \boldsymbol{n}_{\boldsymbol{k}_1} $, and dependence on them arises through the transversality condition. Without loss of generality, one can assume that in the experiment the incident photons propagate in opposite directions with the fixed polarization vectors $ \boldsymbol{e}_{1} $ and  $ \boldsymbol{e}_{2} $, while the outgoing photon has the polarization $ \boldsymbol{e}_3 $ and fixed direction $ \boldsymbol{n}_{\boldsymbol{k}_3} $. 
	Then, denoting the angles between any pair of two vectors as $ \theta_{ij} $ ($ i,\,j=1,\,2,\,3 $), the interference contribution in Eq. (\ref{eq101}) corresponds to the situation, similar to experiments based on the one-photon scattering process (the angle between incident photons can be set equal to $\pi$ or zero) \cite{PhysRevA.95.052503,PhysRevA.90.012512} .  
	
	The resonant transition frequency $\omega_{\mathrm{res}}$ can be determined from $d\sigma_{if}(\omega)$ using the condition 
	\begin{eqnarray}
		\label{max}
		\frac{d\sigma_{if}(\omega)}{d\omega}=0
		.
	\end{eqnarray}
	In the resonant approximation (i.e. retaining only first term in Eq. (\ref{eq101})) we immediately find $ \omega_{\mathrm{res}}=\omega_{\mathrm{max}}=\omega_{ri}/2=(E_{n_rl_rj_rF_r}-E_{n_il_ij_iF_i})/2 $. However, keeping the last interference term in Eq. (\ref{eq101}), setting $\Gamma_r=\Gamma_{r'}\equiv\Gamma$, $\Delta=E_{r'} - E_r$ and solving Eq. (\ref{15}) with respect to $  \omega$ we arrive at the definition $ \omega_{\mathrm{max}} $
	\begin{eqnarray}
		\label{omegamax}
		\omega_{\mathrm{max}}=(\omega_{ai}-\delta_{\mathrm{NR}})/2
		,
	\end{eqnarray}
	where
	\begin{eqnarray}
		\label{eq104}
		\delta_{\mathrm{NR}}=\frac{\sum_{f}f^{(2\gamma)}_{fi}(r,r')}{\sum_{f}f^{(2\gamma)}_{fi}(r,r)}\frac{\Gamma^2}{4\Delta}.
	\end{eqnarray}
	Similar to the one-photon scattering case discussed in section \ref{angleone} (see Eq. (\ref{24})), NR correction Eq. (\ref{eq104}) is obtained as the leading term of the expansion in $ \Gamma/\Delta $ when this parameter is small \cite{Solovyev2020}. The angular correlations in Eq. (\ref{eq104}) are obtained from the ratio $ f^{(2\gamma)}_{fi}(r,r')/f^{(2\gamma)}_{fi}(r,r)$ and represent the dependence on the experimental setup, i.e. on the angles between each pair of vectors $ \boldsymbol{n}_{k_1} $, $\boldsymbol{e}_2 $ and  $\boldsymbol{e}_3 $.

	\subsection{Two-photon spectroscopy of hydrogen}
	\label{twophotonHexamples}
	
	In this section, we consider particular examples of NR corrections to the two-photon $2s-ns/nd$  (with $ n=4,\,6,\,8,\,12 $) transition frequencies in hydrogen. Considering first  $ 2s_{1/2}^{F=0}\rightarrow ns_{1/2}^{F=0} $ and $ 2s_{1/2}^{F=1}\rightarrow ns_{1/2}^{F=1} $ transitions, we assume that the hyperfine structure of the initial $ 2s $ state is resolvable in experiments \cite{Science,deB-0,PhysRevLett.82.4960,Schwob,PhysRevLett.84.5496}. According to the two-photon selection rules the spin-flip electric dipole two-photon transitions $ 2s_{1/2}^{F=0}\rightarrow ns_{1/2}^{F=1} $ or  $ 2s_{1/2}^{F=1}\rightarrow ns_{1/2}^{F=0} $ are strongly suppressed \cite{PhysRevA.91.033417,PhysRevA.93.012510,zalialiutdinov2017generalized}. Therefore, the interference with close $ nd $ states is only possible.  As a result $ 2s_{1/2}^{F=1}\rightarrow ns_{1/2}^{F=1} $ transition branch interfere with $ 2s_{1/2}^{F=1}\rightarrow nd_{3/2}^{F=1} $, $ 2s_{1/2}^{F=1}\rightarrow nd_{3/2}^{F=2} $, $ 2s_{1/2}^{F=1}\rightarrow nd_{5/2}^{F=2} $ and $ 2s_{1/2}^{F=1}\rightarrow nd_{5/2}^{F=3} $ two-photon absorption branches \cite{science2020}. Then for $ 2s_{1/2}^{F=1}\rightarrow ns_{1/2}^{F=1} $ we set in all equations $ n_{i} l_{i}=2s $, $ j_i=1/2 $, $ F_{i}=1 $, $ n_{r} l_{r}=ns $, $ n_{r'} l_{r'}=nd $ (with $ n_{r}=n_{r'}=4,\,6,\,8,\,12 $), $ j_{r}=1/2 $, $ F_{r}=1 $. 
	
\begin{figure}[H]
		\center{\includegraphics[width=0.6\linewidth]{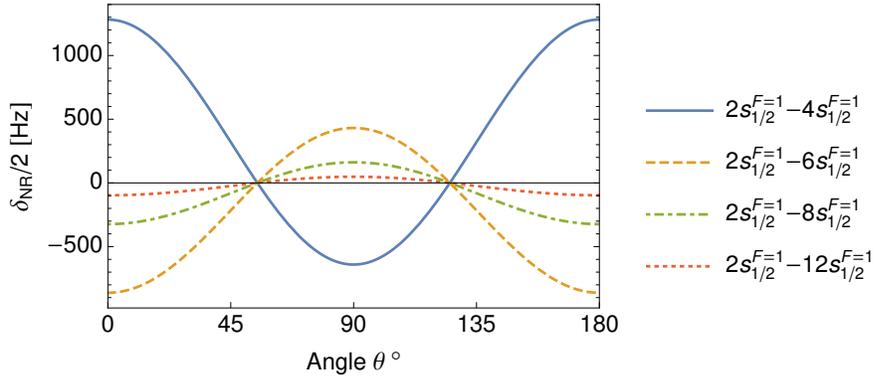}}
		\caption{NR corrections $ \delta_{\mathrm{NR}}/2 $ (in Hz) for the measurement of $2s_{1/2}^{F=1}-ns_{1/2}^{F=1}$ ($ n=4,\,6,\,8,\,12 $) transition frequencies in hydrogen in dependence on angle between polarization vector $ \boldsymbol{e}_1 $ of incident photon (or $ \boldsymbol{e}_2 $, since in experiments $ \boldsymbol{e}_1||\boldsymbol{e}_2 $) and propagation direction $ \boldsymbol{n}_{\boldsymbol{k}_3} $ of outgoing photon.}
		\label{2sns_fig1}
	\end{figure}

	In the experiments \cite{PhysRevA.90.012512, PhysRevA.95.052503} the polarization of incident laser photons $ \boldsymbol{e}_1 $ and $ \boldsymbol{e}_2 $ were fixed as parallel to each other. Then the NR correction, Eq. (\ref{eq104}), depends only on one angle between polarization of outgoing photon $ \boldsymbol{e}_3 $ and one of the two parallel vectors $ \boldsymbol{e}_1 $ or $ \boldsymbol{e}_2 $. Summation over the polarization $ \boldsymbol{e}_3 $ leads to an implicit dependence on the vector of propagation direction $ \boldsymbol{n}_{k_3} $. Denoting the angle between vectors $ \boldsymbol{e}_1 $ (or $ \boldsymbol{e}_2 $) and $ \boldsymbol{n}_{k_3} $  as $ \theta $, substituting all numerical values of levels widths, energy differences into Eq. (\ref{eq104}) and evaluating the sum over entire spectrum in Eq. (\ref{12}) (see details in Appendix \ref{ang2} and \ref{ang2explicit}) we find that for $2s_{1/2}^{F=1}\rightarrow ns_{1/2}^{F=1}$ transition frequencies the NR correction Eq. (\ref{eq104}) is proportional to $(1+3\cos(\theta))$. The corresponding numerical results are presented in Fig. \ref{2sns_fig1}. As in the case of NR corrections to one-photon transition frequencies there are "magic angles", at which NR correction Eq. (\ref{eq104}) vanishes: $ \theta=54.7^{\circ}  $ and $ \theta=125.3^{\circ} $. 

\begin{figure}[H]
		\caption{NR correction $ \delta_{\mathrm{NR}}/2 $ to the $ 1s_{1/2}^{F=1}\rightarrow 3s_{1/2}^{F=1} $ transition frequency in hydrogen (in Hz). Notations are the same as for Fig.~\ref{2sns_fig1}.}
		\center{\includegraphics[width=0.45\linewidth]{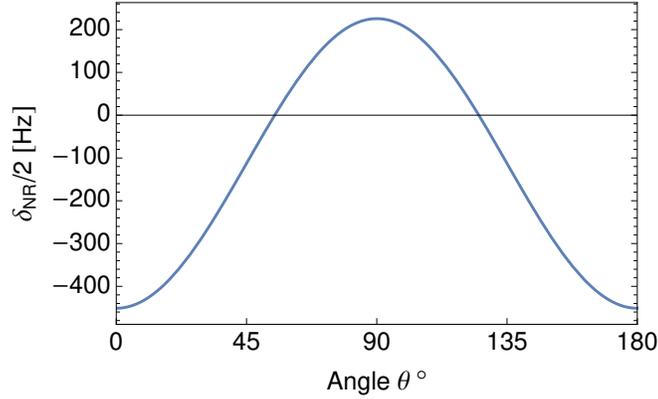}}
		\label{1s3s_fig2}
	\end{figure}

	Recently the similar interference effects in $ 1s-3s $ spectroscopy of hydrogen were studied in \cite{PhysRevA.95.052503}. It was found that for two-photon laser induced transition $ 1s_{1/2}^{F=1}\rightarrow 3s_{1/2}^{F=1} $ the NR correction due to the interference with four neighbouring transitions $ 1s_{1/2}^{F=1}\rightarrow 3d_{3/2}^{F=1} $, $ 1s_{1/2}^{F=1}\rightarrow 3d_{3/2}^{F=2} $, $ 1s_{1/2}^{F=1}\rightarrow 3d_{5/2}^{F=2} $ and $ 1s_{1/2}^{F=1}\rightarrow 3d_{5/2}^{F=3} $ is less than experimental uncertainty. Equation (\ref{eq104}) can be easily extended to the calculation of NR correction to $ 1s_{1/2}^{F=1}\rightarrow 3s_{1/2}^{F=1} $ transition frequency by replacing $ 2s_{1/2}^{F=1}\leftrightarrow 1s_{1/2}^{F=1} $ and setting $ n_{r}=n_{r'}=3 $. 
	Then the NR correction (in Hz) can be found as
	\begin{eqnarray}
		\label{nr1s3s}
		\delta_{\mathrm{NR}}(1s_{1/2}^{F=1}-3s_{1/2}^{F=1} )=-225.61(1+3\cos(\theta))
		.
	\end{eqnarray}
	The corresponding angular correlation is shown in Fig. \ref{1s3s_fig2}.
	
\begin{figure}[H]
		\caption{NR corrections $ \delta_{\mathrm{NR}}/2 $ to the $ 2s_{1/2}^{F=0}\rightarrow ns_{1/2}^{F=0} $ ($ n=4,\,6,\,8,\,12 $) transition frequencies in hydrogen (in Hz). The notations are used as in the previous figures.}
		\center{\includegraphics[width=0.6\linewidth]{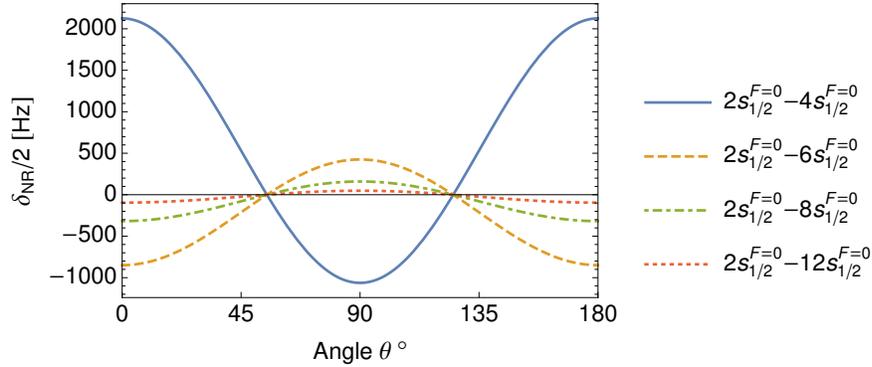}}
		\label{2sns_fig2}
	\end{figure}

	For other transitions, $ 2s_{1/2}^{F=0}\rightarrow ns_{1/2}^{F=0} $, there is interference with $ 2s_{1/2}^{F=0}\rightarrow nd_{3/2}^{F=2} $ and $ 2s_{1/2}^{F=0}\rightarrow nd_{5/2}^{F=2} $ two-photon absorption branches. The results of evaluations are presented in Fig. \ref{2sns_fig2}.
	
	Now we turn to evaluation of NR correction to $ 2s_{1/2}^{F=1}\rightarrow nd_{3/2}^{F=2} $ ($ n=4,\,6,\,8,\,12 $) transition frequencies with the account for neighbouring $ nd_{5/2}^{F=2} $ level. For this purpose we should set in all equations $ n_{i} l_{i}=2s $, $ j_i=1/2 $, $ F_{i}=1 $, $ n_{r} l_{r}=n_{r'} l_{r'}=nd $ (with $ n_{r}=4,\,6,\,8,\,12 $), $ j_{r}=3/2 $, $ F_{r}=2 $, $ j_{r'}=5/2 $, $ F_{r'}=2 $. Numerical evaluation is carried out similarly to the previous results. Finally, we arrive at the NR corrections which in contrast to previous cases do not depend on the angles between the vectors $ \boldsymbol{n}_{k_1} $, $ \boldsymbol{e}_2 $ and $ \boldsymbol{e}_3 $ (in Hz) 
	\begin{eqnarray}
		\label{e4}
		\delta_{\mathrm{NR}}(2s_{1/2}^{F=1}-4d_{3/2}^{F=2} )=967.75
	\end{eqnarray}
	\begin{eqnarray}
		\label{e6}
		\delta_{\mathrm{NR}}(2s_{1/2}^{F=1}-6d_{3/2}^{F=2} )=296.48
	\end{eqnarray}
	\begin{eqnarray}
		\label{e8}
		\delta_{\mathrm{NR}}(2s_{1/2}^{F=1}-8d_{3/2}^{F=2} )=127.31
	\end{eqnarray}
	\begin{eqnarray}
		\label{e12}
		\delta_{\mathrm{NR}}(2s_{1/2}^{F=1}-12d_{3/2}^{F=2} )=38.38
		.
	\end{eqnarray}
	
\begin{figure}[H]
		\caption{Total frequency shift $ \delta_{\mathrm{NR}}/2 $ (in Hz) to the $2s_{1/2}^{F=1}-nd_{3/2}^{F=2}$ ($ n=4,\,6,\,8,\,12 $) transition frequencies in hydrogen, see Eqs. (\ref{e4})-(\ref{e12}) and Eqs. (\ref{h4})-(\ref{h12}).  All notations are the same as in Fig. \ref{2sns_fig2}.}
		\center{\includegraphics[width=0.6\linewidth]{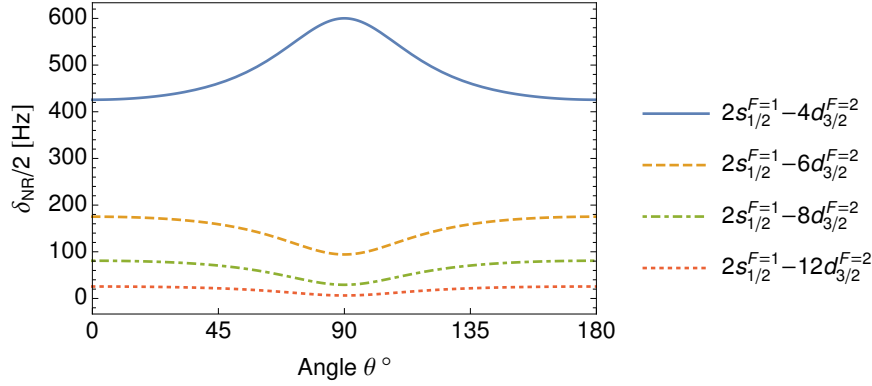}}
		\label{fig_total2snd}
	\end{figure}

	Besides the corrections Eqs. (\ref{e4})-(\ref{e12}) arising due to the neighbouring $ nd_{3/2}^{F=2} $ and $ nd_{5/2}^{F=2} $ states, the quantum interference between $ 2s_{1/2}^{F=1}\rightarrow nd_{3/2}^{F=2} $ and $ 2s_{1/2}^{F=1}-ns_{1/2}^{F=1}$ absorption branches should be considered \cite{PhysRevA.103.022833}. Then, following Eq. (\ref{eq104}) we can write
	\begin{eqnarray}
		\label{nrsumnd22}
		\delta_{\mathrm{NR}}(2s_{1/2}^{F=1}-nd_{3/2}^{F=2})=\frac{\sum\limits_{j_{f}F_{f}}f_{\mathrm{nr}}({nd_{3/2}^{F=2},ns_{1/2}^{F=1}})}{\sum\limits_{j_{f}F_{f}}f_{\mathrm{res}}(nd_{3/2}^{F=2},nd_{3/2}^{F=2})}
		\frac{\Gamma_{nd_{3/2}}^2}{4\Delta'''}
		,
	\end{eqnarray} 
	where $ \Delta'''= E_{nd_{3/2}^{F=2}}-E_{ns_{1/2}^{F=1}}$. Using Eq. (\ref{nrsumnd3R}) in Appendix \ref{ang2explicit}, the NR corrections are (in Hz)
	\begin{eqnarray}
		\label{h4}
		\delta_{\mathrm{NR}}(2s_{1/2}^{F=1}-4d_{3/2}^{F=1} )=
		-232.602
		\frac{1+3\cos(\theta)}{5+3\cos(\theta)}
		,
	\end{eqnarray} 
	\begin{eqnarray}
		\label{h6}
		\delta_{\mathrm{NR}}(2s_{1/2}^{F=1}-6d_{3/2}^{F=1} )=
		107.937
		\frac{1+3\cos(\theta)}{5+3\cos(\theta)}
		,
	\end{eqnarray} 
	\begin{eqnarray}
		\label{h8}
		\delta_{\mathrm{NR}}(2s_{1/2}^{F=1}-8d_{3/2}^{F=1} )=
		68.697
		\frac{1+3\cos(\theta)}{5+3\cos(\theta)}
		,
	\end{eqnarray} 
	\begin{eqnarray}
		\label{h12}
		\delta_{\mathrm{NR}}(2s_{1/2}^{F=1}-12d_{3/2}^{F=1} )=
		25.582
		\frac{1+3\cos(\theta)}{5+3\cos(\theta)}
		.
	\end{eqnarray}
	The total NR correction to the  $ 2s_{1/2}^{F=1}\rightarrow nd_{3/2}^{F=2} $  transition frequencies is given by the sum of constant contributions, Eqs. (\ref{e4})-(\ref{e12}), and corresponding angular-dependent contributions, Eqs. (\ref{h4})-(\ref{h12}). The total frequency shifts, $ \delta_{\mathrm{NR}}/2 $ are depicted in Fig. \ref{fig_total2snd}. It is seen that the denominator of Eqs. (\ref{e4})-(\ref{e12}) is always non-zero and positive while the numerator still turns the NR correction to zero at "magic angles". 
	
	\subsection{Experiments based on $2s$ state quenching}
	
	There is another type of spectroscopic experiments that uses the two-photon absorption process and is based on the quenching rate of the $2s$ state \cite{PhysRevLett.82.4960,deB-2, deB-0,Schwob, Anikin2021}. For such experiments the initially prepared meta-stable state of hydrogen atoms is excited to the $ns/nd $ ($n=4,\,6,\,8,\,12$) states by absorbing two laser photons. In this case, only a part of the atoms in the 2s state is excited in the atomic beam. By applying an external uniform static electric field downstream of the excitation region, the levels of opposite parity $2s$ and $2p$ are mixed, whereupon the luminescence Lyman alpha $2p-1s $ line occurs. Experimentally, the dependence of the intensity of this line on the frequency of the absorbed photons can be observed. The Lyman alpha line is absent if two-photon resonance is reached. 
	
	In experiments of this type, it suffices to consider only the two-photon excitation process itself and not to take into account the subsequent emission process, in contrast to the method where the fluorescence signal is recorded. Following \cite{PhysRevLett.82.4960,deB-2,deB-0,Schwob} the amplitude of the process can be considered only as absorption part of scattering amplitude Eq. (\ref{eq97}) \cite{Anikin2021}
	\begin{eqnarray}
		\label{eq96ver2}
		U^{\mathrm{abs}}_{ni}=e^2\frac{2\pi\sqrt{\omega_{1}\omega_{2}}}{E_{i}+\omega_1+\omega_2-E_{n}}
		\sum\limits_{k}
		\left(
		\frac{
			\langle  n |
			\boldsymbol{e}_1\boldsymbol{r}
			| k \rangle
			\langle k |
			\boldsymbol{e}_2\boldsymbol{r}
			| i \rangle
		}
		{E_{n} - E_{k} - \omega_1}
		+
		\frac{
			\langle  n |
			\boldsymbol{e}_2\boldsymbol{r}
			| k \rangle
			\langle k |
			\boldsymbol{e}_1\boldsymbol{r}
			| i \rangle
		}
		{E_{i} - E_{k} + \omega_1}
		\right).
	\end{eqnarray}
	The formula (\ref{eq96ver2}) is written in a general form, but with a discarded factor corresponding to the de-excitation process. Within the resonance approximation this is justified, since the radiation matrix element enters the amplitude (\ref{eq96ver2}) as a common factor \cite{ANDREEV2008135}. It should also be noted that in experiments of this type, the directions and (or) polarizations of absorbed photons are fixed.
	
	Omitting for brevity the intermediate calculations involving integration over angles and summation over projections, each term in (\ref{eq96ver2}) can be reduced to \cite{Anikin2021}
	\begin{eqnarray}
		\label{eq5ver2}
		\sum\limits_{k}\frac{\langle a| \boldsymbol{e}_2\boldsymbol{r}|k\rangle\langle k|\boldsymbol{e}_1\boldsymbol{r}|i\rangle}{E_i+\omega_1-E_k(1-i0)} 
		= (- 1)^{l_k+l_i+j_a+2j_k+F_a+j_i+F_k} \Pi_{l_i}\Pi_{l_k}\Pi_{j_i}\Pi_{j_a}\Pi_{j_k}
		\\\nonumber
		\times
		\Pi_{F_{k}}\Pi_{F_i}C^{l_a0}_{l_k010}C^{l_k0}_{l_i010}\left\{
		\begin{array}{ccc}
			l_k & s & j_k\\
			j_a & 1 & l_a
		\end{array}\right\}
		\left\{
		\begin{array}{ccc}
			l_i & s & j_i\\
			j_k & 1 & l_k
		\end{array}\right\}
		\left\{
		\begin{array}{ccc}
			j_k & I & F_k\\
			F_a & 1 & j_a
		\end{array}\right\}
		\left\{
		\begin{array}{ccc}
			j_i & I & F_i\\
			F_k & 1 & j_k
		\end{array}\right\} 
		\\
		\nonumber
		\times\sum\limits_{q_1,q_2}(-1)^{q_1+q_2}C^{F_aM_a}_{F_kM_k1-q_1}C^{F_kM_k}_{F_iM_i1-q_2}e_{1_{q_1} }e_{2_{q_2}}g_{l_k}(E_i+\omega).
	\end{eqnarray}
	Here summation over $k$ on the left side of the expression means all necessary summations over quantum numbers that do not appear on the right side of the expression, $e_{1(2)_q}$ are the spherical components of the photon polarization vectors.
	
	The differential absorption probability can be obtained using the relation 	$dW_{ai}^{\mathrm{abs}} = \frac{d^3\boldsymbol{k}_1}{(2\pi)^3}\frac{d^3\boldsymbol{k}_2}{(2\pi)^3}\left|U_{ai}^{\mathrm{abs}}\right|^2$. According to \cite{Jent-NR}, the most significant nonresonant contribution arises when the fine structure of the excited levels is taken into account. Then the amplitude (\ref{4}) takes into account states with the same orbital momentum, but with the different total angular momentum (for example, $nd_{3/2}$ and $nd_{5/2}$ levels in hydrogen). Leaving only these terms \cite{Anikin2021} in the amplitude, the absorption probability can be written as:
	\begin{eqnarray}
		\label{eq6ver2}
		\frac{dW_{ai}^{\mathrm{abs}}}{d\omega d\Omega_1d\Omega_2} \sim \frac{C_a }{(2\omega-\omega_0)^2+\frac{1}{4 }\Gamma^2_a} +
		\frac{C_b}{(2\omega-\omega_0-\Delta_{fs})^2+\frac{1}{4}\Gamma^2_b}+\frac{C_{ab}}{(2\omega -\omega_0)^2+\frac{1}{4}\Gamma^2_a}
		\frac{2(2\omega-\omega_0)}{2\omega-\omega_0-\Delta_{fs}}
		.
	\end{eqnarray}
	Here $\Omega_i$, $i=1,2$ are the solid angles in the phase spaces of the incident photons, $\Gamma_{a(b)}$ is the natural line width of the corresponding state, 
	$\Delta_{fs}$ denotes the fine structure energy interval, $\omega_0 = E_a-E_i$. Coefficients $C_a$, $C_b$ and $C_{ab}$ should be calculated according to expressions (\ref{eq96ver2}), (\ref{eq5ver2}). The first two terms are the resonant line profiles for two neighboring transitions, and the third term represents the interference contribution.
	
	Considering the interfering $2s^{F=1}_{1/2}\rightarrow nd^{F=2}_{3/2}$ and $2s^{F=1}_{1/2}\rightarrow nd^{F=2}_{5/2}$ transitions with $n=4,6,8,12$, so $\Delta = E_{nd^{F=2}_{3/2}} - E_{nd^{F=2}_{5/2}}$ and again assuming that the natural line widths of the $r$ and $r'$ levels are approximately equal to each other, denoting them as $\Gamma_{nd}$, the results of the calculated NR corrections are given in Table~\ref{gamH}.
	\begin{table}[hbtp]
		\caption{Nonresonant corrections (fourth column) in Hz for interfering transitions $2s^{F=1}_{1/2}\rightarrow nd^{F=2}_{3/2}$ and $2s^{F=1}_{1/2}\rightarrow nd^{F=2}_{5/2}$ with $n=4,\,6,\,8,\,12$. Fine structure splitting energies in Hz are given in the second column, natural line widths in Hz are listed in the third column.
		}
		\begin{center}
			\begin{tabular}{c | c | c | c  }
				\hline
				\hline
				state & $\Delta_{fs}$, Hz & $\Gamma_{nd}$, Hz & $\delta_{\mathrm{NR}}$
				\\
				\hline
				
				$4d$ & $4.557026\times 10^8$ & $4.40503\times 10^6$ & $-8691.82$ 
				\\
				
				$6d$ & $1.350231\times 10^8$ & $1.33682\times 10^6$ & $-2701.67$ 
				\\
				
				$8d$ & $5.69628\times 10^7$ & $5.72382\times 10^5$ & $-1174.02$ 
				\\
				
				$12d$ & $1.68779\times 10^7$ & $1.72261\times 10^5$ & $-358.88$ 
				\\
				
				\hline\hline
			\end{tabular}
		\end{center}
		\label{gamH}
	\end{table}
	
	As can be seen from Table \ref{gamH}, the NR corrections are of the order of experimental uncertainty, see \cite{Mohr,Mohr-2016} and decrease with increasing principal quantum numbers $n$. Based on the results of the previous section, we can also conclude that the NR corrections for this type of experiment are larger than for the same transitions in experiments where the $nd-2p$ fluorescent signal is detected, see Figs. \ref{2sns_fig2}-\ref{fig_total2snd}. It is important to point out that the recent new experimental measurement of the $2s_{1/2}-8d_{5/2}$ transition frequency is based on the second type of experiment discussed above. As stated in \cite{matveevPRLnew}, the QIE is negligible. However, we find that NR is $-1174.02$ Hz and has an experimental error order of 2 kHz \cite{matveevPRLnew}. A simple analysis of the involvement of NR corrections, listed in Table \ref{gamH}, in determining the proton charge radius and the Rydberg constant can be found in \cite{Anikin2021}.

	\subsection{Two-photon spectroscopy of helium}
	\label{twophotonHe}
	
	Significant progress in the spectroscopy of one-electron systems has spurred on the study of nonresonant corrections to the transition energies in many-electron systems \cite{PhysRevA.92.022514,PhysRevLett.107.023001,PhysRevA.87.032504,doi:10.1063/1.4922796}. Although helium has been studied theoretically and experimentally for many years, accounting for the nonresonant effect \cite{Jent-NR} and the QIE as a part of them in spectroscopic measurements of the transition frequencies has not been considered until recently \cite{PhysRevA.92.022514,PhysRevA.87.032504}.
	
	The energies of atomic levels in helium are conventionally expressed as the sum of nonrelativistic energies, lowest-order relativistic corrections, Lamb shift, etc., which includes quantum electrodynamics corrections (QED) and higher-order relativistic terms. Recent calculations of QED effects at the $ \alpha^7m $ level have improved theoretical predictions for the energies of helium atomic levels, leading to complete agreement with the measured $2^3S -2^3P$ transition frequency \cite{PhysRevA.103.042809}. However, as found in \cite{PhysRevA.103.042809}, such calculations do not eliminate the discrepancy between theoretical predictions and the experimental result for the $2^3S_{1} -3^3D_{1}$ transition discussed earlier in \cite{yerokhin2020}. 
	
	Using the results of previous sections we analyse here the nonresonant effects arising from neighbouring fine structure sublevels, when measuring the energy of the $2^3S_{1} -3^3D_{1}$ transition in the helium atom. In the experiment reported in \cite{PhysRevLett.78.3658}, helium atoms in an atomic beam are prepared in metastable $ 2^3S_{1} $ state and then excited into the $ 3^3D_{1} $ state by absorbing two photons with equal frequencies, $ \omega_{1}=\omega_{2} $ having parallel polarizations $ \textbf{e}_1 $ and $ \textbf{e}_2 $, and propagating in opposite directions \cite{Schwob}. Detection of the excited fraction of $ 3^3D_{1} $ atoms is observed by fluorescence (i.e., decay into $ 2^3P $  states) with the emission of a photon with frequency $ \omega_{3} $, polarization $ \textbf{e}_3^* $ in the direction of $ \boldsymbol{n}_{k_3} $. Accordingly, interference should occur between the sublevels of the fine structure $ 3^3D_{1} $, $ 3^3D_{2} $ and $ 3^3D_{3} $. This experimental situation is similar to that previously discussed in section \ref{twophotonHexamples} and was recently investigated in \cite{PhysRevA.103.022833}. 
	Therefore, following the experimental setting \cite{PhysRevLett.78.3658}, one should calculate the nonresonant corrections due to interference between different fine sublevels, $3^3D_{J_{n}}$, for the scattering process $2^3S_{1}+2\gamma(E1)\rightarrow 3^3D_{J_{n}}\rightarrow 2^3P_{J_{f}}+\gamma(\mathrm{E1})$, where $ J_{n}=1,\,2,\,3 $ and the frequency of absorbed photons $ \omega_{1}=\omega_{2} = (E_{3^3D_{J_{n}}}-E_{2^3S_{1}})/2$.
	
	
	Repeating the derivations presented in the previous sections for the helium atom, one can find similar expressions for the nonresonant correction. Then, substituting the values of the natural line width $ \Gamma_{3^3D_1}= 11.35(6)$ MHz \cite{1991} and the corresponding energy differences $\Delta_{12} =1325.025(33) $ MHz and $ \Delta_{13} =  1400.290(33)$ MHz \cite{yerokhin2020} in Eq. (\ref{eq104}), the NR correction to the transition frequency $2^3S_{1} -3^3D_{1}$ is
	\begin{eqnarray}
		\label{nr2}
		\delta_{\mathrm{NR}}=0.0124(4)\,\,\, {\mathrm{MHz}}.
	\end{eqnarray}
	
	An analytical evaluation of Eqs. (\ref{eq101})-(\ref{eq104}), shows that similarly to the NR correction to the $ 2s_{1/2}^{F=1}\rightarrow nd_{3/2(5/2)}^{F=2} $ (see Eqs. (\ref{e4})-(\ref{e12})) transition frequencies in hydrogen \cite{PhysRevA.103.022833,Anikin2021}, the considered correction also does not depend on the angles between any pair of the vectors $ \boldsymbol{n}_{k_3} $, $\boldsymbol{e}_1 $ and $ \boldsymbol{e}_2 $. Therefore, the asymmetry of observed line profile cannot be eliminated by choosing "magic angles" or experimental geometry as in \cite{Science,PhysRevA.90.012512}. An important outcome of these calculations, involving the natural level width, is that the magnitude of effect is at the level of experimental uncertainty $0.056$ MHz \cite{PhysRevLett.78.3658}. 
	
	However, the actual experimental width $ \Gamma^{\mathrm{exp}} $ of the observed profile differs significantly from the natural one $ \Gamma^{\mathrm{nat}} $ due to different broadening mechanisms \cite{riehle}. In fact, the level width in Eq. (\ref{eq104}) have to be associated with an experimental value \cite{deB-2}. In \cite{PhysRevLett.78.3658}, the main broadening effects are due to pressure and transit time. Denoting two latter contributions as  $ \Gamma^{\mathrm{pb}} $ and $ \Gamma^{\mathrm{tt}} $, respectively, the full width at half maximum in the experiment \cite{PhysRevLett.78.3658} can be expressed as a sum of three contributions  
	\begin{eqnarray}
		\label{gamexp}
		\Gamma^{\mathrm{exp}}=\Gamma^{\mathrm{nat}}+\Gamma^{\mathrm{pb}}+\Gamma^{\mathrm{tt}},
	\end{eqnarray}
	where $\Gamma^{\mathrm{nat}}$ denotes the natural level width.
	
	According to \cite{PhysRevLett.78.3658}, the pressure broadening is parametrized as $ \Gamma^{\mathrm{pb}}/p=35.7(1.7)\;[\mathrm{MHz/Torr}] $, where $ p $ is the pressure in Torr. Typically, the absorption signal is measured at various values of $ p $ and then the result is extrapolated to zero pressure (in \cite{PhysRevLett.78.3658} the $p$ values were used in the range of $ 0.05-0.5 $ Torr). The value of transit time broadening $ \Gamma^{\mathrm{tt}} $ is not presented in the experiment \cite{PhysRevLett.78.3658}. However, we can roughly estimate $ \Gamma^{\mathrm{tt}} $ as the difference between the experimental width extrapolated to zero pressure $ \Gamma^{\mathrm{exp}}=11.33(19)$ MHz and the natural width $ \Gamma^{\mathrm{nat}}=11.26$ MHz calculated theoretically \cite{drake2006springer}: $ \Gamma^{\mathrm{tt}}=0.07(19) $ MHz. 
	Finally, for the pressures $p$ in the range from  $ 0.05$ to $0.5 $ Torr the experimental width Eq. (\ref{gamexp}) of observed profile belongs to the interval $ \Gamma^{\mathrm{exp}}\in [13.2(4),29.2(4)] $ MHz. Substitution of these values into Eq. (\ref{eq104}) leads to the NR corrections $ \delta_{\mathrm{NR}}$ to transition frequency in the range from $0.016(1)$ to $0.082(19)$ MHz. The latter value partly removes the current discrepancy between theoretical and experimental value of $2^3S_{1} -3^3D_{1}$ transition frequency which is of about $ 0.5 $ MHz \cite{PhysRevA.103.042809}.
	
\begin{figure}[hbtp]
		\caption{Extrapolations of line position for $2^3S_{1}-3^3D_{1}$ transition versus the pressure. The blue points take into account the corresponding NR corrections to the original data (red dots). The corresponding dashed blue line represents the extrapolation to zero pressure through the points  that take into account the NR effects.  Likewise, the solid red line extrapolates uncorrected points. Uncertainties are also taken from \cite{PhysRevLett.78.3658}. The frequency points take into account the second order Doppler effect, which turned out to be 8.1 KHz at T=300 K \cite{PhysRevLett.78.3658}.  The least squares approach is used for extrapolation.}
		\centering
		\includegraphics[scale=1.2]{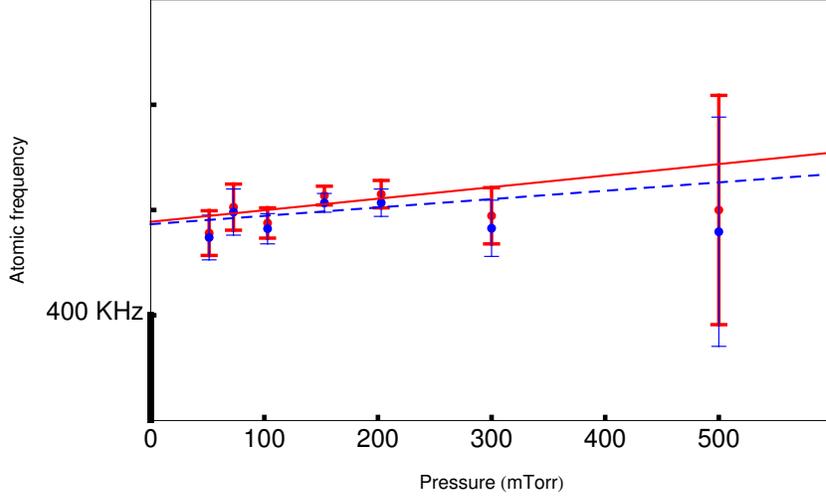}
		\label{figHe}
	\end{figure}

	The observed fluorescent signal was also fitted at a pressure of $ p=0.151 $ Torr (see Fig. 1  in \cite{PhysRevLett.78.3658}). The NR correction corresponding to $ p=0.151 $ Torr is $ \delta_{\mathrm{NR}}=0.027 $ MHz. This value still does not remove the discrepancy between theoretical calculations and experiment found in \cite{PhysRevA.103.042809}, however reaches the level of experimental uncertainty $E^{\mathrm{exp}}(2^3S_{1}-3^3D_{1})=786\,823\,850.002(56) $ MHz \cite{PhysRevLett.78.3658}. It is important to note that the resulting experimental value of $2^3S_1-3^3D_1$ transition frequency in \cite{PhysRevLett.78.3658} was obtained by extrapolating the line position to zero pressure, i.e. to the case when $ \Gamma^{\mathrm{exp}} $ approximately equal $ \Gamma^{\mathrm{nat}} $. Then repairing the center of the line by NR correction at each value of pressure in accordance with Eq. (\ref{omegamax}), the result is expected to be different (see Fig. 2b in \cite{PhysRevLett.78.3658}), eliminating the current discrepancy with theory, at least in part. We schematically reproduce fit given in \cite{PhysRevLett.78.3658} and draw a new fit accounting for transition frequency corrected to $ \delta_{\mathrm{NR}}$ at each value of pressure according to $\omega'_{0}=\omega_{0}-\delta_{\mathrm{NR}} $, see Fig. \ref{figHe} (blue points and corresponding line).
	The result of such extrapolation to zero pressure for shifted value $\omega'_{0}$ is, however, on the level of experimental uncertainty. Subtracting $\delta_{\mathrm{NR}}$ from experimental points $ \omega_{0} $ in Fig. \ref{figHe} (blue points) we proceed from the fact that in experiment \cite{PhysRevLett.78.3658} the nonresonant corrections were not taken into account, which means that they are included in the measured value (red points in Fig. \ref{figHe} ) and do not correspond exactly to the theoretical one \cite{PhysRevA.103.042809}. 
	
	Similar calculations of NR corrections can be done for the measurement of $2^3S_1-4^3D_1$ and $2^3S_1-5^3D_1$ transition frequencies reported in \cite{PhysRevLett.50.328}. These experiments were carried out using the same technique as \cite{PhysRevLett.78.3658} resulting to the values $ \Delta E^{\mathrm{exp}}(2^3S_{1}-4^3D_{1})=947\,000\,197.11(1.8)  $ MHz and $ \Delta E^{\mathrm{exp}}(2^3S_{1}-5^3D_{1})=102\,112\,869\,7.31(2.4)  $ MHz.  The pressure-broadening coefficients of $2^3S_1-4^3D_1$ and $2^3S_1-5^3D_1$ lines were determined as $ \Gamma^{\mathrm{pb}}/p=68.1(2.7)[\mathrm{MHz/Torr}] $ and $ \Gamma^{\mathrm{pb}}/p=78.5(2.7)[\mathrm{MHz/Torr}] $, respectively (see Table III in \cite{PhysRevLett.50.328}), while the transit-time effects are supposed to be negligible. Then using the values of the corresponding fine structure intervals $ \Delta_{12}=E_{4^3D_{1}}-E_{4^3D_{2}}=555.231(7) $ MHz, $\Delta_{13}=E_{4^3D_{1}}-E_{4^3D_{3}}=591.253(6)$ MHz \cite{doi:10.1139/p06-009}, and the natural level width $ \Gamma_{4^3D_{1}}^{\mathrm{nat}}=4.96274 $ MHz \cite{Theodosiou}, we find the NR correction to the transition frequency $2^3S_1-4^3D_1$ in the range from $0.350(23)$ to $2.65(2)$ MHz for pressure $ p=0.5-1.5 $ Torr. The indicated corrections are at the level or exceed the experimental uncertainty for $ 2^3S_{1}-4^3D_{1} $ transition frequency which is $ 1.8 $ MHz. Likewise, the NR correction values for the $ 2^3S_{1}-5^3D_{1} $ transition frequency start at $0.79(5)$ and end at $6.6(4)$ MHz for the energy intervals $ \Delta_{12}=E_{5^3D_{1}}-E_{5^3D_{2}}=283.560(8)  $ MHz, $\Delta_{13}=E_{5^3D_{1}}-E_{5^3D_{3}}=302.781(8) $ MHz \cite{doi:10.1139/p06-009}, and the natural level width $ \Gamma_{5^3D_{1}}^{\mathrm{nat}}=2.61381  $ MHz \cite{Theodosiou}. The results for all considered examples are summarized in Table \ref{tab1HeHe}. 
	\begin{center}
		\begin{table}[h]
			\caption{Range of nonresonant corrections to $ 2^3S_{1}-n^3D_{1} $ ($ n=3,\;4,\;5 $) (5th column) transition frequencies (2nd column) at different ranges of experimental transition width (4th column). All values are in MHz. Uncertainties are given in brackets.}
			\begin{tabular}{c c c c l}
				\hline
				\hline
				Transition & Experiment, MHz \cite{PhysRevLett.78.3658,PhysRevLett.50.328} & Theory, MHz \cite{yerokhin2020,PhysRevA.103.042809} & Exp. widths $ \Gamma^{\mathrm{exp}} $, MHz & $ \delta_{\mathrm{NR}} $, MHz \\
				\hline
				$ 2^3S_{1}-3^3D_{1} $ & $ 786\,823\,850.002(56)    $ & $ 786\,823\,849.540(57) $ & $ 13.2(4)-29.2(4) $ & $ 0.016(1)-0.082(19) $\\
				$ 2^3S_{1}-4^3D_{1} $ & $ 947\,000\,197.11(1.8)    $ & $947\,000\,194.44(5) $ & $ 39(1.4)-107(4)$ & $ 0.350(23)-2.65(2) $\\
				$ 2^3S_{1}-5^3D_{1} $ & $ 102\,112\,869\,7.31(2.4) $ & $ 102\,112\,869\,8.36(5) $ & $ 41.9(1.4)-120(4)$ & $ 0.79(5)-6.6(4) $\\
				\hline
			\end{tabular}
			\label{tab1HeHe}
		\end{table}
	\end{center}
	
	Despite the fact that the presented analysis cannot completely eliminate the current discrepancies with the theoretical predictions of the transition frequencies, it is necessary to accurately take into account nonresonant corrections in experiments like \cite{PhysRevLett.78.3658, PhysRevLett.50.328}. Then we can expect that the reconciliation between experiment and theory is most likely related to the issue of more accurate measurement and appropriate fitting, see Fig.~\ref{figHe}, or precise QED calculations of unaccounted contributions.

	\subsection{Effect of thermal line broadening}
	\label{BBR}
	
	In view of the discussion in the previous section \ref{twophotonHe}, the observed line width plays a decisive role in determining the transition frequency. This conclusion follows directly from the expression for the nonresonant correction, Eq. (\ref{eq104}). The proportionality of the NR correction to $\Gamma^2$ makes it sensitive to the width of the spectral line observed in the experiment. There are a number of contributions leading to the spectral line broadening, e.g., transit time, pressure, residual electric and magnetic fields, light shift, photo-ionisation and saturation, collision and motion effects, etc. In principle, all these effects require the use of a particular line profile, but in the simplest way such consideration can be reduced to the profile width represented by the sum of the respective contributions, see Eq. (\ref{gamexp}). So, diminishing the role of mentioned effects can make it possible to more accurately determine the transition frequencies by narrowing the observed spectral bands to lines with natural widths, but is accompanied by a significant complication of the experiment. Another obvious manner to avoid the corresponding error in comparative analysis is to make appropriate theoretical calculations of these "tiny" effects.
	
	In this section, we turn our attention to one more line broadening effect, which is also well known and has been considered by many authors. It is the thermal induced broadening. The theoretical description of this effects is reduced to the calculation of the stimulated transition rates induced by blackbody radiation (BBR). The latter can be performed in the nonrelativistic limit and in the dipole approximation with the following equation (see, e.g., \cite{PhysRevA.23.2397,PhysRevA.92.022508,solovyev2020thermal}):
	\begin{eqnarray}
		\label{eq142}
		\Gamma^{\beta}_a=\frac{4}{3}e^2\sum_n\big{|} \langle a | \boldsymbol{r} | n \rangle \big{|}^2 n_{\beta}\big{(} \omega_{an} \big{)} \omega^3_{an},
	\end{eqnarray}
	where $\omega_{ab}=E_a - E_b$ is the energy difference between the atomic states $a$ and $b$ (resonant transition frequency), $k_B$ is the Boltzmann constant, $T$ is the temperature in kelvin, $n_{\beta}(\omega)$ is the Planck's distribution function
	\begin{eqnarray}
		\label{eq143}
		n_{\beta}(\omega)=\frac{1}{e^{\frac{\omega}{k_BT}}-1}.
	\end{eqnarray}
	Summation in (\ref{eq142}) runs over all spectrum of Schr\"odinger equation. At low temperatures, the partial bound-free contributions in Eq. (\ref{eq143}) are insignificant. At room temperature, the numerical values of the BBR-induced widths for $ns/nd$ ($n=4,\,6,\,8,\,12$) levels in hydrogen are shown in Table \ref{tab1q}, where for comparison natural widths are also given.
	\begin{table}[H]
		\begin{center}
			\caption{BBR-induced and natural line widths for $ns/nd$ states in the hydrogen atom at $T=300$ K. All values are in Hz. The summation over $n$ in (\ref{eq142}) is only limited by the discrete spectrum and $n=300$, which is justified by the sufficiently small coefficient $k_BT\approx 9.5\times 10^{-4} $ in atomic units and the fast convergence of matrix elements (like $1/n^3$ for squared). }
			\begin{tabular}{c|c|c|c|c|c|c|c|c}
				\hline\hline
				\multirow{2}{*}{n} & \multicolumn{4}{c|}{$ns$}                                                             & \multicolumn{4}{c}{$nd$}                                                                                   \\ \cline{2-9} 
				& $\Gamma^{\beta}$, \cite{PhysRevA.23.2397} & $\Gamma^{\beta}$ & $\Gamma^{\mathrm{nat}}$ & $x=2\Gamma^{\beta}/\Gamma^{\mathrm{nat}}$ & $\Gamma^{\beta}$, \cite{PhysRevA.23.2397} & $\Gamma^{\beta}$ & $\Gamma^{\mathrm{nat}}$   & $x=2\Gamma^{\beta}/\Gamma^{\mathrm{nat}}$                    \\ \hline
				4                  & 2.54966                                        & 2.53778          & $7.03\times 10^5$ & $7.22\times 10^{-6}$      & 4.30037                                        & 4.2809           & $4.41\times 10^6$ & $1.94\times 10^{-6}$\\
				6                  & 1120.13                                        & 1119.48          & $2.98\times 10^5$  & $0.075$       & 1530.75                                        & 1529.73          & $1.34\times 10^6$  & $0.0023$ \\
				8                  & 4036.17                                        & 4036.93          & $1.44\times 10^5$  &  $0.056$    & 5026.11                                        & 5027.23          & $5.72\times 10^5$   &   $0.018$                        \\
				12                 & 5718.44                                        & 5721.15          & $4.77\times 10^4$  &   $0.24$   & 6434.63                                        & 6437.0            & $1.72 \times 10^5$    &    $0.075$                      \\ \hline\hline
			\end{tabular}
			\label{tab1q}
		\end{center}
	\end{table}
	The effect of thermal line broadening on the NR corrections can be taken into account by substituting the total widths of atomic level, $\Gamma_{a}^{\mathrm{tot}}\equiv \Gamma^{\beta}_{a} + \Gamma^{\mathrm{nat}}_a$, into Eq. (\ref{eq104}) for the resonance state $a$. Since $\Gamma^{\beta}$ is smaller than $\Gamma^{\mathrm{nat}}$, the order of magnitude of the thermal induced nonresonant correction can be estimated as the ratio $x=2\Gamma^{\beta}/\Gamma^{\mathrm{nat}}$ with respect to the value given by the natural level width. The values of $x$ are given in Table \ref{tab1q} and can be used by multiplying the results of the previous sections by this factor. In particular, it follows from the values of $x$ that the thermal induced broadening becomes significant for highly excited $ns$ states, leading to an additional contribution at the level of a quarter of the "natural" one.
	
	
	\section{Conclusions and outlook}
	\label{theendfinal}
	
	Modern trends in atomic spectroscopy lead us to believe that experimental observations and theoretical studies of the spectral line profile will play an increasingly important role in the coming decades. With a detailed introduction (section \ref{intro}), the authors intend to provide the reader with the actual situation, the areas of validity, and the mutual connections between theoretical QED approaches and precise spectroscopic experiments that can help them in realizing the role of the spectral line profile approach in the framework of bound-state QED. We have shown that the asymmetry of the line profile, closely related to the concept of atomic levels characterized by the energy and width of the atomic state, becomes insufficient when the distortions caused by nonresonant effects can no longer be neglected. Under these conditions, only the line profile remains as a quantum mechanical quantity that can be measured and compared with experimental data.
	
	In this review, we have investigated the asymmetry of the line profile in precision one- and two-photon spectroscopy of hydrogen and helium atoms within the framework of a rigorous QED approach. A detailed analysis of the angular correlations of the quantum interference effect has been carried out using various examples. Nonresonant effects are also considered in relation to some astrophysical problems. In particular, a rigorous QED derivation of the nonresonant extension for the Lorentz line profile is given using the Ly$_{\alpha}$ transition as an example; such a QED derivation has been lacking in the literature.
	
	\section{Acknowledgements}
	This work was supported by Russian Science Foundation (Grant No. 20-72-00003). The results obtained in section \ref{BBR} were supported by Russian Foundation for Basic Research (Grant No. 20-02-00111).

	\appendix

\begin{appendices}
\renewcommand{\theequation}{A\arabic{equation}}
\setcounter{equation}{0}
		\section{Scattering amplitude in nonrelativistic limit and dipole approximation}
	\label{appendix1}
	In this Appendix, we present a derivation of nonrelativistic equation for the one-photon scattering amplitude given by the fully relativistic result Eq.(\ref{eq5}). Following \cite{A-B-QED} we consider first a part of Eq. (\ref{eq5}) with negative-energy intermediate states 
	\begin{align}
		\label{eqA1}
		U^{(-)}_{fi}=\frac{2\pi e^2}{\sqrt{\omega_1\omega_2}}\Bigg{[} \sum_{E_n<0}
		\frac{
			\Big{(} \boldsymbol{\alpha}\boldsymbol{e}_1^* e^{i\boldsymbol{k}_1\boldsymbol{r}} \Big{)}_{fn} \Big{(} \boldsymbol{\alpha}\boldsymbol{e}_2 e^{-i\boldsymbol{k}_2\boldsymbol{r}} \Big{)}_{ni}}{E_n(1-i0) - E_i - \omega_1} + \sum_{E_n<0} 
		\frac{
			\Big{(} 
			\boldsymbol{\alpha}\boldsymbol{e}_2 e^{-i\boldsymbol{k}_2\boldsymbol{r}} 
			\Big{)}_{fn} 
			\Big{(} 
			\boldsymbol{\alpha}
			\boldsymbol{e}_1^*
			e^{i\boldsymbol{k}_1\boldsymbol{r}} \Big{)}_{ni}}{E_n(1-i0) - E_f + \omega_1} 
		\Bigg{]}.
	\end{align}
It is convenient to write energies of positive (+) and negative (-) defined energy states as $E^{(\pm)}=\pm (m+\varepsilon)$, where $\varepsilon$ is the binding energy in the field of the nucleus. As well as the transition frequency, parameter $\varepsilon$ is of order $m(\alpha Z)^2$ (in relativistic units).  

Then taking into account that initial and final states of the process under consideration being positive-energy states, the following estimations are valid 
	\begin{align}
			\label{eqA2}
			\frac{1}{E^{(-)}_n(1-i0) - E^{(+)}_i - \omega_1} = \mathcal{O}\Big{(}\frac{1}{2m}\Big{)},
		\end{align}
	\begin{align}
			\label{eqA3}
			\frac{1}{E^{(-)}_n(1-i0) - E^{(+)}_f + \omega_1} = \mathcal{O}\Big{(}\frac{1}{2m}\Big{)},
		\end{align}
\begin{align}
	\label{eqA4}
	\psi^{(-)}_n(\boldsymbol{r})=\frac{m-\hat{H}}{m-E^{(-)}_n}\psi^{(-)}_n(\boldsymbol{r})\approx\frac{m-\hat{H}}{2m}\psi^{(-)}_n(\boldsymbol{r}),
\end{align}
where $\hat{H}$ is the electron Hamiltonian. In the approximation we are interested $\hat{H}=\beta m$,  $\hat{H}\psi^{(\pm)}_n(\boldsymbol{r})=\pm\mathcal{O}(m)\psi^{(\pm)}_n(\boldsymbol{r})$, where $\psi^{(+)}_n$ is a wave function of positive energy state Eq. (\ref{eqA4}) can be reduced to
	\begin{align}
	\label{eqA5}
	\frac{m-\hat{H}}{2m}\psi^{(-)}_n(\boldsymbol{r})\approx\psi^{(-)}_n(\boldsymbol{r}),
\end{align}
\begin{align}
	\label{eqA6}
	\frac{m-\hat{H}}{2m}\psi^{(+)}_n(\boldsymbol{r})\approx0.
\end{align}
Then summation over $E_n <0$ in (\ref{eqA1}) can be extended over the states with $E_n >0$ as well. Taking into account (\ref{eqA2}), (\ref{eqA3}), (\ref{eqA5}) and (\ref{eqA6}) one can rewrite (\ref{eqA1}) in the form 
\begin{align}
	\label{eqA7}
	U^{(-)}_{fi}=-\frac{\pi e^2}{2m\sqrt{\omega_1\omega_2}}\sum_n\Big{[} \langle f| \boldsymbol{\alpha}\boldsymbol{e}_1^*e^{i\boldsymbol{k}_1\boldsymbol{r}}(1-\beta)|n\rangle \langle n| \boldsymbol{\alpha}\boldsymbol{e}_2e^{-i\boldsymbol{k}_2\boldsymbol{r}} |i\rangle + \langle f| \boldsymbol{\alpha}\boldsymbol{e}_2e^{-i\boldsymbol{k}_2\boldsymbol{r}}(1-\beta)|n\rangle \langle n| \boldsymbol{\alpha}\boldsymbol{e}_1^*e^{i\boldsymbol{k}_1\boldsymbol{r}} |i\rangle \Big{]}.
\end{align}
Summation in this expression runs over the entire nonrelativistic spectrum. Using the completeness condition $\sum_n |n\rangle \langle n|=1$ one obtains
\begin{align}
	\label{eqA8}
	U^{(-)}_{fi}=-\frac{\pi e^2}{2m\sqrt{\omega_1\omega_2}}\langle f| \boldsymbol{\alpha}\boldsymbol{e}_1^*e^{i\boldsymbol{k}_1\boldsymbol{r}}(1-\beta) \boldsymbol{\alpha}\boldsymbol{e}_2e^{-i\boldsymbol{k}_2\boldsymbol{r}} + \boldsymbol{\alpha}\boldsymbol{e}_2e^{-i\boldsymbol{k}_2\boldsymbol{r}}(1-\beta) \boldsymbol{\alpha}\boldsymbol{e}_1^*e^{i\boldsymbol{k}_1\boldsymbol{r}} |i\rangle.
\end{align}
Using the properties of Pauli matrices Eq. (\ref{eqA8}) can be reduced to \cite{A-B-QED}
\begin{align}
	\label{eqA9}
	U^{(-)}_{fi}=-\frac{2\pi e^2}{m\sqrt{\omega_1\omega_2}}(\boldsymbol{e}^*_1\boldsymbol{e}_2)\langle f| e^{i\boldsymbol{r}(\boldsymbol{k}_1-\boldsymbol{k}_2)} | i\rangle.
\end{align}
	
In relativistic units the characteristic scale for radius vector of the atomic electron is $|\boldsymbol{r}|=\frac{1}{m\alpha Z}$, and the photon momentum is $|\boldsymbol{k}|=\omega=\mathcal{O}(E_f - E_i)=m(\alpha Z)^2$, so that $\boldsymbol{k}\boldsymbol{r}=\alpha Z$. Then the exponent in this expression could be replaced by $1$
	\begin{align}
		\label{eqA10}
		U^{(-)}_{fi}=-\frac{2\pi e^2}{m\sqrt{\omega_1\omega_2}}(\boldsymbol{e}^*_1\boldsymbol{e}_2)\langle f| i\rangle=-\frac{2\pi e^2}{m\sqrt{\omega_1\omega_2}}(\boldsymbol{e}^*_1\boldsymbol{e}_2)\delta_{fi}.
	\end{align}
	
	Consider now a part of Eq. (\ref{eq5}) with positive-defined intermediate energy states. In this case in nonrelativistic limit for the large (upper) $\varphi$ and small (lower) component $\chi$ of electron wave function the following equality is valid
	\begin{align}
		\label{eqA11}
		\chi\approx\frac{\boldsymbol{\sigma}\boldsymbol{p}}{2m}\varphi,
	\end{align}
	see, for example \cite{A-B-QED}. Then the matrix element between two postive states $a$and $b $ is reduced to
	\begin{align}
		\label{eqA12}
		\psi^{\dag}_a \boldsymbol{\alpha}\boldsymbol{e} \psi_b = \varphi^{\dag}_a \boldsymbol{\sigma}\boldsymbol{e}\chi_b + \chi^{\dag}_a \boldsymbol{\sigma}\boldsymbol{e} \varphi_b \approx \frac{1}{2m} \varphi^{\dag}_a \left[ (\boldsymbol{\sigma}\boldsymbol{e}) \left( \boldsymbol{\sigma}\boldsymbol{p} \right) + \left( \boldsymbol{\sigma}\boldsymbol{p} \right)(\boldsymbol{\sigma}\boldsymbol{e}) \right] \varphi_b=\frac{1}{m}\varphi^{\dag}_a\boldsymbol{p}\boldsymbol{e}\varphi_b,
	\end{align}
	where the equality $\left( \boldsymbol{\sigma} \boldsymbol{a} \right)\left( \boldsymbol{\sigma} \boldsymbol{b} \right)=\left( \boldsymbol{a} \boldsymbol{b} \right) + i \boldsymbol{\sigma} \left[ \boldsymbol{a} \times \boldsymbol{b} \right]$ was used. Functions $\varphi$ here are solutions of the Pauli equation. Thus, scattering amplitude (\ref{eq5}) in nonrelativistic limit takes the form
	\begin{align}
		\label{eqA13}
		U^{(2)}_{fi}=\frac{2\pi e^2}{\sqrt{\omega_1\omega_2}}\Bigg{[}\frac{1}{m^2} \sum_n \frac{\Big{(}\boldsymbol{p}\boldsymbol{e}^*_1 \Big{)}_{fn} \Big{(} \boldsymbol{p}\boldsymbol{e}_2 \Big{)}_{ni}}{E_n(1-i0) - E_i - \omega_1} + \frac{1}{m^2}\sum_n \frac{\Big{(} \boldsymbol{p}\boldsymbol{e}_2 \Big{)}_{fn} \Big{(} \boldsymbol{p}\boldsymbol{e}^*_1 \Big{)}_{ni}}{E_n(1-i0) - E_f + \omega_1} - \frac{1}{m}(\boldsymbol{e}^*_1\boldsymbol{e}_2)\delta_{fi} \Bigg{]}.
	\end{align}
	Matrix elements in this expression are written in "velocity form". One can rewrite them in "length form" using the identity
	\begin{eqnarray}
	\label{relation}
	 \frac{\mathrm{i}}{m}(\boldsymbol{p})_{ab}=\big{(}[\boldsymbol{r},\hat{H}\big]{)}_{ab}=\left( E_b - E_a \right)(\boldsymbol{r})_{ab},
	\end{eqnarray}
	where $\hat{H}$ is the nonrelativistic Hamiltonian. To do this one have to note that the last term in square brackets of Eq. (\ref{eqA13}) can be reduced as follows
	\begin{align}
		\label{A14}
			(\boldsymbol{e}^*_1\boldsymbol{e}_2)\delta_{fi}=i\langle f| \big{[} (\boldsymbol{p}\boldsymbol{e}^*_1), (\boldsymbol{r}\boldsymbol{e}_2) \big{]}|i\rangle=m\langle f| \Big{[}\big{[} (\boldsymbol{r}\boldsymbol{e}^*_1), \hat{H} \big{]}, (\boldsymbol{r}\boldsymbol{e}_2)\Big{]}|i\rangle =
			\\
			\nonumber
			=m\sum_n \Big{[} E_n \big{(} \langle f | (\boldsymbol{r}\boldsymbol{e}^*_1)|n\rangle \langle n |(\boldsymbol{r}\boldsymbol{e}_2) |i\rangle + \langle f| (\boldsymbol{r}\boldsymbol{e}_2)|n\rangle \langle n |(\boldsymbol{r}\boldsymbol{e}^*_1) \big{)} | i \rangle \big{)} - E_i \langle f | (\boldsymbol{r}\boldsymbol{e}_2)| n \rangle \langle n |(\boldsymbol{r}\boldsymbol{e}^*_1) | i \rangle - E_f \langle f | (\boldsymbol{r}\boldsymbol{e}^*_1)|n\rangle \langle n |(\boldsymbol{r}\boldsymbol{e}_2) | i \rangle  \Big{]},
	\end{align}
	where the equality $\sum_n |n\rangle \langle n |=1$ was used. Summation in Eq. (\ref{A14}) runs over the spectrum of the Pauli equation. Then taking into account Eq. (\ref{A14}) and performing algebraic transformation the last term in Eq. (\ref{eqA13}) takes the form
	\begin{align}
		\label{eqA15}
		(\boldsymbol{e}^*_1\boldsymbol{e}_2)\delta_{fi}= \frac{1}{2}\big{(} (\boldsymbol{e}^*_1\boldsymbol{e}_2) + (\boldsymbol{e}_2\boldsymbol{e}^*_1) \big{)}\delta_{fi} =\frac{m}{2}\sum_{n}\Big{(} 2E_n - E_f - E_i \Big{)} \Big{(} \langle f| (\boldsymbol{r}\boldsymbol{e}^*_1) |n\rangle \langle n|(\boldsymbol{r}\boldsymbol{e}_2) |i\rangle + \langle f| (\boldsymbol{r}\boldsymbol{e}_2) |n\rangle \langle n|(\boldsymbol{r}\boldsymbol{e}^*_1) |i\rangle \Big{)}.
	\end{align}
	Substitution of Eqs. (\ref{relation}) and (\ref{eqA15}) in Eq. (\ref{eqA8}) yields
	\begin{align}
		\nonumber
		U^{(2)}_{fi}=-\frac{2\pi e^2}{\sqrt{\omega_1\omega_2}}\Bigg{[} \sum_n \frac{\big{(} E_f - E_n\big{)}\big{(} E_n - E_i \big{)}\Big{(}\boldsymbol{r}\boldsymbol{e}^*_1 \Big{)}_{fn} \Big{(} \boldsymbol{r}\boldsymbol{e}_2 \Big{)}_{ni}}{E_n(1-i0) - E_i - \omega_1} + \sum_n \frac{\big{(} E_f - E_n\big{)}\big{(} E_n - E_i \big{)}\Big{(} \boldsymbol{r}\boldsymbol{e}_2 \Big{)}_{fn} \Big{(} \boldsymbol{r}\boldsymbol{e}^*_1 \Big{)}_{ni}}{E_n(1-i0) - E_f + \omega_1} +
	\end{align}
	\begin{align}
		\label{eqA16}
		+\frac{1}{2}\sum_{n}\Big{(}  2E_n - E_f - E_i \Big{)} \Big{(} \langle f| (\boldsymbol{r}\boldsymbol{e}^*_1) |n\rangle \langle n|(\boldsymbol{r}\boldsymbol{e}_2) |i\rangle + \langle f| (\boldsymbol{r}\boldsymbol{e}_2) |n\rangle \langle n|(\boldsymbol{r}\boldsymbol{e}^*_1) |i\rangle \Big{)} \Bigg{]}.
	\end{align}
	Adding to this expression equality
	\begin{align}
		\label{eqA17}
		\frac{\omega_2}{2}\sum_n\Big{(} \langle f| (\boldsymbol{r}\boldsymbol{e}^*_1) |n\rangle \langle n| (\boldsymbol{r}\boldsymbol{e}_2) | i\rangle -\langle f| (\boldsymbol{r}\boldsymbol{e}_2) |n\rangle \langle n| (\boldsymbol{r}\boldsymbol{e}^*_1) | i\rangle \Big{)}=0,
	\end{align}
	and taking into account that $\omega_2 = E_i + \omega_1 - E_f$ we finally find	\begin{align}
		\label{eqA19}
		U^{(2)}_{fi}=2\pi e^2\sqrt{\omega_1\omega_2}\Bigg{[} \sum_n \frac{\Big{(}\boldsymbol{r}\boldsymbol{e}^*_1 \Big{)}_{fn} \Big{(} \boldsymbol{r}\boldsymbol{e}_2 \Big{)}_{ni}}{E_n(1-i0) - E_i - \omega_1} + \sum_n \frac{\Big{(} \boldsymbol{r}\boldsymbol{e}_2 \Big{)}_{fn} \Big{(} \boldsymbol{r}\boldsymbol{e}^*_1 \Big{)}_{ni}}{E_n(1-i0) - E_f + \omega_1}\Bigg{]},
	\end{align}
	where summation runs over positive-energy states, related to the spectrum of nonrelativistic hamiltonian.

	\renewcommand{\theequation}{B\arabic{equation}}
	\setcounter{equation}{0}
	\section{Angular algebra for the process of two-photon scattering}
	\label{ang2}
	
	Below we present the derivation of the most general expression for the photon scattering cross section on the hydrogen atom. In this expression, in addition to the fine splitting, the hyperfine structure of the atomic levels is explicitly taken into account.
	The closed-form expression for the cross section is written in terms of $ 6j $- symbols and simple radial integrals, with the angular calculations performed using the book \cite{VMK}. 
	
	Our aim is to obtain an analytical expression for the differential cross section with an explicit dependence on the angles between the polarization vectors of three photons. Another type of correlation involving the dependence on the propagation directions of the photons can be easily obtained from this expression by summing over polarization.
	
	First, we present the relations that will be useful for the subsequent derivations, see \cite{VMK}. Summation over photon polarizations can be performed with the use of the formula \cite{LabKlim}.
	\begin{eqnarray}
		\label{r1}
		\sum_{\textbf{e}}(\textbf{e}^{*}\textbf{a})(\textbf{e}\,\textbf{b})=(\boldsymbol{n}_{k}\times \bm{a})(\boldsymbol{n}_{k}\times \bm{b}),
	\end{eqnarray}
	where $ \textbf{a} $, $ \textbf{b} $ are two arbitrary vectors and $\boldsymbol{n}_{k}$ is photon propagation vector. We denote the vector components in a cyclic basis as $ (\textbf{a})_{q} $, with $ q=0,\;\pm 1 $. In general, we will use irreducible tensors $ a_{p} $ of the rank $ p $ with the components $ a_{pq} $. The first lower index denotes the rank and the second one denotes the component. The irreducible tensor $ a_{1} $ of the rank 1 with the components $ a_{1q} $ correspond to the vector $ \textbf{a} $ and cyclic vector component $ (\textbf{a})_{q} $. The vector component $ (\textbf{a})_{q} $ is equal to the tensor component $ a_{1q} $. 
	
	The scalar product of two arbitrary vectors $ \textbf{a} $ and $ \textbf{b} $ can be written in terms of cyclic components as
	\begin{eqnarray}
		\label{cyclic}
		\textbf{a}\textbf{b}=\sum\limits_{q}(-1)^qa_{q}b_{-q}
		.
	\end{eqnarray}
	The irreducible tensor product of two polarization vectors $ \textbf{e}_3 $ and $ \textbf{e}_2 $ can be expressed as follows
	\begin{eqnarray}
		\label{r2}
		\lbrace \textbf{e}_{3}^*\otimes \textbf{e}_{2} \rbrace_{x\xi}=\sum\limits_{q_3q_2}C^{x\xi}_{1q_31q_2}(e^*_{1})_{q_{3}}(e_{2})_{q_2}
		=(-1)^{\xi}\Pi_{x}
		\sum\limits_{q_1q_2}
		\begin{pmatrix}
			1 & 1 & x \\
			q_3 & q_2 & -\xi
		\end{pmatrix}
		(e^*_{3})_{q_{3}}(e_{2})_{q_2}
		,
	\end{eqnarray}
	where $ C^{LM}_{l_1m_1l_2m_2} $ is the Clebsch-Gordan coefficient.
	The irreducible tensor product $\lbrace \textbf{e}_1^*\otimes\textbf{e}_2\rbrace_{x\xi}$ satisfies the relation \cite{rappoport}:
	\begin{eqnarray}
		\label{r3}
		\lbrace \textbf{e}_3^*\otimes\textbf{e}_2\rbrace_{x\xi}^{*}=(-1)^{x-\xi}\lbrace \textbf{e}_3\otimes\textbf{e}_2^*\rbrace_{x-\xi}
		.
	\end{eqnarray}
	The scalar product of two irreducible tensors of rank $ x $ is
	\begin{eqnarray}
		\label{r4}
		\sum_{\xi}a_{x\xi}b^*_{x\xi}=(a_x\cdot b_{x})
		=(-1)^{-\xi}\sqrt{2x+1}\left\lbrace a_x\otimes b_{x}\right\rbrace_{00}
		.
	\end{eqnarray}
	
	Permutation of three first rank tensors in mixed irreducible tensor product satisfies the following relation  \cite{VMK}:
	\begin{eqnarray}
		\label{r5}
		\lbrace
		\lbrace \textbf{e}_{i}\otimes \textbf{e}_{j}^* \rbrace_{x}
		\otimes
		\textbf{e}^*_{k}
		\rbrace
		_{g}
		=(-1)^{x+1+g}\sum_{h}
		\Pi_{xh}
		\begin{Bmatrix}
			1 & 1 & x\\
			g & 1 & h 
		\end{Bmatrix}
		\lbrace
		\textbf{e}_{j}^*
		\otimes
		\lbrace \textbf{e}_{i}\otimes \textbf{e}_{k}^* \rbrace_{h}
		\rbrace
		_{g}
		=
		\sum_{h}
		(-1)^{x+h}
		\Pi_{xh}
		\begin{Bmatrix}
			1 & 1 & x\\
			g & 1 & h 
		\end{Bmatrix}
		\\\nonumber
		\times
		\lbrace
		\lbrace \textbf{e}_{i}\otimes \textbf{e}_{k}^* \rbrace_{h}
		\otimes
		\textbf{e}_{j}^*
		\rbrace
		_{g}
		.
	\end{eqnarray}
	Here indices $ i,\,j,\,k=1,\,2,\,3 $ denotes the corresponding photons. Permutation of two first rank tensors in irreducible tensor product obeys the equality
	\begin{eqnarray}
		\label{r5a}
		\lbrace \textbf{e}_{i}\otimes \textbf{e}_{j} \rbrace_{z}=(-1)^{z}\lbrace \textbf{e}_{j}\otimes \textbf{e}_{i} \rbrace_{z}
		.
	\end{eqnarray}
	
	The matrix element of the cyclic component of radius vector is given by \cite{VMK}
	\begin{eqnarray}
		\langle n'l'j'F'M_{F'}|r_{q}|nljFM_{F}\rangle=(-1)^{F'-M_{F'}}
		\begin{pmatrix}
			F' & 1 & F \\
			-M_{F'} & q & M_{F}
		\end{pmatrix}
		\langle n'l'j'F'||r||nljF\rangle
	\end{eqnarray}
	where the reduced matrix element is
	\begin{gather}
		\label{red}
		\langle n'l'j'F'||r||nljF\rangle= (-1)^{j'+j+I+l'+1/2+F}
		\Pi_{j'jF'F}
		\begin{Bmatrix}
			j' & F' & I \\
			F  & j  & 1
		\end{Bmatrix}
		\begin{Bmatrix}
			l' & j' & 1/2 \\
			j  & l  & 1
		\end{Bmatrix}
		\langle n' l' || r|| nl \rangle
		.
	\end{gather}
	Here $ I $ is the nuclear spin ($ I=1/2 $ for hydrogen atom) and
	\begin{eqnarray}
		\label{rad}
		\langle n' l' || r || nl \rangle = (-1)^{l'}\Pi_{l'l}
		\begin{pmatrix}
			l & 1 & l'\\
			0 & 0 & 0
		\end{pmatrix}
		\int_{0}^{\infty}r^3 R_{n'l'}R_{nl}dr.
	\end{eqnarray}
	In Eq. (\ref{rad}) $ R_{nl} $ denotes the radial part of hydrogen wave function.

	The nonrelativistic cross section with account of QIE is given by Eq. (\ref{eq101}) together with notations defined by Eqs. (\ref{eq102}) and (\ref{eq103}) in the main text.  For the evaluation of angular correlations it is convenient to consider the product $ T_{r}T_{r'}^* $, which reduces to the first and second terms of Eq. (\ref{eq102}) at certain values of $ r $ and $ r' $
	\begin{gather}
		\label{s1}
		\sum\limits_{M_{F_i}M_{F_f}}T_{r}T_{r'}^*
		=
		\omega^3\omega_{rf}^{3/2}\omega_{r'f}^{3/2}\left(\omega_{ri}-\omega \right)^{3/2}
		\left(\omega_{r'i}-\omega \right)^{3/2}
		\\\nonumber
		\times
		\sum\limits_{\substack{M_{F_r}M_{F_{r'}}\\M_{F_k}M_{F_{k'}}\\M_{F_i}M_{F_f}}}
		\sum\limits_{\substack{n_{k}l_kj_kF_k\\n_{k'}l_{k'}j_{k'}F_{k'}}}
		\left\lbrace
		\frac{
			P_{rk}(321)
		}
		{
			\omega_{ki}-\omega
		}
		+
		\frac{
			P_{rk}(312)
		}
		{
			\omega_{kr}+\omega
		}
		\right\rbrace
		\left\lbrace
		\frac{
			P_{r'k'}(321)
		}
		{
			\omega_{k'i}-\omega
		}
		+
		\frac{
			P_{r'k'}(312)
		}
		{
			\omega_{k'r'}+\omega
		}
		\right\rbrace^*
		,
	\end{gather}
	where 
	\begin{gather}
		\label{prk}
		P_{rk}(\lambda_{3}\lambda_{2}\lambda_{1})=
		\langle n_{f}l_{f}j_{f}F_{f}M_{F_f} | \textbf{e}_{\lambda_{3}}^*\textbf{r} | n_{r}l_{r}j_{r}F_{r}M_{F_r} \rangle
		\langle n_{r}l_{r}j_{r}F_{r}M_{F_r} | \textbf{e}_{\lambda_{2}}\textbf{r} | n_{k}l_{k}j_{k}F_{k}M_{F_k} \rangle
		\\\nonumber
		\times
		\langle n_{k}l_{k}j_{k}F_{k}M_{F_k} | \textbf{e}_{\lambda_{1}}\textbf{r}  | n_{i}l_{i}j_{i}F_{i}M_{F_i} \rangle
		.
	\end{gather}
	Then, expanding the curly braces in Eq. (\ref{s1}), we arrive at
	
	\begin{eqnarray}
		\label{s2}
		\sum\limits_{M_{F_i}M_{F_f}}T_{r}T_{r'}^*=
		\omega^3\omega_{rf}^{3/2}\omega_{r'f}^{3/2}\left(\omega_{ri}-\omega \right)^{3/2}
		\left(\omega_{r'i}-\omega \right)^{3/2}
		\sum\limits_{\substack{M_{F_r}M_{F_{r'}}\\M_{F_k}M_{F_{k'}}\\M_{F_i}M_{F_f}}}
		\sum\limits_{\substack{n_{k}l_kj_kF_k\\n_{k'}l_{k'}j_{k'}F_{k'}}}
		\\\nonumber
		\times
		\left\lbrace
		\frac{P_{rk}(321)P^*_{r'k'}(321)}
		{
			(\omega_{ki}-\omega)(\omega_{k'i}-\omega)
		}
		+
		\frac{P_{rk}(312)P^*_{r'k'}(321)}
		{
			(\omega_{kr}+\omega)(\omega_{k'i}-\omega)
		}
		+
		\frac{P_{rk}(321)P^*_{r'k'}(312)}
		{
			(\omega_{ki}-\omega)(\omega_{k'r'}-\omega)
		}
		+
		\frac{P_{rk}(312)P^*_{r'k'}(312)}
		{
			(\omega_{kr}-\omega)(\omega_{k'r'}-\omega)
		}
		\right\rbrace
		.
	\end{eqnarray}

	Summation over projections can be performed separately for each term. Using Eq. (\ref{cyclic}) and the Eckart-Wigner theorem for the first summand in curly brackets in Eq. (\ref{s2}) we obtain
	\begin{align}
		\label{sb3}
		P_{rk}(321)P^*_{r'k'}(321)=
		\sum\limits_{\substack{q_3q_2q_1\\q_3'q_2'q_1'}}
		(-1)^{\phi+F_{i}+F_{f}+F_{r}+F_{r'}}
		(-1)^{F_{k}+F_{k'}-M_{F_i}-M_{F_f}-M_{F_{r}}-M_{F_{r'}}-M_{F_{k}}-M_{F_{k'}}}
		\\\nonumber
		\times
		\begin{pmatrix}
			F_f & 1 & F_r \\
			-M_{F_f} & -q_3 & M_{F_r}
		\end{pmatrix}
		\begin{pmatrix}
			F_r & 1 & F_{k} \\
			-M_{F_r} & -q_2 & M_{F_{k}}
		\end{pmatrix}
		\begin{pmatrix}
			F_{k} & 1 & F_i \\
			-M_{F_{k}} & -q_1 & M_{F_i}
		\end{pmatrix}
		\begin{pmatrix}
			F_i & 1 & F_{k'} \\
			-M_{F_i} & -q_1' & M_{F_{k'}}
		\end{pmatrix}
		\\\nonumber
		\times
		\begin{pmatrix}
			F_{k'} & 1 & F_{r'} \\
			-M_{F_{k'}} & -q_2' & M_{F_{r'}}
		\end{pmatrix}
		\begin{pmatrix}
			F_{r'} & 1 & F_f \\
			-M_{F_{r'}} & -q_3' & M_{F_f}
		\end{pmatrix}
		(e_3^*)_{q_3}(e_2)_{q_2}(e_1)_{q_1}
		\left[(e_1^*)_{q'_1}(e_2^*)_{q'_2}(e_3)_{q'_3}\right]
		P_{\mathrm{red}}
		,
	\end{align}
	where 
		$\phi=q_1+q_2+q_3+q_1'+q_2'+q_3'$
	and $ P_{\mathrm{red}} $ denotes the product of reduced matrix elements 
	\begin{eqnarray}
		\label{pred}
		\left.
		\begin{aligned}
			P_{\mathrm{red}}=
			\langle n_{f}l_{f}j_{f}F_{f}|| r || n_{r}l_{r}j_{r}F_{r}\rangle
			\langle n_{r}l_{r}j_{r}F_{r}|| r ||n_{k}l_{k}j_{k}F_{k}\rangle
			\langle n_{k}l_{k}j_{k}F_{k}|| r || n_{i}l_{i}j_{i}F_{i}\rangle
			\\\nonumber
			\times
			\langle n_{i}l_{i}j_{i}F_{i}||r ||  n_{k}l_{k'}j_{k'}F_{k'} \rangle
			\langle n_{k'}l_{k'}j_{k'}F_{k'}|| r ||n_{r'}l_{r'}j_{r'}F_{r'}\rangle
			\langle n_{r'}l_{r'}j_{r'}F_{r'}|| r ||  n_{f}l_{f}j_{f}F_{f}\rangle
		\end{aligned}
		\right
		.
	\end{eqnarray} 
	Then summation over projections corresponding to the initial ($ M_{F_i} $), final ($ M_{F_f} $) and intermediate ($ M_{F_r} $, $ M_{F_{r'}} $, $ M_{F_k} $, $ M_{F_{k'}} $) states in Eq. (\ref{s2}) can be done using equality (see Eq. (29) page 392 in \cite{VMK}):
	\begin{align}
		\label{s3}
		\sum\limits_{\substack{\mathrm{all}\\\mathrm{projections} }}
		P_{rk}(321)P^*_{r'k'}(321)=
		\sum\limits_{\substack{q_3q_2q_1\\q_3'q_2'q_1'}}
		\sum_{\substack{xyz \\ \xi\eta\zeta}}\Pi^2_{xyz}
		(-1)^{\phi}
		\begin{pmatrix}
			1   & 1   & x \\
			q_3 & q_2 & \xi
		\end{pmatrix}
		\begin{pmatrix}
			1   & 1    & y \\
			q_1 & q'_1 & \eta
		\end{pmatrix}
		\begin{pmatrix}
			1 & 1 & z \\
			q'_2 & q'_3 & \zeta
		\end{pmatrix}
		\\\nonumber
		\times
		\begin{pmatrix}
			x & y & z \\
			-\xi & -\eta & -\zeta
		\end{pmatrix}
		\begin{Bmatrix}
			1 & x & 1 \\
			F_k & F_r & F_f
		\end{Bmatrix}
		\begin{Bmatrix}
			1 & y & 1 \\
			F_{k'} & F_i & F_k
		\end{Bmatrix}
		\begin{Bmatrix}
			1 & z & 1 \\
			F_f & F_{r'} & F_{k'}
		\end{Bmatrix}
		\begin{Bmatrix}
			x & y & z \\
			F_{k'} & F_f & F_k
		\end{Bmatrix}
		\\\nonumber
		\times
		(e_3^*)_{q_3}(e_2)_{q_2}(e_1)_{q_1}
		\left[(e_1^*)_{q'_1}(e_2^*)_{q'_2}(e_3)_{q'_3}\right]
		P_{\mathrm{red}}.
	\end{align}
	
	Remaining summation over $ q_i $ and $ q_j' $ ($ i,\,j=1,\,2,\,3 $) in Eq. (\ref{s3}) with the use of Eqs. (\ref{r2}) and (\ref{r3}) yields 
	\begin{align}
		\label{s4}
		\sum\limits_{\substack{\mathrm{all}\\\mathrm{projections} }}
		P_{rk}(321)P^*_{r'k'}(321)=
		\sum_{\substack{xyz \\ \xi\eta\zeta}}(-1)^{y-\eta}
		(-1)^{x+z+\eta}
		\Pi_{xyz} 
		\lbrace e_{3}^*\otimes e_{2} \rbrace_{x\xi}
		\lbrace e_{2}^*\otimes e_{3}\rbrace_{z\zeta}
		\lbrace e_{1}\otimes e_{1}^*\rbrace_{y-\eta}
		P_{\mathrm{red}}
		\\\nonumber
		\times
		\begin{pmatrix}
			x & z & y \\
			\xi & \zeta & -\eta
		\end{pmatrix}
		\begin{Bmatrix}
			1 & x & 1 \\
			F_k & F_r & F_f
		\end{Bmatrix}
		\begin{Bmatrix}
			1 & y & 1 \\
			F_{k'} & F_i & F_k
		\end{Bmatrix}
		\begin{Bmatrix}
			1 & z & 1 \\
			F_f & F_{n'} & F_{k'}
		\end{Bmatrix}
		\begin{Bmatrix}
			x & y & z \\
			F_{k'} & F_f & F_k
		\end{Bmatrix}.
	\end{align}
	Then, by the definition of the irreducible tensor product, the expression (\ref{s4}) recasts to
	\begin{eqnarray}
		\label{s5}
		\sum\limits_{\substack{\mathrm{all}\\\mathrm{projections} }}
		P_{rk}(321)P^*_{r'k'}(321)=
		\sum_{\substack{xyz \\ \eta}}(-1)^{y-\eta}
		\Pi_{xz}
		\begin{Bmatrix}
			1 & x & 1 \\
			F_k & F_r & F_f
		\end{Bmatrix}
		\begin{Bmatrix}
			1 & y & 1 \\
			F_{k'} & F_i & F_k
		\end{Bmatrix}
		\begin{Bmatrix}
			1 & z & 1 \\
			F_f & F_{r'} & F_{k'}
		\end{Bmatrix}
		\begin{Bmatrix}
			x & y & z \\
			F_{k'} & F_f & F_k
		\end{Bmatrix}
		\\\nonumber
		\times
		\lbrace
		\lbrace e_{3}^*\otimes e_{2} \rbrace_{x}
		\otimes
		\lbrace e_{2}^*\otimes e_{3}\rbrace_{z}
		\rbrace
		_{y\eta}
		\lbrace e_{1}\otimes e_{1}^*\rbrace_{y-\eta}
		P_{\mathrm{red}}
		.
	\end{eqnarray}
	
	According to Eq. (\ref{r4}) the sum over $ \eta $ in Eq. (\ref{s5}) could be rewritten as the scalar product of two tensors of rank $ y $
	\begin{eqnarray}
		\label{s6}
		\sum\limits_{\substack{\mathrm{all}\\\mathrm{projections} }}
		P_{rk}(321)P^*_{r'k'}(321)=
		\sum_{xyz}\Pi_{xz}
		\begin{Bmatrix}
			1 & x & 1 \\
			F_k & F_r & F_f
		\end{Bmatrix}
		\begin{Bmatrix}
			1 & y & 1 \\
			F_{k'} & F_i & F_k
		\end{Bmatrix}
		\begin{Bmatrix}
			1 & z & 1 \\
			F_f & F_{r'} & F_{k'}
		\end{Bmatrix}
		\begin{Bmatrix}
			x & y & z \\
			F_{k'} & F_f & F_k
		\end{Bmatrix}
		\\\nonumber
		\times
		\lbrace
		\lbrace e_{3}^*\otimes e_{2} \rbrace_{x}
		\otimes
		\lbrace e_{2}^*\otimes e_{3}\rbrace_{z}
		\rbrace
		_{y}
		\cdot
		\lbrace e_{1}\otimes e_{1}^*\rbrace_{y}
		P_{\mathrm{red}}
		,
	\end{eqnarray}
	or, in equivalent form,
	\begin{eqnarray}
		\label{s6a}
		\sum\limits_{\substack{\mathrm{all}\\\mathrm{projections} }}
		P_{rk}(321)P^*_{r'k'}(321)=
		\sum_{xyz}\Pi_{xyz}
		\begin{Bmatrix}
			1 & x & 1 \\
			F_k & F_r & F_f
		\end{Bmatrix}
		\begin{Bmatrix}
			1 & y & 1 \\
			F_{k'} & F_i & F_k
		\end{Bmatrix}
		\begin{Bmatrix}
			1 & z & 1 \\
			F_f & F_{r'} & F_{k'}
		\end{Bmatrix}
		\begin{Bmatrix}
			x & y & z \\
			F_{k'} & F_f & F_k
		\end{Bmatrix}
		\\\nonumber
		\times
		\lbrace
		\lbrace
		\lbrace e_{3}^*\otimes e_{2} \rbrace_{x}
		\otimes
		\lbrace e_{2}^*\otimes e_{3}\rbrace_{z}
		\rbrace
		_{y}
		\otimes
		\lbrace e_{1}\otimes e_{1}^*\rbrace_{y}
		\rbrace_{00}
		U_{xzg}
		P_{\mathrm{red}}
		.
	\end{eqnarray}
	Conversing the coupling scheme in the tensor product in Eq. (\ref{s6a}) with the use of Eq. (\ref{r5a}), we get
	\begin{eqnarray}
		\label{o1}
		\sum\limits_{\substack{\mathrm{all}\\\mathrm{projections} }}
		P_{rk}(321)P^*_{r'k'}(321)
		=
		\sum_{xyzg}(-1)^{\psi}
		\Pi_{xz}
		\Pi_{y}^2
		\begin{Bmatrix}
			x & z & y \\
			1 & 1 & g
		\end{Bmatrix}
		\begin{Bmatrix}
			1 & x & 1 \\
			F_k & F_r & F_f
		\end{Bmatrix}
		\begin{Bmatrix}
			1 & y & 1 \\
			F_{k'} & F_i & F_k
		\end{Bmatrix}
		\begin{Bmatrix}
			1 & z & 1 \\
			F_f & F_{r'} & F_{k'}
		\end{Bmatrix}
		\\\nonumber
		\times
		\begin{Bmatrix}
			x & y & z \\
			F_{k'} & F_f & F_k
		\end{Bmatrix}
		U_{xzg}
		P_{\mathrm{red}}\equiv O^{321}_{321}
		,
	\end{eqnarray}
	where $ \psi = y+g+1 $, and the tensor product $U_{xzg}$ is
		\begin{eqnarray}
			\label{Uxzg}
			U_{xzg}\equiv
			\lbrace
			\lbrace e_{3}^*\otimes e_{2} \rbrace_{x}
			\otimes
			e_{1}
			\rbrace
			_{g}
			\cdot
			\lbrace
			\lbrace e_{3}\otimes e_{2}^* \rbrace_{z}
			\otimes e_{1}^*
			\rbrace_{g}.\qquad
		\end{eqnarray}
	All the necessary information about the angular correlations is contained in the expression above. An explicit dependence on the angle between polarizations arises from the relations \cite{VMK}:
	\begin{gather}
		\label{udefmain}
		U_{001}=\frac{\cos^2{\theta_{32}}}{3},\\
		\nonumber
		U_{110}=-\frac{1}{6}\sin^2{\theta_{32}}\cos^2{\theta_{321}},\\
		\nonumber
		U_{111}=\frac{1}{4}(\cos^2{\theta_{31}} - 2\cos{\theta_{32}}\cos{\theta_{31}}\cos{\theta_{21}}
		+ \cos^2{\theta_{21}}),\\
		\nonumber
		U_{221}=\frac{1}{60}(4\cos^2{\theta_{32}} + 9\cos^2{\theta_{31}} + 9\cos^2{\theta_{21}}
		- 6\cos{\theta_{32}}\cos{\theta_{31}}\cos{\theta_{21}}),\\
		\nonumber
		U_{021}=U_{201}=\frac{1}{6\sqrt{5}}(6\cos{\theta_{32}}\cos{\theta_{31}}\cos{\theta_{21}} 
		- 2\cos^2{\theta_{32}}),\\
		\nonumber
		U_{121}=U_{211}=-\frac{1}{4\sqrt{15}}(3\cos^2{\theta_{21}} - 3\cos^2{\theta_{31}})
		,
	\end{gather}
	where $ \theta_{32} $ is the angle between vectors $ \textbf{e}_{3}$ and $ \textbf{e}_{2}$,  $ \theta_{31} $ is the angle between vectors $ \textbf{e}_{3}^*$ and $ \textbf{e}_{1}$, $ \theta_{21} $ is the angle between vectors $ \textbf{e}_{2}$ and $ \textbf{e}_{1}$ and $\theta_{321}$ is the angle between vector product $[\textbf{e}_3^*\times \textbf{e}_2]$ and $\textbf{e}_1$.
	
	In the case of two parallel polarizations of incident laser photons, when the vector $ \textbf{e}_{2}$ is parallel to the vector $ \textbf{e}_1 $ and, therefore, $ \theta_{21}=0 $ and $ \theta_{32}=\theta_{31}\equiv \theta $, the following nonzero contributions of $ U_{xzg} $ are
	\begin{eqnarray}
		\label{udef}
		\begin{aligned}
			&
			U_{001}=\frac{\cos ^2\theta }{3},\\
			&
			U_{111}=\frac{\sin ^2\theta }{4},\\
			&
			U_{221}=\frac{7 \cos ^2\theta }{60}+\frac{3}{20},\\
			&
			U_{021}=U_{201}=\frac{2 \cos ^2\theta }{3 \sqrt{5}},\\
			&
			U_{121}=U_{211}=-\frac{1}{4} \sqrt{\frac{3}{5}} \sin ^2\theta 
			.
			&
		\end{aligned}
	\end{eqnarray}
	
	Summation over projections in the remaining three terms in Eq. (\ref{s2}) can be performed in the same way. This yields
	\begin{eqnarray}
		\label{o2}
		\sum\limits_{\substack{\mathrm{all}\\\mathrm{projections} }}
		P_{rk}(321)P^*_{r'k'}(312)
		=
		\sum_{xyzgh'}(-1)^{\psi}
		\Pi_{xz}
		\Pi_{y}^2
		(-1)^{z+h'}\Pi_{zh'}
		\begin{Bmatrix}
			x & z & y \\
			1 & 1 & g
		\end{Bmatrix}
		\begin{Bmatrix}
			1 & 1 & z \\
			g & 1 & h'
		\end{Bmatrix}
		\begin{Bmatrix}
			1 & x & 1 \\
			F_k & F_r & F_f
		\end{Bmatrix}
		\\\nonumber
		\times
		\begin{Bmatrix}
			1 & y & 1 \\
			F_{k'} & F_i & F_k
		\end{Bmatrix}
		\begin{Bmatrix}
			1 & z & 1 \\
			F_f & F_{r'} & F_{k'}
		\end{Bmatrix}
		\begin{Bmatrix}
			x & y & z \\
			F_{k'} & F_f & F_k
		\end{Bmatrix}
		U_{xh'g}
		P_{\mathrm{red}}\equiv O^{321}_{312}
		,
	\end{eqnarray}
	\begin{eqnarray}
		\label{o3}
		\sum\limits_{\substack{\mathrm{all}\\\mathrm{projections} }}
		P_{rk}(312)P^*_{r'k'}(321)
		=
		\sum_{xyzgh}(-1)^{\psi}
		\Pi_{xz}
		\Pi_{y}^2
		(-1)^{x+h}\Pi_{xh}
		\begin{Bmatrix}
			x & z & y \\
			1 & 1 & g
		\end{Bmatrix}
		\begin{Bmatrix}
			1 & 1 & x \\
			g & 1 & h
		\end{Bmatrix}
		\begin{Bmatrix}
			1 & x & 1 \\
			F_k & F_r & F_f
		\end{Bmatrix}
		\begin{Bmatrix}
			1 & y & 1 \\
			F_{k'} & F_i & F_k
		\end{Bmatrix}
		\\\nonumber
		\times
		\begin{Bmatrix}
			1 & z & 1 \\
			F_f & F_{r'} & F_{k'}
		\end{Bmatrix}
		\begin{Bmatrix}
			x & y & z \\
			F_{k'} & F_f & F_k
		\end{Bmatrix}
		U_{hzg}
		P_{\mathrm{red}}\equiv O^{312}_{321}
		,
	\end{eqnarray}
	\begin{eqnarray}
		\label{o4}
		\sum\limits_{\substack{\mathrm{all}\\\mathrm{projections} }}
		P_{rk}(312)P^*_{r'k'}(312)
		=
		\sum_{xyzghh'}(-1)^{\psi}
		\Pi_{xz}
		\Pi_{y}^2
		(-1)^{z+x+h+h'}\Pi_{xzhh'}
		\begin{Bmatrix}
			x & z & y \\
			1 & 1 & g
		\end{Bmatrix}
		\begin{Bmatrix}
			1 & 1 & z \\
			g & 1 & h'
		\end{Bmatrix}
		\begin{Bmatrix}
			1 & 1 & x \\
			g & 1 & h
		\end{Bmatrix}
		\\\nonumber
		\times
		\begin{Bmatrix}
			1 & x & 1 \\
			F_k & F_r & F_f
		\end{Bmatrix}
		\begin{Bmatrix}
			1 & y & 1 \\
			F_{k'} & F_i & F_k
		\end{Bmatrix}
		\begin{Bmatrix}
			1 & z & 1 \\
			F_f & F_{r'} & F_{k'}
		\end{Bmatrix}
		\begin{Bmatrix}
			x & y & z \\
			F_{k'} & F_f & F_k
		\end{Bmatrix}
		U_{hh'g}
		P_{\mathrm{red}}\equiv O^{312}_{312}
		,
	\end{eqnarray}
	where we used Eq. (\ref{r5}). Then substitution of Eqs. (\ref{o1}), (\ref{o2}), (\ref{o3}) and (\ref{o4}) into Eq. (\ref{s2}) leads to
	\begin{gather}
		\label{theend}
		\sum\limits_{M_{F_i}M_{F_f}}T_{r}T_{r'}^*=
		\omega^3\omega_{rf}^{3/2}\omega_{r'f}^{3/2}
		\left(\omega_{ri}-\omega \right)^{3/2}
		\left(\omega_{r'i}-\omega \right)^{3/2}
		\sum\limits_{\substack{n_{k}l_kj_kF_k\\n_{k'}l_{k'}j_{k'}F_{k'}}}
		\left\lbrace
		\frac{O^{321}_{321}}
		{
			(\omega_{ki}-\omega)(\omega_{k'i}-\omega)
		}
		+
		\frac{O^{312}_{321}}
		{
			(\omega_{kr}+\omega)(\omega_{k'i}-\omega)
		}
		\right.
		\\\nonumber
		+
		\left.
		\frac{O^{321}_{312}}
		{
			(\omega_{ki}-\omega)(\omega_{k'r'}-\omega)
		}
		+
		\frac{O^{312}_{312}}
		{
			(\omega_{kr}-\omega)(\omega_{k'r'}-\omega)
		}
		\right\rbrace
		.
	\end{gather}
	
	The expression (\ref{theend}) contains all necessary angular correlations for the three-photon scattering cross section given by Eq. (\ref{eq101}) in the main text. Summing over quantum numbers $ l_kj_kF_k$ and $l_{k'}j_{k'}F_{k'} $ in (\ref{theend}) numerically, one can obtain the common angular depended factor for the interfering contributions, Eq. (\ref{theend}). This factor arises finally in the expressions for the NR corrections, see the main text.

	\renewcommand{\theequation}{C\arabic{equation}}
	\setcounter{equation}{0}
	\section{Analytical expressions for NR corrections to $2s-ns/nd$ transitions}
	\label{ang2explicit}
	
	The NR correction to $ 2s_{1/2}^{F=1}\rightarrow ns_{1/2}^{F=1} $ (with $ n=4,\,6,\,8,\,12 $) transition frequencies (see Eq. (\ref{eq104}) in the main text) occurs using Eqs. (\ref{theend}) and (\ref{eq102}). Then summing over quantum numbers $ j_{f}F_{f} $ and $ l_{k}j_{j}F_{k} $ in Eq. (\ref{eq104}), we arrive at
	
	\begin{eqnarray}
		\label{nrsum3}
		\delta_{\mathrm{NR}}(2s_{1/2}^{F=1}-ns_{1/2}^{F=1} )=
		\frac{\Gamma_{ns_{1/2}}^2}{4}
		\left(
		\frac{1}{25\Delta_{1}}
		+
		\frac{1}{25\Delta_{2}}
		+
		\frac{2}{75\Delta_{3}}
		+
		\frac{7}{75\Delta_{4}}
		\right)
		\frac{\beta_{2pnd2s}(\omega_{ns2s}/2)}{\beta_{2pns2s}(\omega_{ns2s}/2)}
		\\
		\nonumber
		\times
		\frac{10 U_{001}+\sqrt{5} U_{021}-15 U_{111}-\sqrt{15} U_{121}+10 \sqrt{5} U_{201}+5 \sqrt{15} U_{211}+5 U_{221}}{ U_{001}+\sqrt{5} U_{021}+3 U_{111}-\sqrt{15}
			U_{121}+\sqrt{5} U_{201}-\sqrt{15} U_{211}+5 U_{221}}
		,
	\end{eqnarray} 
	where functions $ U_{xyz} $ are given by Eq. (\ref{udefmain}) and $ \Delta_1=E_{ns_{1/2}^{F=1}}-E_{nd_{3/2}^{F=1}}$, $ \Delta_2=E_{ns_{1/2}^{F=1}}-E_{nd_{3/2}^{F=2}} $, $ \Delta_3=E_{ns_{1/2}^{F=1}}-E_{nd_{5/2}^{F=2}} $,  $ \Delta_4=E_{ns_{1/2}^{F=1}}-E_{nd_{5/2}^{F=3}} $ and $ \Gamma_{ns_{1/2}}$ is the level width of $ ns_{1/2} $ state (here and below we assume that $ \Gamma_{nlj}=\Gamma_{nljF} $).
	The coefficient $ \beta $ in Eq. (\ref{nrsum3}) is defined by the following equation
	\begin{eqnarray}
		\label{beta}
		\beta_{2pns(nd)2s}(\omega)=
		I_{2pns}
		\sum\limits_{k}
		\left\lbrace
		\frac{
			I_{ns(nd)kp}
			I_{kp2s}
		}
		{
			E_{kp}-E_{2s}-\omega
		}
		\right.
		+
		\left.
		\frac{
			I_{ns(nd)kp}
			I_{kp2s}
		}
		{
			E_{kp}-E_{ns(nd)}+\omega
		}
		\right\rbrace
		,
	\end{eqnarray}
	where 
	\begin{eqnarray}
		I_{n'l'nl}=\int\limits_{0}^{\infty}r^3R_{n'l'}R_{nl}dr
	\end{eqnarray}
	and $ R_{nl} $ represents the corresponding radial part of the Schr\"{o}dinger wave function. Summation over $k$ runs over all the entire spectrum including the continuum. 
	The numerical values of Eq. (\ref{beta}) calculated with the use of B-spline method \cite{dkb} are listed in Table~\ref{tab1beta}. For the particular case of parallel (anti-parallel) polarizations of incident photons, the relations Eqs. (\ref{udef}) should be used. Then, substituting Eqs. (\ref{udef}) into Eq. (\ref{nrsum3}), we obtain the NR corrections presented in Fig. \ref{2sns_fig1}.  
	
	After performing a similar calculation, the NR correction to $ 2s_{1/2}^{F=0}-ns_{1/2}^{F=0} $  ($ n=4,\,6,\,8,\,12 $) transition frequencies the following equation can be found
	
	\begin{eqnarray}
		\label{nrsum5}
		\delta_{\mathrm{NR}}(2s_{1/2}^{F=0}-ns_{1/2}^{F=0} )=
		\frac{\Gamma_{ns_{1/2}}^2}{4}
		\left(
		\frac{2}{25\Delta_{1}'}
		+
		\frac{3}{25\Delta_{2}'}
		\right)
		\frac{\beta_{2snd2p}(\omega_{ns2s}/2)}{\beta_{2sns2p}(\omega_{ns2s}/2)}
		\\
		\nonumber
		\times
		\frac{10 U_{001}+\sqrt{5} U_{021}-15 U_{111}-\sqrt{15} U_{121}+10 \sqrt{5} U_{201}+5 \sqrt{15} U_{211}+5 U_{221}}{U_{001}+\sqrt{5} U_{021}+3 U_{111}-\sqrt{15} U_{121}+\sqrt{5} U_{201}-\sqrt{15} U_{211}+5 U_{221}}
		.
	\end{eqnarray} 
	Here $ \Delta_1'=E_{ns_{1/2}^{F=0}}-E_{nd_{3/2}^{F=2}} $ and $ \Delta_2'=E_{ns_{1/2}^{F=0}}-E_{nd_{5/2}^{F=2}} $. Note that the correlation factor here coincides with Eq. (\ref{nrsum3}).
	
	The situation is different for the NR correction to the $ 2s_{1/2}^{F=1}-nd_{3/2}^{F=2} $ ($ n=4,\,6,\,8,\,12 $) transition frequencies. Performing angular algebra we find
	\begin{eqnarray}
		\label{nrsumnd2}
		\delta_{\mathrm{NR}}(2s_{1/2}^{F=1}-nd_{3/2}^{F=2})=-\frac{\Gamma_{nd_{3/2}}^2}{4\Delta''}
		\frac{1}{11}
		,
	\end{eqnarray} 
	where $ \Delta''=E_{nd_{3/2}^{F=2}}-E_{nd_{5/2}^{F=2}} $.
	This correction is independent on angles, and, therefore, can not be eliminated by choosing the geometry of the experiment, see Fig. \ref{fig_total2snd} in the main text.
	
	The remaining NR correction to $ 2s_{1/2}^{F=1}-nd_{3/2}^{F=1} $ ($ n=4,\,6,\,8,\,12 $) transition frequencies due to the neighbouring $ ns_{1/2}^{F=1} $ state is
	\begin{eqnarray}
		\label{nrsumnd3R}
		\delta_{\mathrm{NR}}(2s_{1/2}^{F=1}-nd_{3/2}^{F=1} )=
		\frac{\Gamma_{nd_{3/2}}^2}{4}
		\left(\frac{50}{33\Delta'''}
		\right)
		\frac{\beta_{2sns2p}(\omega_{2snd}/2)}{\beta_{2snd2p}(\omega_{2snd}/2)}
		\\\nonumber
		\times
		\frac{10 U_{001}+10 \sqrt{5} U_{021}-15 U_{111}+5 \sqrt{15} U_{121}+\sqrt{5} U_{201}-\sqrt{15} U_{211}+5 U_{221}}{20 U_{001}+2 \sqrt{5} U_{021}+15 U_{111}+\sqrt{15} U_{121}+2 \sqrt{5}
			U_{201}+\sqrt{15} U_{211}+U_{221}}
		,
	\end{eqnarray} 
	together with notation $ \Delta'''= E_{nd_{3/2}^{F=2}}-E_{ns_{1/2}^{F=1}}$.

	\begin{table}
		\centering
		\caption{Coefficients $ \beta $ in a.u. and level widths $ \Gamma_{nlj} $ in MHz.}
		\begin{tabular}{c c c c c}
			\hline
			\hline
			$n$ & $ \beta_{2sns2p} $ & $ \beta_{2snd2p} $ & $ \Gamma_{ns_{1/2}} $ & $ \Gamma_{nd_{3/2}} $  \\
			\hline
			4  & -38.1593 & 2449.09 & 0.70 & 4.41\\
			6  &  13.949  & 591.154 & 0.29 & 1.33\\
			8  &  8.41272 & 240.557 & 0.14 & 0.56\\
			12 &  3.10962 & 71.980 & 0.05 & 0.17\\
			\hline
			$n$ & $ \beta_{1sns2p} $ & $ \beta_{1snd2p} $ & $ \Gamma_{ns_{1/2}} $  & $ \Gamma_{nd_{3/2}} $  \\
			\hline
			3  & 187.375 & 1.005 & 1.01 &  10.30\\
			\hline
		\end{tabular} 
		\label{tab1beta}
	\end{table} 
\end{appendices}	
	
	\newpage
	\addcontentsline{toc}{section}{List of Literature}
	\bibliographystyle{ieeetr}  
	\bibliography{bib}    
	
\end{document}